\documentclass{article}
\usepackage{arxiv}
\usepackage[T1]{fontenc}
\usepackage{graphicx,amsmath,booktabs,longtable,array,caption,float,placeins,url,listings}
\usepackage[natbibapa]{apacite}
\usepackage{hyperref}
\AtBeginDocument{}
\hypersetup{colorlinks=true,linkcolor=black,citecolor=black,urlcolor=blue}
\hypersetup{
  pdftitle={From language-model stock rankings to testable economic rules: A computational audit},
  pdfauthor={Shuai Wu, Xue Li, Zhijun Wang, Bolun Liu, Weilin Cai},
  pdfsubject={Research preprint}
}
\title{From language-model stock rankings to\\testable economic rules: A computational audit}
\author{
  Shuai Wu\thanks{Corresponding author: shuai.wu@colorado.edu.\newline ORCID: Shuai Wu (0009-0007-7657-6208); Xue Li (0009-0002-7480-723X); Zhijun Wang (0009-0000-0742-6985).}\\
  \normalfont\footnotesize University of Colorado Boulder
  \And
  Xue Li\\
  \normalfont\footnotesize Beijing Union University
  \And
  Zhijun Wang\\
  \normalfont\footnotesize University of North Dakota
  \And
  Bolun Liu\\
  \normalfont\footnotesize City University of Macau
  \AND
  Weilin Cai\\
  \normalfont\footnotesize The University of Sydney
}
\date{September 2026}
\renewcommand{\shorttitle}{From Language-Model Stock Rankings to Economic Rules}
\begin{document}
\shortcites{r2,r20}
\maketitle
\begin{abstract}
We test the stability, reproducibility and investment outcomes of language-model stock rankings. Four models and five numerical comparators share a portfolio engine over 72 monthly holding periods in the Shanghai Stock Exchange (SSE) 50, China Securities Index (CSI) 300 and CSI 500. Rankings use nine characteristics, and five repeated SSE 50 runs measure variation under identical inputs. Linear rules fitted to development-period model preferences are frozen before unseen-month, larger-pool and controlled-intervention tests. Their mean Spearman agreement with model rankings is 0.923-0.984 in the SSE 50 and 0.795-0.985 after transfer. Aggregate rank-change error falls relative to a zero-change prediction in twelve archived feature-group comparisons and eight matched single-feature comparisons, with Holm adjustments applied in separate nine-plus-three and six-plus-two families. Prediction of individual entries and exits remains weak (event Jaccard 0.000-0.125). Historical mean model compound annual growth rates range from 6.26\% to 11.17\%. At 10 basis points per side and six-month blocks, the twelve-comparison model-minus-rule return family and factor-controlled associations yield no adjusted finding. Three higher-cost, twelve-month-block comparisons favor a Terra rule within their twelve-test slices, with no adjusted finding across the full 144-test sensitivity grid. Two input-intervention batches totaling 5,184 responses supply paired intervention-return tests. The three-model batch has no bootstrap-adjusted finding at the primary block length; a Luna row-order effect appears under heteroskedasticity- and autocorrelation-consistent (HAC) adjustment within its three-test family but not in a pooled 24-test adjustment. Compact rules approximate aggregate rankings; return conclusions depend on comparison families, uncertainty methods and tie priorities.
\end{abstract}
\FloatBarrier
\section{Introduction}
Language models can turn a cross-section of firm characteristics into stock rankings, but a realized return path reveals little about the choices that produced it. A model may change its rankings when given identical inputs, follow a familiar factor rule or respond unpredictably to an input change. Evaluating its investment outcomes therefore also requires testing the stability and reproducibility of its rankings.

Characteristic-based investing supplies transparent numerical benchmarks \citep{r1}, while supervised machine learning estimates flexible combinations of firm attributes \citep{r2}. Financial applications of language models study news forecasts, extrapolation from return histories and joint numerical-textual signals \citep{r3,r5,r6,r7}. Repeated recommendations also reveal substantial response variation \citep{r20}. Recent work uses interpretable surrogates to study investment advice \citep{r18} and probes internal economic representations \citep{r19}. We combine tests of repeatability, rule reproduction, controlled responses and realized investment value in one policy-level audit.

We conduct that audit with four models, three Chinese equity index universes and one trading engine. Each ranking method receives the same nine contemporaneous stock-characteristic percentiles; the eligible-universe comparator holds every candidate at equal target weight. All methods share signal dates and transaction costs and are evaluated in the same execution engine; ranked selections use a common tie rule. Five complete Shanghai Stock Exchange (SSE) 50 runs measure identical-input variation. Linear rules fitted to development-period preferences are frozen before evaluation, transferred without refitting to the China Securities Index (CSI) 300 and 500, and challenged with controlled feature interventions.

Repeated runs show how much rankings vary under identical inputs. We then test frozen rules on unseen months, new stock pools and feature changes, and use matched portfolios to measure returns associated with preferences beyond those rules. These tests assess whether a rule predicts behavior outside the observations used to fit it.

Frozen linear rules attain SSE 50 rank correlations of 0.923-0.984 and retain correlations of 0.795-0.985 in the larger stock pools. Under the primary six-month block analysis, aggregate rank-change prediction error falls relative to a zero-change prediction in all twelve feature-group comparisons, adjusted in separate nine-test and Astra three-test families. The matched experiment with 864 responses finds improvements in all eight single-feature comparisons, adjusted in separate six-test and two-test families. At the primary 10-bp-per-side cost and six-month block length, the twelve-test model-minus-rule return family and the factor-controlled associations have no adjusted finding. Three Terra effects at higher costs and twelve-month blocks are exploratory sensitivity results. The rules approximate aggregate ranking responses more closely than they predict individual stock entries and exits.

\FloatBarrier
\section{Computational investment environment}
\FloatBarrier
\subsection{Data, samples and timing}
Daily quotations, daily stock indicators, adjustment factors, historical index membership, and daily price limits come from the authorized Tushare data interface \citep{r13,r14,r15,r16,r17}. Query parameters, fields, retrieval times, response sources, and coverage checks are archived with the study. Historical industry assignments use the dated intervals in the official Shenwan classification history. These are the historical records available at retrieval; the analysis does not establish that every vendor record is an unrevised publication-time vintage.

For each signal, membership uses the latest index snapshot dated no later than the preceding calendar month end. A candidate requires a traded signal-day bar, a daily-indicator record, adjusted-price history spanning 253 exchange sessions, and at least 200 actual trading bars in that window. Eligibility uses signal-time observations and is not conditioned on a later exit trade. The monthly candidate set therefore follows recorded constituent additions, removals, and available trading history.

Development signals run from December 2014 through November 2019, and test signals from December 2019 through November 2025. The first test portfolio enters on 03 January 2020 and the final test portfolio exits on 06 January 2026. Annual reporting assigns a holding interval to its entry month. Supervised training admits a return label only when its exit date is strictly before the current signal. Development-period labels still awaiting realization are consequently excluded from the first applicable training windows.

\begingroup\small\setlength{\tabcolsep}{3pt}\renewcommand{\arraystretch}{1.17}
\begin{longtable}{@{}>{\raggedright\arraybackslash}p{\dimexpr 0.264706\linewidth-7.94118pt\relax}>{\raggedright\arraybackslash}p{\dimexpr 0.147059\linewidth-4.41176pt\relax}>{\raggedright\arraybackslash}p{\dimexpr 0.147059\linewidth-4.41176pt\relax}>{\raggedright\arraybackslash}p{\dimexpr 0.147059\linewidth-4.41176pt\relax}>{\raggedright\arraybackslash}p{\dimexpr 0.147059\linewidth-4.41176pt\relax}>{\raggedright\arraybackslash}p{\dimexpr 0.147059\linewidth-4.41176pt\relax}@{}}
\caption{Historical sample coverage}\label{tab:1}\\
\toprule
\textbf{Universe} & \textbf{Historical members} & \textbf{All stock-months} & \textbf{Test stock-months} & \textbf{Test stocks} & \textbf{Monthly test candidates}\\\midrule\endfirsthead
\multicolumn{6}{l}{\textbf{Table 1 (continued)}}\\\toprule
\textbf{Universe} & \textbf{Historical members} & \textbf{All stock-months} & \textbf{Test stock-months} & \textbf{Test stocks} & \textbf{Monthly test candidates}\\\midrule\endhead
\midrule\multicolumn{6}{r}{Continued on next page}\\\endfoot
\bottomrule\endlastfoot
SSE 50 & 129 & 6,363 & 3,563 & 97 & 47-50\\
CSI 300 & 667 & 37,562 & 21,446 & 487 & 292-300\\
CSI 500 & 1,323 & 61,262 & 35,806 & 995 & 490-500\\
\end{longtable}
\noindent\footnotesize Table 1 defines the opportunity sets. All stock-months cover 132 signals, and the historical test covers 72. Counts are reported within each universe. Stocks and calendar months can overlap across universes; thus, these rows are not independent samples.\par\endgroup
Table 1 defines the three opportunity sets. They share market months and may share constituent stocks; inference preserves the common calendar rather than treating the universes as independent replications. Model requests were made after the historical holding periods. Anonymous identifiers withheld company names, dates, and stock codes. The historical analysis is an ex post policy evaluation. The matched-intervention inputs and final inference specification were fixed before the responses were collected.

\FloatBarrier
\subsection{Inputs to the ranking methods}
Every ranking policy uses the same nine characteristics in Table 2, calculated from observations available at or before the signal close. Larger percentiles have the economic interpretations shown in the table.

\begingroup\small\setlength{\tabcolsep}{3pt}\renewcommand{\arraystretch}{1.17}
\begin{longtable}{@{}>{\raggedright\arraybackslash}p{\dimexpr 0.208333\linewidth-2.50000pt\relax}>{\raggedright\arraybackslash}p{\dimexpr 0.479167\linewidth-5.75000pt\relax}>{\raggedright\arraybackslash}p{\dimexpr 0.312500\linewidth-3.75000pt\relax}@{}}
\caption{Input characteristics}\label{tab:2}\\
\toprule
\textbf{Characteristic} & \textbf{Raw definition} & \textbf{Larger percentile represents}\\\midrule\endfirsthead
\multicolumn{3}{l}{\textbf{Table 2 (continued)}}\\\toprule
\textbf{Characteristic} & \textbf{Raw definition} & \textbf{Larger percentile represents}\\\midrule\endhead
\midrule\multicolumn{3}{r}{Continued on next page}\\\endfoot
\bottomrule\endlastfoot
Book to price & Reciprocal of positive price-to-book ratio & Higher book value per market value\\
Earnings yield & Reciprocal of nonzero trailing price-to-earnings ratio & Higher trailing earnings yield\\
Dividend yield & Trailing dividend yield from daily indicators & Higher dividend yield\\
12-1 momentum & Adjusted price at t-21 divided by price at t-252, minus 1 & Stronger intermediate momentum\\
6-1 momentum & Adjusted price at t-21 divided by price at t-126, minus 1 & Stronger intermediate momentum\\
1-month reversal & Negative adjusted return over the latest 21 sessions & Weaker recent performance\\
Low volatility & Negative 63-return sample standard deviation times sqrt(252) & Lower historical volatility\\
Small size & Negative log of floating market capitalization & Smaller floating capitalization\\
Low turnover & Negative turnover rate on the signal date & Lower current turnover\\
\end{longtable}
\noindent\footnotesize The time index denotes exchange sessions. Prices incorporate the corresponding adjustment factors. Each characteristic is measured at or before the signal close.\par\endgroup
Observed values are converted to within-month fractional ranks, with average ranks for ties. Missing characteristic values receive 0.5. Every input is rounded to four decimals. All ranking methods receive this same matrix for a given universe, month, and experimental condition. The factor transformations use the eligible cross-section before any portfolio selection.

\FloatBarrier
\subsection{Numerical comparators and model requests}
Equal-factor selection averages the nine characteristic percentiles, assigning each a weight of 1/\allowbreak{}9. The factor-category comparator first averages within valuation, momentum, reversal, volatility, size, and turnover, then assigns each of these six categories a weight of 1/\allowbreak{}6. Comparing these fixed rules distinguishes equal weighting of individual inputs from equal weighting of economic categories. The eligible-universe comparator holds every candidate at equal target weight.

Ridge regression and histogram gradient boosting predict the within-month rank of the next holding-period stock return, centered at 0.5. Both use an expanding training sample whose holding periods end strictly before the current signal. For each universe and each factor-ablation condition, hyperparameters are selected from development data and the supervised models are fitted again using the corresponding inputs.

All stock-level forward-return labels use the same zero-cost economic-wealth calculation between the entry and exit 09:30 endpoints, including share exchanges, locked entitlements and board transfers. The initial mark uses the actual opening quotation or the last earlier observed close during suspension. These labels measure the return on a held economic claim. Strategy purchases additionally require an executable opening quotation and satisfy the engine's price-limit checks. Supervised return targets, selection hit rates and factor-controlled return associations use this common stock-wealth definition.

Development-only temporal validation selects the supervised hyperparameters using mean monthly rank correlation with completed forward returns. Validation uses 2017-2019 signals and label exits no later than 30 November 2019. The parameter grid, selected fits and expanding training samples are reported in Supplementary Material. Supervised return predictors and preference-reproducing proxies have different targets and are evaluated separately.

GPT-5.6 Luna, GPT-5.6 Terra, GPT-5.6 Sol and GPT-6 Astra use the official OpenAI Responses API at https:/\allowbreak{}/\allowbreak{}api.openai.com/\allowbreak{}v1/\allowbreak{}responses. Historical rankings, matched single-feature responses and repeated input-intervention responses use medium reasoning, independent contexts and no external tools. Model names are the service labels returned by the interface. Supplementary Material records each batch's configuration and collection dates.

At each universe-month signal, a fixed seed assigns anonymous stock identifiers and determines their display order. The identifiers, numeric matrix, and display order remain identical across the repeated primary runs of each universe. Company names, real stock codes, calendar dates, and realized returns remain outside the transmitted prompt. The prompt supplies characteristic definitions and asks for one desirability score between zero and one for every candidate, expressing relative investment preference for the next monthly holding period. A structured JSON schema specifies all required candidate identifiers.

Stocks are ranked by descending score. Equal scores are resolved by ascending anonymous identifier, using the same monthly mapping for LLM and conventional top-fraction portfolios. Each prediction uses the first complete response satisfying the score schema. Score ties and selection-boundary ties are reported separately.

The response archive contains 10,104 valid responses: 3,696 historical rankings, 864 matched single-feature responses, 360 development-planning responses and 5,184 repeated input-intervention responses. The last group consists of an Astra batch with 1,296 responses and a Luna, Terra and Sol batch with 3,888. Each intervention batch covers six SSE 50 conditions with three complete runs per model and uses its own primary-condition controls and multiplicity families. The development-planning sample supplies return-precision inputs. All test portfolios share the same 72 market months.

\FloatBarrier
\subsection{Portfolio construction and execution}
For N eligible stocks, each active top-fraction strategy selects k = ceil(0.2N) stocks and assigns equal target weights. The eligible-universe comparator assigns 1/\allowbreak{}N to every candidate. Signals are formed after the last trading-day close of each month; portfolios rebalance at the second trading-day open of the following month and hold until the corresponding opening event one month later. Each complete strategy run starts with capital of one and carries its own actual holdings and cash through the sample.

The engine sells positions above target before purchasing target shortfalls. Cash finances purchases and their costs, with proportional scaling when cash is scarce. Actual A-share opening quotations are required for trades; upper-limit opens block purchases and lower-limit opens block sales. Unfilled orders leave holdings or cash in the portfolio. Each security is marked at an observed adjusted opening price, or at the last close known before that opening when an opening quotation is absent. Daily close marks use observations available by that close. Fractional positions, long-only holdings and fixed notional trading costs apply to all strategies.

The principal cost is 10 basis points per unit of bought or sold notional. Scenarios of 0, 25 and 50 basis points use the same rankings and trading rules, with costs paid from each portfolio. Terminal wealth is measured at 09:30 on 06 January 2026 after feasible A-share liquidation, with locked claims and unsold board-transfer shares retained as residual assets. That opening valuation uses only previously observed board quotations. The engine reconciles compounded monthly returns with terminal wealth and checks self-financing at each trading event. Daily marked values determine maximum drawdown.

Share exchanges preserve the investor's economic entitlement across security codes. At the record-date close, the engine converts held adjusted units to original shares using that date's adjustment factor, applies the announced exchange ratio and records a locked claim on the successor. The registration date records completion of registration; the claim becomes tradable at 09:30 on its announced availability date. Where the successor already has an observed market price, locked claims are valued using contemporaneous successor quotations. Claims on a successor awaiting listing retain the known value of the surrendered holding until a market quote becomes available. Successor adjustment factors convert original shares into adjusted accounting units without changing the announced share entitlement.

This common implementation maps intended rankings into simulated holdings. Selection statistics concern target stocks, while returns include trading costs, unfilled orders, cash and retained economic claims from share exchanges or board transfers. Corporate-event dates, exchange ratios and availability rules follow archived exchange and issuer announcements. Fractional positions and fixed notional costs simplify execution; round-lot sizing and time-varying market impact are outside the simulation.

\FloatBarrier
\section{Computational audit design}
\FloatBarrier
\subsection{Decision stability and return benchmark}
The benchmark endpoint is the time mean of 72 paired monthly net-return differences between a model and equal-factor selection at 10 basis points per side. Model returns are averaged within month over the five complete SSE 50 runs, or the single run in each larger universe. The resulting twelve model comparisons form one adjustment family, while 48 comparisons for the other four numerical policies form another. Supplementary Material maps the complete testing families.

An intercept-only regression estimates each mean contrast using Newey-West heteroskedasticity- and autocorrelation-consistent (HAC) standard errors with three lags and a finite-sample correction \citep{r10}. Two-sided tests retain both positive and negative differences. A circular block bootstrap draws 10,000 samples, using six-month blocks for the main analysis and three- and twelve-month blocks for sensitivity \citep{r11}. Within each draw, complete LLM run trajectories are sampled with replacement and averaged; the same sampled month blocks are applied to this mean trajectory and its comparator. A common date mapping across models and universes preserves their shared market months. Percentile intervals describe uncertainty from months and repeated runs; bootstrap p-values compare the observed mean with the centered bootstrap distribution.

Holm correction controls multiplicity within each declared family \citep{r12}. Confidence intervals are pointwise and accompany adjusted p-values. Economic equivalence is tested separately with two one-sided normal tests using the HAC standard error and a monthly mean-return bound of 0.001, or 0.10 percentage points, in either direction. The larger boundary-test p-value enters the twelve-comparison equivalence family. Evidence of a small economic difference therefore has its own test and is distinguished from a nonsignificant test of zero difference.

Each complete run also yields compound annual growth rate (CAGR), annualized monthly volatility, zero-reference-rate Sharpe, daily maximum drawdown, positive-return frequency, and trading intensity. CAGR is calculated for each run before reporting its distribution. Selection hit rates are computed within month as the fraction of target stocks with positive returns or returns above the eligible-universe average, then averaged over months. Monthly score-rank Spearman correlation uses all candidates. Selection measures characterize intended choices, while net portfolio outcomes incorporate execution and costs.

The five SSE 50 primary runs create ten pairwise comparisons per model at every test month. Full-cross-section score-rank Spearman correlation, selected-set overlap, Jaccard similarity, and identical-order frequencies quantify agreement. Constant-score correlations are recorded as undefined; prediction validity and defined-pair counts accompany the summaries. The distribution of complete-run returns connects ranking differences to investment outcomes.

\FloatBarrier
\subsection{Frozen-rule reproduction and transfer}
For each model in the SSE 50, one development-period LLM score set per month supplies within-month fractional preference ranks, centered at 0.5. Ridge and Lasso predict these preference targets from the same nine inputs. Training weights are inversely proportional to each month's number of stocks and normalized to average one, giving development months equal aggregate weight.

Expanding validation folds use signal years 2017, 2018, and 2019, training on earlier development signals with at least twelve earlier valid months. Validation loss is mean squared preference-rank error averaged equally over months. Ridge penalties are 0.1, 1, 10, and 100; Lasso penalties are 0.0001, 0.001, and 0.01. The best parameter within each family is fitted to all available development targets and its coefficients are frozen. The family with the lowest development validation error defines the development-selected proxy comparator. Both families retain their coefficients, test-period agreement, and portfolio outcomes. Proxy choice never uses test returns.

All learned preference rules target the within-month average-tie percentile rank of the model score minus 0.5, rather than stock returns. We compare an equal-weight nine-factor average, a development-selected single factor and orientation, a frozen linear proxy and a bounded random forest. Astra uses the same development-only Ridge/\allowbreak{}Lasso selection procedure on 60 SSE 50 months; its coefficients are frozen before test and intervention evaluation. Forests use 200 trees, depths 2, 3 or 4 and minimum leaves of 20 or 40 observations. Expanding validation in 2017-2019 selects parameters by mean monthly squared error, weighting months equally. Single-factor selection uses the development sample.

Evaluation reports monthly Spearman correlation and overlap of selected top-20\% stocks. The primary complexity contrast is the forest-minus-linear correlation, first averaged over five answers within month and then over 72 months. Agreement among all ten same-input run pairs provides a reference for variability, not a mathematical ceiling. Spearman correlation is not a fraction of explained decisions or variance. Conclusions about nonlinear rules refer to the declared bounded forest family.

The selected linear coefficients are applied unchanged to the CSI 300 and CSI 500 test matrices. Agreement is reported by stock pool and holding year. Both the model and its frozen rule continue to use the same displayed factor percentiles and selection convention. The incremental-return evaluation uses one twelve-comparison family covering the four models and all three universes.

\FloatBarrier
\subsection{Controlled response tests}
Factor ablations replace valuation inputs, momentum inputs, or risk-and-trading inputs with 0.5. Valuation comprises book to price, earnings yield, and dividend yield; momentum comprises both intermediate-return horizons; risk and trading comprises low volatility, small size, and low turnover. One-month reversal remains available in all three conditions. Conventional methods receive the same altered matrices, and supervised baselines are retuned and refitted for each ablation. Every altered LLM trajectory is compared with the prespecified primary run 1 across all 72 months.

Removing feature semantics retains the numerical matrix but replaces names with F1-F9 and removes definitions and economic-judgment instructions. Reordering rows preserves stock identifiers and all numerical values. These two conditions complement the three numerical group interventions. For historical return diagnostics, altered-input run 1 is paired with primary run 1; the baseline-adjusted factor diagnostic also subtracts the corresponding change in equal-factor returns.

The archived interventions neutralize all valuation columns, both momentum columns, or the risk/\allowbreak{}trading columns, hide feature meanings, or reorder displayed stocks. A frozen linear proxy receives the exact displayed numerical matrices. Predicted response is its altered-input percentile rank minus its original-input percentile rank. Actual response uses the corresponding LLM ranks, pairing each archived intervention run with original run 1. Numerical group interventions form a nine-comparison family and a three-comparison Astra family. Semantics and row-order changes leave proxy inputs unchanged and are descriptive controls.

The primary response statistic is the mean squared error (MSE) of a zero-change prediction minus the MSE of the frozen predicted change. Positive values indicate predictive improvement. We aggregate across stocks within each month before inference. Direction agreement is accompanied by coverage because zero changes provide no directional prediction. Selection entry/\allowbreak{}exit statistics describe discrete portfolio changes separately. Group neutralizations remove different numbers of columns and therefore do not by themselves establish a ranking of feature importance.

The matched experiment selects 24 months equally across six holding years. Eight stocks per month have both valuation and momentum percentiles in [0.2,0.8]. Four receive +0.2 and four -0.2 on a single feature, using identical stocks and signs for book-to-price and 12-to-1-month momentum. Every other numerical cell, identifier, row order and prompt component is held fixed. The conditions are original input, valuation shift and momentum shift, with three independent responses per model and condition, totaling 864 valid responses across four models (648 primary responses and 216 Astra replication responses). Selection and call order are seeded. These are controlled numerical input changes, not simulated changes in company fundamentals.

We determined the matched-experiment sample size from five-run variability using a predeclared precision target for conditional measurement noise in signed rank responses. This calculation does not estimate financial-return power or total heterogeneity of treatment responses. Ranks are averaged over three responses per condition before computing changes. Inference resamples month clusters within holding-year strata, keeping all responses, models and conditions together within each sampled month.

\FloatBarrier
\subsection{Economic value beyond the frozen rule}
The economic comparator is the original frozen linear proxy transferred from the SSE 50. We measure incremental net investment return as LLM monthly return minus proxy monthly return under matched trading rules and costs. We define remaining preference as the centered LLM percentile rank minus the proxy raw prediction of that same target. We examine its consistency across repeated responses, its stock-return rank correlation, and its partial Pearson association with forward returns after projecting both variables off an intercept and the nine original factor columns.

Averaging ranks across runs precedes the partial-correlation calculation. Because the proxy lies in the factor-control span, the partial association tests whether model preferences beyond that linear span are associated with forward returns; it does not measure the quality of a particular coefficient vector. The monthly regression is a conditional association diagnostic, not a newly fitted trading strategy. Constant responses that add no variation beyond controls have undefined partial correlation. They remain in the 72-month calendar, with defined-month denominators reported explicitly; executable portfolio-return comparisons retain all months.

Rule-complexity and archived response-prediction intervals use 10,000 circular six-month block-bootstrap draws, with three- and twelve-month sensitivity analyses \citep{r11}. Scheduled answers are first averaged within month. Incremental-return and partial-correlation tests resample whole trajectories when repeated runs exist, while matched responses use holding-year-stratified month clusters. Holm adjustment is applied within the declared families listed in Supplementary Material \citep{r12}. Intervals are pointwise, and adjusted p-values determine familywise findings. The benchmark equivalence test in Section 3.1 is separate from the rule comparisons.

\FloatBarrier
\subsection{Robustness and precision}
Two robustness exercises test whether exposure constraints or overlapping index membership drive the comparisons. Risk-matched portfolios use a common equal-weight-universe reference and constrain stock weights, industry deviations, size, volatility and beta. Disjoint-stock diagnostics remove overlapping constituents in a fixed allocation order while preserving each model's original parent-pool scores. Both exercises retain the shared market calendar. Supplementary Material reports their objectives, tolerances and complete results.

Separate archived calculations determine precision for financial returns and for the matched response experiment. The return calculation evaluates a monthly effect of 0.10 percentage points under 80\% planned power and correction for three CSI 300 comparisons. Its fixed rule selects 36 months when the 24-month option misses that target; the selected duration also falls below 80\% for the three comparisons. The response calculation uses repeated-answer measurement noise to select 24 stratified months. Supplementary Material reports the achieved planning power and interval precision for each calculation.

Selection-boundary sensitivity holds the stored model scores, candidate stocks and market observations fixed while varying priority within exactly tied scores. Fifty fixed hash seeds define alternative priorities, shared across models for each stock-month; ascending anonymous identifiers define the reference priority. All priorities use the same self-financing rebalance engine across the three stock pools and four trading-cost scenarios. Reference-priority monthly returns agree with the stored returns to numerical precision. Variation across priorities measures the sensitivity of simulated portfolios to the tie rule (Supplementary Material, Section S17).

\FloatBarrier
\subsection{Repeated input-intervention design}
The repeated input-intervention design covers the 72 SSE 50 test months under primary inputs, valuation neutralization, momentum neutralization, risk/\allowbreak{}trading neutralization, hidden semantics and reordered rows. Each model has three complete runs per condition. Astra contributes 1,296 responses, and Luna, Terra and Sol contribute 3,888 responses across 432 input matrices and 54 trajectories. The two batches share the six-condition design and input matrices. For Luna, Terra and Sol, the analysis specification was locked before collection. The intervention-return analyses are exploratory. Supplementary Material, Sections S14-S15, reports batch-specific results and collection records.

Within each run and month, an altered-condition net return is paired with its replication primary-condition return. Inference resamples whole paired trajectories and common circular month blocks in 10,000 draws, using six-month blocks for the primary analysis and three- and twelve-month blocks for sensitivity. HAC tests use three lags. Holm adjustment applies separately to nine numerical-intervention contrasts, three semantics contrasts, three row-order contrasts and nine contrasts adjusted for the corresponding equal-factor return change. Both procedures use these families. The primary cost is 10 basis points per side. This batch evaluates intervention returns and repeated-ranking agreement; the frozen-rule prediction tests use the archived and matched experiments.

\FloatBarrier
\subsection{Reproducibility}
The research archive retains numerical inputs, prompts, model responses, frozen coefficients, portfolio records and analysis specifications, allowing numerical reproduction without new model requests. Supplementary Material supplies the detailed results and procedures; Python and Matplotlib generate the figures from archived results. Shuai Wu led the manuscript integration and checked the analyses, citations and final text. All authors reviewed the manuscript within their areas of expertise and remain responsible for their contributions.

\FloatBarrier
\clearpage
\section{Results}
\FloatBarrier
\subsection{Stability and reproducibility of ranking policies}
\begingroup\small\setlength{\tabcolsep}{3pt}\renewcommand{\arraystretch}{1.17}
\begin{longtable}{@{}>{\raggedright\arraybackslash}p{\dimexpr 0.310345\linewidth-7.44828pt\relax}>{\raggedright\arraybackslash}p{\dimexpr 0.172414\linewidth-4.13793pt\relax}>{\raggedright\arraybackslash}p{\dimexpr 0.172414\linewidth-4.13793pt\relax}>{\raggedright\arraybackslash}p{\dimexpr 0.172414\linewidth-4.13793pt\relax}>{\raggedright\arraybackslash}p{\dimexpr 0.172414\linewidth-4.13793pt\relax}@{}}
\caption{Stability across all ten pairs of complete main runs}\label{tab:3}\\
\toprule
\textbf{Universe /\allowbreak{} model} & \textbf{Valid /\allowbreak{} planned pairs} & \textbf{Mean rank correlation} & \textbf{Mean selected overlap} & \textbf{Identical selected sets}\\\midrule\endfirsthead
\multicolumn{5}{l}{\textbf{Table 3 (continued)}}\\\toprule
\textbf{Universe /\allowbreak{} model} & \textbf{Valid /\allowbreak{} planned pairs} & \textbf{Mean rank correlation} & \textbf{Mean selected overlap} & \textbf{Identical selected sets}\\\midrule\endhead
\midrule\multicolumn{5}{r}{Continued on next page}\\\endfoot
\bottomrule\endlastfoot
SSE 50 /\allowbreak{} Luna & 720 /\allowbreak{} 720 & 0.892 & 83.36\% & 132\\
SSE 50 /\allowbreak{} Terra & 720 /\allowbreak{} 720 & 0.966 & 90.50\% & 249\\
SSE 50 /\allowbreak{} Sol & 720 /\allowbreak{} 720 & 0.971 & 91.43\% & 270\\
SSE 50 /\allowbreak{} Astra & 720 /\allowbreak{} 720 & 0.991 & 94.63\% & 395\\
\end{longtable}
\noindent\footnotesize Each universe-model row contains ten run pairs across 72 shared months. These are dependent descriptive comparisons, not 720 independent market observations. Correlations omit undefined constant-score cases; overlap uses the common top ceil(0.2N) selections. All five altered-input conditions are archived at monthly resolution.\par\endgroup
Table 3 shows that identical-input rankings are highly repeatable without being deterministic. Mean selected-stock overlap is 83.36\% for Luna, 90.50\% for Terra, 91.43\% for Sol and 94.63\% for Astra; mean rank correlations are 0.892, 0.966, 0.971 and 0.991. The ten run pairs within each of the same 72 months provide the behavioral reference for the frozen rules.

\begingroup\small\setlength{\tabcolsep}{3pt}\renewcommand{\arraystretch}{1.17}
\begin{longtable}{@{}>{\raggedright\arraybackslash}p{\dimexpr 0.375000\linewidth-6.75000pt\relax}>{\raggedright\arraybackslash}p{\dimexpr 0.208333\linewidth-3.75000pt\relax}>{\raggedright\arraybackslash}p{\dimexpr 0.208333\linewidth-3.75000pt\relax}>{\raggedright\arraybackslash}p{\dimexpr 0.208333\linewidth-3.75000pt\relax}@{}}
\caption{Preference reproduction in the SSE 50}\label{tab:4}\\
\toprule
\textbf{Model} & \textbf{Rule} & \textbf{Spearman} & \textbf{Overlap (\%)}\\\midrule\endfirsthead
\multicolumn{4}{l}{\textbf{Table 4 (continued)}}\\\toprule
\textbf{Model} & \textbf{Rule} & \textbf{Spearman} & \textbf{Overlap (\%)}\\\midrule\endhead
\midrule\multicolumn{4}{r}{Continued on next page}\\\endfoot
\bottomrule\endlastfoot
Luna & Equal factor & 0.827 & 75.4\\
Luna & Frozen linear & 0.923 & 85.8\\
Luna & Random forest & 0.845 & 77.9\\
Luna & Single factor & 0.774 & 68.5\\
Sol & Equal factor & 0.947 & 84.9\\
Sol & Frozen linear & 0.975 & 90.3\\
Sol & Random forest & 0.863 & 78.8\\
Sol & Single factor & 0.749 & 65.2\\
Terra & Equal factor & 0.909 & 79.7\\
Terra & Frozen linear & 0.962 & 87.3\\
Terra & Random forest & 0.880 & 79.5\\
Terra & Single factor & 0.809 & 68.0\\
Astra & Equal factor & 0.894 & 80.1\\
Astra & Frozen linear & 0.984 & 91.3\\
Astra & Random forest & 0.904 & 79.4\\
Astra & Single factor & 0.695 & 62.6\\
\end{longtable}
\noindent\footnotesize Each value is the mean across 72 months after averaging the five original responses within month. Overlap refers to the common top-20\% stock selection.\par\endgroup
Table 4 and Figure 1 show that frozen linear rules approximate SSE 50 rankings with mean correlations of 0.923-0.984. Forest-minus-linear differences range from -0.112 to -0.077 and are significant within the declared families. The linear rules have higher rank agreement with the observed choices than the bounded forest.

\begin{figure}[H]\centering
\includegraphics[width=\linewidth]{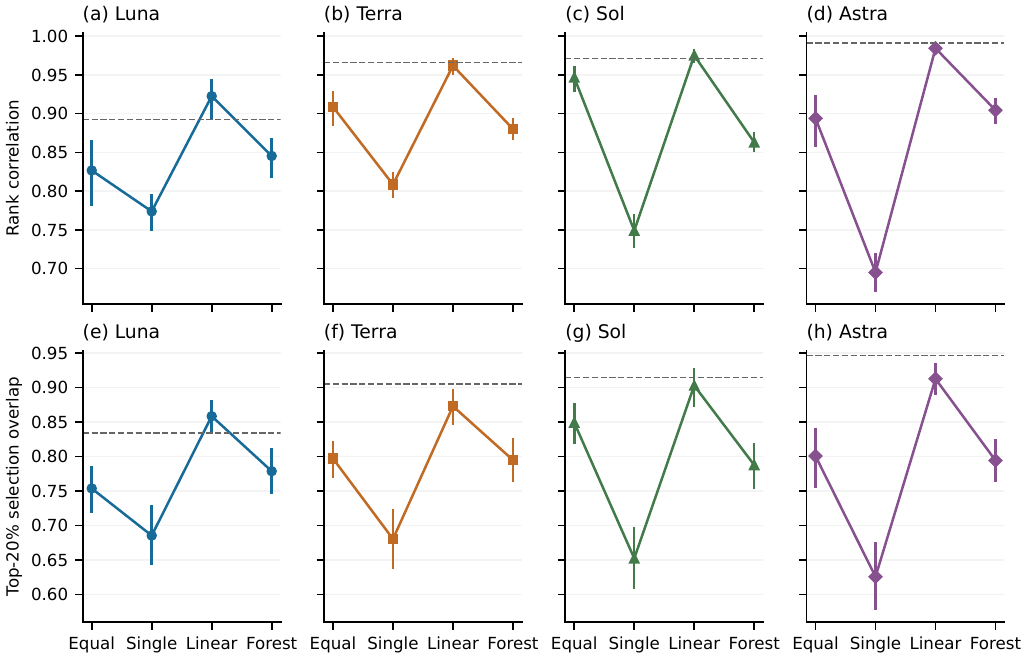}
\caption{Preference reproduction by rule family. Panels a-d show rank correlation and panels e-h show top-20\% selection overlap for Luna, Terra, Sol and Astra, respectively. Equal, Single, Linear and Forest denote equal-factor, selected single-factor, frozen linear and random-forest rules. Points are monthly means; bars are descriptive 95\% six-month block-bootstrap intervals. Dashed lines mark agreement among identical-input model repeats, a variability reference rather than a mathematical upper bound}\label{fig:1}
\end{figure}
\begingroup\small\setlength{\tabcolsep}{3pt}\renewcommand{\arraystretch}{1.17}
\begin{longtable}{@{}>{\raggedright\arraybackslash}p{\dimexpr 0.375000\linewidth-6.75000pt\relax}>{\raggedright\arraybackslash}p{\dimexpr 0.208333\linewidth-3.75000pt\relax}>{\raggedright\arraybackslash}p{\dimexpr 0.208333\linewidth-3.75000pt\relax}>{\raggedright\arraybackslash}p{\dimexpr 0.208333\linewidth-3.75000pt\relax}@{}}
\caption{Direct transfer of the frozen SSE 50 linear rules}\label{tab:5}\\
\toprule
\textbf{Pool /\allowbreak{} model} & \textbf{Spearman} & \textbf{Overlap (\%)} & \textbf{Defined months}\\\midrule\endfirsthead
\multicolumn{4}{l}{\textbf{Table 5 (continued)}}\\\toprule
\textbf{Pool /\allowbreak{} model} & \textbf{Spearman} & \textbf{Overlap (\%)} & \textbf{Defined months}\\\midrule\endhead
\midrule\multicolumn{4}{r}{Continued on next page}\\\endfoot
\bottomrule\endlastfoot
CSI 300 /\allowbreak{} Luna & 0.870 & 82.3 & 72\\
CSI 300 /\allowbreak{} Sol & 0.926 & 85.3 & 72\\
CSI 300 /\allowbreak{} Terra & 0.917 & 84.8 & 72\\
CSI 500 /\allowbreak{} Luna & 0.795 & 70.2 & 68\\
CSI 500 /\allowbreak{} Sol & 0.933 & 83.6 & 72\\
CSI 500 /\allowbreak{} Terra & 0.892 & 75.2 & 65\\
CSI 300 /\allowbreak{} Astra & 0.985 & 92.6 & 72\\
CSI 500 /\allowbreak{} Astra & 0.978 & 90.1 & 72\\
\end{longtable}
\noindent\footnotesize No fitting or parameter reselection in the transfer pools. Undefined rank correlations from constant scores do not remove those months from selection or portfolio-return summaries.\par\endgroup
Across the two larger stock pools, mean frozen-linear rank correlations range from 0.795 to 0.985 (Table 5). Annual summaries and disjoint-stock diagnostics retain the same coefficients and original parent-pool feature percentiles. Selection overlap and defined-month counts show how this agreement varies across pools and when constant-score responses leave correlations undefined.

\FloatBarrier
\subsection{Prediction of controlled responses}
\begingroup\small\setlength{\tabcolsep}{3pt}\renewcommand{\arraystretch}{1.17}
\begin{longtable}{@{}>{\raggedright\arraybackslash}p{\dimexpr 0.473684\linewidth-5.68421pt\relax}>{\raggedright\arraybackslash}p{\dimexpr 0.263158\linewidth-3.15789pt\relax}>{\raggedright\arraybackslash}p{\dimexpr 0.263158\linewidth-3.15789pt\relax}@{}}
\caption{Archived numerical intervention response prediction}\label{tab:6}\\
\toprule
\textbf{Model /\allowbreak{} intervention} & \textbf{MSE improvement [95\% CI]} & \textbf{Holm p}\\\midrule\endfirsthead
\multicolumn{3}{l}{\textbf{Table 6 (continued)}}\\\toprule
\textbf{Model /\allowbreak{} intervention} & \textbf{MSE improvement [95\% CI]} & \textbf{Holm p}\\\midrule\endhead
\midrule\multicolumn{3}{r}{Continued on next page}\\\endfoot
\bottomrule\endlastfoot
Luna /\allowbreak{} Momentum & 15.156 [12.870, 17.534] & <0.001\\
Luna /\allowbreak{} Risk/\allowbreak{}trading & 5.658 [1.438, 10.692] & 0.016\\
Luna /\allowbreak{} Valuation & 52.763 [42.505, 64.459] & <0.001\\
Sol /\allowbreak{} Momentum & 24.858 [21.591, 28.285] & <0.001\\
Sol /\allowbreak{} Risk/\allowbreak{}trading & 12.879 [8.485, 18.078] & <0.001\\
Sol /\allowbreak{} Valuation & 46.366 [34.186, 61.217] & <0.001\\
Terra /\allowbreak{} Momentum & 17.090 [14.599, 19.961] & <0.001\\
Terra /\allowbreak{} Risk/\allowbreak{}trading & 3.204 [2.145, 4.483] & <0.001\\
Terra /\allowbreak{} Valuation & 55.958 [42.799, 70.796] & <0.001\\
Astra /\allowbreak{} Valuation & 29.017 [23.086, 35.709] & <0.001\\
Astra /\allowbreak{} Momentum & 41.406 [34.206, 49.770] & <0.001\\
Astra /\allowbreak{} Risk/\allowbreak{}trading & 5.275 [4.352, 6.277] & <0.001\\
\end{longtable}
\noindent\footnotesize Squared percentile-rank units multiplied by 1,000. Positive gain favors the frozen predicted change over no change. Six-month blocks; the nine historical comparisons and the three Astra comparisons retain separate Holm families. Each altered condition has one historical response per month. CI = confidence interval. MSE = mean squared error.\par\endgroup
Under the primary six-month block analysis, all twelve archived numerical contrasts in Table 6 show positive MSE improvement and pass the adjusted 0.05 criterion within their declared families. On average, the fixed rules predict preference changes under altered numerical inputs more accurately than a zero-change prediction. With twelve-month blocks, the Luna risk-and-trading contrast has adjusted p=0.052; Supplementary Material, Table S24 reports the complete sensitivity analysis.

\begingroup\small\setlength{\tabcolsep}{3pt}\renewcommand{\arraystretch}{1.17}
\begin{longtable}{@{}>{\raggedright\arraybackslash}p{\dimexpr 0.473684\linewidth-5.68421pt\relax}>{\raggedright\arraybackslash}p{\dimexpr 0.263158\linewidth-3.15789pt\relax}>{\raggedright\arraybackslash}p{\dimexpr 0.263158\linewidth-3.15789pt\relax}@{}}
\caption{Matched single-feature response prediction}\label{tab:7}\\
\toprule
\textbf{Model /\allowbreak{} intervention} & \textbf{MSE improvement [95\% CI]} & \textbf{Holm p}\\\midrule\endfirsthead
\multicolumn{3}{l}{\textbf{Table 7 (continued)}}\\\toprule
\textbf{Model /\allowbreak{} intervention} & \textbf{MSE improvement [95\% CI]} & \textbf{Holm p}\\\midrule\endhead
\midrule\multicolumn{3}{r}{Continued on next page}\\\endfoot
\bottomrule\endlastfoot
Luna /\allowbreak{} Valuation shift & 0.438 [0.182, 0.687] & 0.001\\
Luna /\allowbreak{} Momentum shift & 0.476 [0.178, 0.842] & 0.006\\
Terra /\allowbreak{} Valuation shift & 0.358 [0.169, 0.550] & 0.002\\
Terra /\allowbreak{} Momentum shift & 0.268 [0.126, 0.416] & 0.002\\
Sol /\allowbreak{} Valuation shift & 0.248 [0.094, 0.408] & 0.003\\
Sol /\allowbreak{} Momentum shift & 0.731 [0.547, 0.915] & <0.001\\
Astra /\allowbreak{} Valuation shift & 0.174 [0.095, 0.261] & <0.001\\
Astra /\allowbreak{} Momentum shift & 1.245 [0.940, 1.559] & <0.001\\
\end{longtable}
\noindent\footnotesize Squared percentile-rank units multiplied by 1,000. Three responses per condition and month. Holding-year-stratified month-cluster intervals; the six historical and two Astra comparisons retain separate Holm families. CI = confidence interval. MSE = mean squared error.\par\endgroup
Table 7 reports matched single-feature MSE improvements of 0.174 to 1.245, in squared percentile-rank units multiplied by 1,000. All eight comparisons show positive improvement with adjusted p below 0.05, using the six-comparison family and the two-comparison Astra family. The number of affected stocks and perturbation amplitude are fixed across valuation and momentum.

Frozen predictions reduce aggregate squared rank-change error by 5.6\%-55.7\% relative to a zero-change prediction. Signed rank responses among perturbed stocks average 3.8-8.3 percentile points across the eight comparisons. Event Jaccard for predicted versus actual entries and exits is 0.000-0.125; the aggregate error improvement does not imply accurate prediction of each discrete selection change (Supplementary Material, Table S29).

Hiding economic feature names and changing stock display order produce exactly zero predicted numerical-proxy change. The archived model responses can still change under these conditions. Their monthly changes and the original repeated-input reference are reported descriptively; a single archived altered response cannot independently separate every prompt effect from answer variability. Directional response agreement and discrete changes in portfolio membership are also reported separately.

Figure 2 compares the two intervention designs. Under both sets of numerical changes, the frozen mapping lowers aggregate error in predicting preference changes relative to a zero-change prediction. Historical return effects of factor neutralization, hidden semantics and reordered rows are reported in Supplementary Material and evaluated separately from behavioral response prediction.

\begin{figure}[H]\centering
\includegraphics[width=\linewidth]{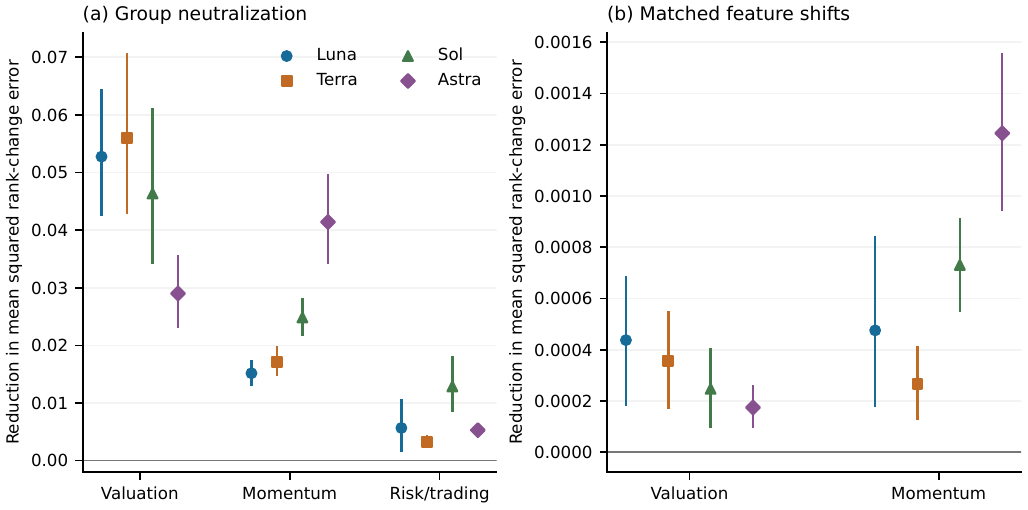}
\caption{Prediction of ranking responses. Panel a neutralizes feature groups; panel b shifts a single feature for matched stocks. Panel scales differ because the interventions have different magnitudes and estimands. Symbols identify the models. Intervals are pointwise 95\% intervals; adjusted decisions appear in Tables 6 and 7}\label{fig:2}
\end{figure}
\FloatBarrier
\subsection{Investment performance under common execution}
Table 8 compares all nine policies in each universe at the primary transaction cost. For repeated model runs, compound annual growth rates (CAGR) are computed separately and summarized by their mean and observed range. The twelve universe-model mean CAGRs range from 6.26\% to 11.17\%, compared with 8.56\%-10.39\% for equal-factor selection. Figure 3 displays every complete model trajectory.

\begingroup\small\setlength{\tabcolsep}{3pt}\renewcommand{\arraystretch}{1.17}
\begin{longtable}{@{}>{\raggedright\arraybackslash}p{\dimexpr 0.224490\linewidth-6.73469pt\relax}>{\raggedright\arraybackslash}p{\dimexpr 0.217687\linewidth-6.53061pt\relax}>{\raggedright\arraybackslash}p{\dimexpr 0.136054\linewidth-4.08163pt\relax}>{\raggedright\arraybackslash}p{\dimexpr 0.095238\linewidth-2.85714pt\relax}>{\raggedright\arraybackslash}p{\dimexpr 0.136054\linewidth-4.08163pt\relax}>{\raggedright\arraybackslash}p{\dimexpr 0.190476\linewidth-5.71429pt\relax}@{}}
\caption{Net portfolio and selection outcomes at 10 bp}\label{tab:8}\\
\toprule
\textbf{Universe /\allowbreak{} method} & \textbf{CAGR mean [range]} & \textbf{Volatility} & \textbf{Sharpe} & \textbf{Max drawdown} & \textbf{Positive /\allowbreak{} relative hit}\\\midrule\endfirsthead
\multicolumn{6}{l}{\textbf{Table 8 (continued)}}\\\toprule
\textbf{Universe /\allowbreak{} method} & \textbf{CAGR mean [range]} & \textbf{Volatility} & \textbf{Sharpe} & \textbf{Max drawdown} & \textbf{Positive /\allowbreak{} relative hit}\\\midrule\endhead
\midrule\multicolumn{6}{r}{Continued on next page}\\\endfoot
\bottomrule\endlastfoot
SSE 50 /\allowbreak{} Equal factor & 8.56\% & 16.46\% & 0.58 & -19.85\% & 52.08\% /\allowbreak{} 51.11\%\\
SSE 50 /\allowbreak{} Equal factor-group & 5.27\% & 15.32\% & 0.41 & -25.45\% & 50.56\% /\allowbreak{} 49.72\%\\
SSE 50 /\allowbreak{} Ridge & 8.75\% & 15.23\% & 0.63 & -26.78\% & 54.17\% /\allowbreak{} 53.06\%\\
SSE 50 /\allowbreak{} Gradient boosting & -1.43\% & 17.39\% & 0.00 & -51.48\% & 47.64\% /\allowbreak{} 46.67\%\\
SSE 50 /\allowbreak{} Eligible universe & 3.07\% & 20.10\% & 0.25 & -34.62\% & 47.62\% /\allowbreak{} 46.11\%\\
SSE 50 /\allowbreak{} Luna & 7.74\% [6.88\%, 8.83\%] & 15.88\% & 0.55 & -21.26\% & 53.47\% /\allowbreak{} 51.50\%\\
SSE 50 /\allowbreak{} Terra & 8.51\% [7.34\%, 10.51\%] & 15.92\% & 0.59 & -17.81\% & 53.17\% /\allowbreak{} 51.39\%\\
SSE 50 /\allowbreak{} Sol & 9.83\% [8.79\%, 10.90\%] & 15.63\% & 0.68 & -17.69\% & 53.78\% /\allowbreak{} 52.50\%\\
SSE 50 /\allowbreak{} Astra & 8.80\% [8.07\%, 9.50\%] & 16.25\% & 0.60 & -19.76\% & 52.94\% /\allowbreak{} 52.06\%\\
CSI 300 /\allowbreak{} Equal factor & 8.99\% & 16.79\% & 0.60 & -18.11\% & 50.83\% /\allowbreak{} 49.51\%\\
CSI 300 /\allowbreak{} Equal factor-group & 7.58\% & 17.59\% & 0.50 & -16.13\% & 49.87\% /\allowbreak{} 48.11\%\\
CSI 300 /\allowbreak{} Ridge & 7.08\% & 16.07\% & 0.51 & -25.48\% & 50.44\% /\allowbreak{} 48.65\%\\
CSI 300 /\allowbreak{} Gradient boosting & 6.89\% & 17.21\% & 0.47 & -32.29\% & 49.74\% /\allowbreak{} 48.18\%\\
CSI 300 /\allowbreak{} Eligible universe & 4.90\% & 23.21\% & 0.31 & -34.47\% & 46.70\% /\allowbreak{} 44.85\%\\
CSI 300 /\allowbreak{} Luna & 11.17\% & 17.26\% & 0.70 & -17.04\% & 50.65\% /\allowbreak{} 50.01\%\\
CSI 300 /\allowbreak{} Terra & 10.66\% & 17.00\% & 0.68 & -17.61\% & 50.57\% /\allowbreak{} 49.99\%\\
CSI 300 /\allowbreak{} Sol & 10.39\% & 17.36\% & 0.66 & -17.98\% & 50.92\% /\allowbreak{} 49.90\%\\
CSI 300 /\allowbreak{} Astra & 8.17\% & 16.52\% & 0.56 & -19.32\% & 50.46\% /\allowbreak{} 49.34\%\\
CSI 500 /\allowbreak{} Equal factor & 10.39\% & 20.12\% & 0.59 & -18.35\% & 49.15\% /\allowbreak{} 44.44\%\\
CSI 500 /\allowbreak{} Equal factor-group & 9.82\% & 21.34\% & 0.54 & -21.38\% & 49.18\% /\allowbreak{} 44.45\%\\
CSI 500 /\allowbreak{} Ridge & 9.10\% & 19.46\% & 0.54 & -19.92\% & 49.85\% /\allowbreak{} 44.87\%\\
CSI 500 /\allowbreak{} Gradient boosting & 9.96\% & 21.06\% & 0.55 & -22.76\% & 48.93\% /\allowbreak{} 44.42\%\\
CSI 500 /\allowbreak{} Eligible universe & 9.43\% & 25.28\% & 0.48 & -31.81\% & 47.18\% /\allowbreak{} 42.94\%\\
CSI 500 /\allowbreak{} Luna & 11.13\% & 20.88\% & 0.61 & -30.98\% & 48.47\% /\allowbreak{} 45.60\%\\
CSI 500 /\allowbreak{} Terra & 6.26\% & 19.86\% & 0.40 & -26.20\% & 48.17\% /\allowbreak{} 42.94\%\\
CSI 500 /\allowbreak{} Sol & 10.72\% & 20.17\% & 0.60 & -16.85\% & 49.22\% /\allowbreak{} 44.66\%\\
CSI 500 /\allowbreak{} Astra & 10.51\% & 19.59\% & 0.61 & -21.23\% & 49.20\% /\allowbreak{} 44.74\%\\
\end{longtable}
\noindent\footnotesize LLM rows average the locked complete runs: five in the SSE 50 and one in each of the CSI 300 and CSI 500. Numerical comparators have one deterministic run. Brackets give the minimum and maximum of the five separately compounded CAGRs. Volatility is annualized from monthly returns; Sharpe uses a zero reference rate. Drawdown uses daily net asset values. Hit rates concern target selections, while net returns include actual execution. The 72 holding months are common across universes.\par\endgroup
\begin{figure}[H]\centering
\includegraphics[width=\linewidth]{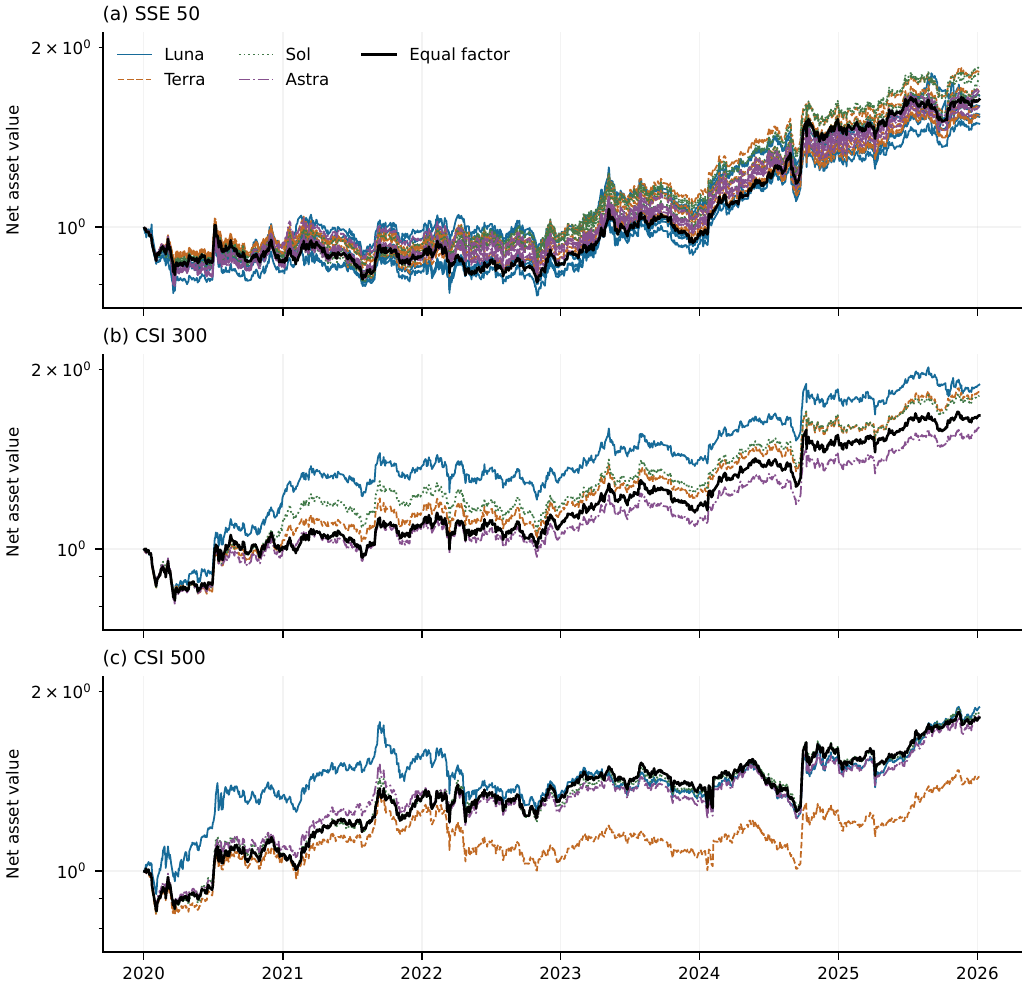}
\caption{Net asset values at 10 basis points per side for every complete model run and equal-factor selection. Panels a-c show the SSE 50, CSI 300 and CSI 500. Line style and color identify models; all five SSE 50 trajectories are displayed with the same model style. Logarithmic scales show compounded growth. Every trajectory starts from unit capital; no run is selected by its realized return}\label{fig:3}
\end{figure}
\begingroup\small\setlength{\tabcolsep}{3pt}\renewcommand{\arraystretch}{1.17}
\begin{longtable}{@{}>{\raggedright\arraybackslash}p{\dimexpr 0.310345\linewidth-7.44828pt\relax}>{\raggedright\arraybackslash}p{\dimexpr 0.172414\linewidth-4.13793pt\relax}>{\raggedright\arraybackslash}p{\dimexpr 0.172414\linewidth-4.13793pt\relax}>{\raggedright\arraybackslash}p{\dimexpr 0.172414\linewidth-4.13793pt\relax}>{\raggedright\arraybackslash}p{\dimexpr 0.172414\linewidth-4.13793pt\relax}@{}}
\caption{Twelve model-minus-equal-factor return comparisons}\label{tab:9}\\
\toprule
\textbf{Universe /\allowbreak{} model} & \textbf{Mean difference pp} & \textbf{Pointwise 95\% interval (pp)} & \textbf{HAC adjusted p} & \textbf{Bootstrap adjusted p}\\\midrule\endfirsthead
\multicolumn{5}{l}{\textbf{Table 9 (continued)}}\\\toprule
\textbf{Universe /\allowbreak{} model} & \textbf{Mean difference pp} & \textbf{Pointwise 95\% interval (pp)} & \textbf{HAC adjusted p} & \textbf{Bootstrap adjusted p}\\\midrule\endhead
\midrule\multicolumn{5}{r}{Continued on next page}\\\endfoot
\bottomrule\endlastfoot
SSE 50 /\allowbreak{} Luna & -0.070 & [-0.521, 0.331] & 1.000 & 1.000\\
SSE 50 /\allowbreak{} Terra & -0.010 & [-0.420, 0.342] & 1.000 & 1.000\\
SSE 50 /\allowbreak{} Sol & 0.088 & [-0.231, 0.401] & 1.000 & 1.000\\
SSE 50 /\allowbreak{} Astra & 0.018 & [-0.352, 0.338] & 1.000 & 1.000\\
CSI 300 /\allowbreak{} Luna & 0.172 & [-0.232, 0.724] & 1.000 & 1.000\\
CSI 300 /\allowbreak{} Terra & 0.130 & [-0.069, 0.372] & 1.000 & 1.000\\
CSI 300 /\allowbreak{} Sol & 0.115 & [-0.191, 0.563] & 1.000 & 1.000\\
CSI 300 /\allowbreak{} Astra & -0.067 & [-0.231, 0.113] & 1.000 & 1.000\\
CSI 500 /\allowbreak{} Luna & 0.065 & [-0.623, 0.890] & 1.000 & 1.000\\
CSI 500 /\allowbreak{} Terra & -0.323 & [-0.733, 0.014] & 1.000 & 1.000\\
CSI 500 /\allowbreak{} Sol & 0.026 & [-0.160, 0.211] & 1.000 & 1.000\\
CSI 500 /\allowbreak{} Astra & 0.001 & [-0.264, 0.257] & 1.000 & 1.000\\
\end{longtable}
\noindent\footnotesize Monthly percentage-point differences at 10 basis points per side. Pointwise 95\% intervals use whole-run trajectories and six-month calendar blocks. HAC and bootstrap p-values are separately Holm-adjusted across twelve model-universe comparisons. The other-baseline family contains 48 comparisons.\par\endgroup
Table 9 and Figure 4 show mean monthly model-minus-equal-factor differences ranging from -0.323 to 0.172 percentage points. None of the twelve tests passes the Holm-adjusted difference criterion. The intervals are also too wide to establish equivalence within the specified margin.

\begin{figure}[H]\centering
\includegraphics[width=\linewidth]{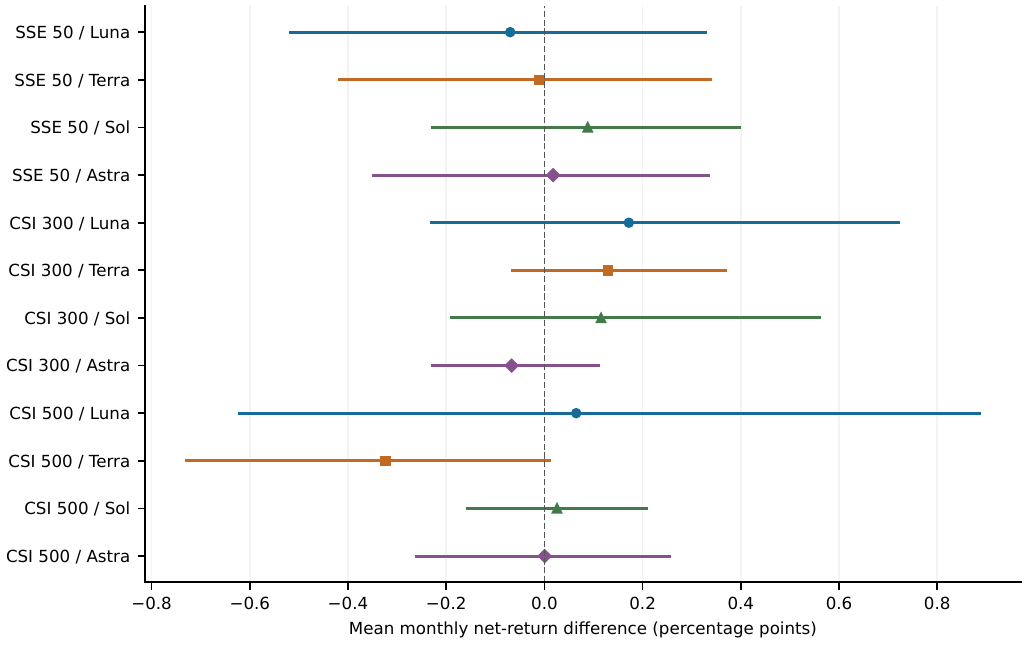}
\caption{Mean monthly model-minus-equal-factor net-return differences and pointwise 95\% bootstrap intervals across 72 shared market months. Values are percentage points per month at 10 basis points per side. Table 9 supplies the separately adjusted tests; interval exclusion alone is not a multiplicity-adjusted finding}\label{fig:4}
\end{figure}
Across the 27 universe-method combinations, moving from zero to 50 basis points changes mean CAGR by -7.25 to -0.73 percentage points. All four cost scenarios retain complete daily values, execution records, transactions, corporate-event records, holdings snapshots and terminal positions, and reuse the same scoring decisions.

Selection-boundary score ties occur in 62 of 72 CSI 500 Luna months and 60 of 72 Terra months; all candidate scores are identical in four and seven of those months, respectively. At 10 basis points per side, the archived CSI 500 Luna CAGR of 11.13\% ranges from 10.09\% to 12.34\% across 50 fixed tie priorities, and Terra's 6.26\% ranges from 5.34\% to 7.50\%. The comparable SSE 50 differences are small. These exploratory ranges quantify dependence on a deterministic choice among tied stocks, rather than uncertainty from new model responses or market periods (Supplementary Material, Table S43).

\FloatBarrier
\subsection{Economic value beyond the frozen rule}
\begingroup\small\setlength{\tabcolsep}{3pt}\renewcommand{\arraystretch}{1.17}
\begin{longtable}{@{}>{\raggedright\arraybackslash}p{\dimexpr 0.473684\linewidth-5.68421pt\relax}>{\raggedright\arraybackslash}p{\dimexpr 0.263158\linewidth-3.15789pt\relax}>{\raggedright\arraybackslash}p{\dimexpr 0.263158\linewidth-3.15789pt\relax}@{}}
\caption{LLM minus frozen-linear net monthly returns}\label{tab:10}\\
\toprule
\textbf{Pool /\allowbreak{} model} & \textbf{Difference [95\% CI], bps} & \textbf{Holm p}\\\midrule\endfirsthead
\multicolumn{3}{l}{\textbf{Table 10 (continued)}}\\\toprule
\textbf{Pool /\allowbreak{} model} & \textbf{Difference [95\% CI], bps} & \textbf{Holm p}\\\midrule\endhead
\midrule\multicolumn{3}{r}{Continued on next page}\\\endfoot
\bottomrule\endlastfoot
SSE 50 /\allowbreak{} Luna & -18.968 [-43.180, 4.296] & 1.000\\
SSE 50 /\allowbreak{} Terra & -20.094 [-39.035, -1.941] & 0.382\\
SSE 50 /\allowbreak{} Sol & -1.529 [-15.449, 10.527] & 1.000\\
CSI 300 /\allowbreak{} Luna & 7.383 [-30.781, 56.811] & 1.000\\
CSI 300 /\allowbreak{} Terra & 2.717 [-11.641, 18.770] & 1.000\\
CSI 300 /\allowbreak{} Sol & 9.034 [-15.677, 45.733] & 1.000\\
CSI 500 /\allowbreak{} Luna & 10.583 [-56.438, 92.149] & 1.000\\
CSI 500 /\allowbreak{} Terra & -34.013 [-66.166, -6.034] & 0.342\\
CSI 500 /\allowbreak{} Sol & 1.550 [-13.075, 16.845] & 1.000\\
SSE 50 /\allowbreak{} Astra & -6.697 [-33.124, 15.971] & 1.000\\
CSI 300 /\allowbreak{} Astra & -6.109 [-25.385, 8.050] & 1.000\\
CSI 500 /\allowbreak{} Astra & -9.161 [-22.649, 4.622] & 1.000\\
\end{longtable}
\noindent\footnotesize Model-minus-frozen-linear net return at 10 basis points per side. Whole-run and six-month calendar-block bootstrap; pointwise 95\% intervals and one twelve-comparison Holm family. CI = confidence interval. bps = basis points.\par\endgroup
At 10 basis points per side and six-month blocks, Table 10 and Figure 5a report twelve mean model-minus-rule net-return differences ranging from -34.01 to 10.58 basis points per month. None meets the Holm-adjusted 0.05 criterion in the twelve-comparison family. Three Terra comparisons at higher costs and twelve-month blocks meet the criterion within their respective twelve-test slices (Supplementary Material, Table S30). Across the complete four-cost, three-block-length grid of 144 comparisons, the smallest Holm-adjusted p-value is 0.101; these sensitivity results are exploratory.

The fixed-score tie-order analysis preserves this primary return finding: none of the 50 alternative priorities produces a Holm-adjusted model-minus-rule difference at 10 basis points per side and six-month blocks in the twelve-comparison family. At 50 basis points, six of the 50 priorities yield a within-slice adjusted finding for CSI 500 Terra at six-month blocks, compared with no such finding for the archived priority. These higher-cost outcomes are exploratory and show that some return inferences depend on how exact score ties are resolved (Supplementary Material, Section S17 and Table S44).

For the three Terra comparisons identified at higher costs with twelve-month blocks, the adjusted finding persists under 26 of 50 fixed priorities for SSE 50 at 25 basis points, all 50 for SSE 50 at 50 basis points, and 40 for CSI 500 at 50 basis points. CSI 500 Terra at 25 basis points also crosses the within-slice threshold under 17 priorities. These counts show sensitivity of the exploratory higher-cost results to the tie rule; they are not additional independent tests (Supplementary Material, Table S45).

\begingroup\small\setlength{\tabcolsep}{3pt}\renewcommand{\arraystretch}{1.17}
\begin{longtable}{@{}>{\raggedright\arraybackslash}p{\dimexpr 0.375000\linewidth-6.75000pt\relax}>{\raggedright\arraybackslash}p{\dimexpr 0.208333\linewidth-3.75000pt\relax}>{\raggedright\arraybackslash}p{\dimexpr 0.208333\linewidth-3.75000pt\relax}>{\raggedright\arraybackslash}p{\dimexpr 0.208333\linewidth-3.75000pt\relax}@{}}
\caption{Conditional return association of remaining preferences}\label{tab:11}\\
\toprule
\textbf{Pool /\allowbreak{} model} & \textbf{Partial r [95\% CI]} & \textbf{Defined months} & \textbf{Holm p}\\\midrule\endfirsthead
\multicolumn{4}{l}{\textbf{Table 11 (continued)}}\\\toprule
\textbf{Pool /\allowbreak{} model} & \textbf{Partial r [95\% CI]} & \textbf{Defined months} & \textbf{Holm p}\\\midrule\endhead
\midrule\multicolumn{4}{r}{Continued on next page}\\\endfoot
\bottomrule\endlastfoot
SSE 50 /\allowbreak{} Luna & -0.017 [-0.056, 0.032] & 72 & 1.000\\
SSE 50 /\allowbreak{} Terra & -0.025 [-0.062, 0.016] & 72 & 1.000\\
SSE 50 /\allowbreak{} Sol & -0.004 [-0.048, 0.043] & 72 & 1.000\\
CSI 300 /\allowbreak{} Luna & -0.012 [-0.033, 0.010] & 72 & 1.000\\
CSI 300 /\allowbreak{} Terra & -0.004 [-0.022, 0.014] & 72 & 1.000\\
CSI 300 /\allowbreak{} Sol & -0.005 [-0.027, 0.015] & 72 & 1.000\\
CSI 500 /\allowbreak{} Luna & -0.009 [-0.020, 0.001] & 68 & 0.962\\
CSI 500 /\allowbreak{} Terra & -0.003 [-0.016, 0.010] & 65 & 1.000\\
CSI 500 /\allowbreak{} Sol & -0.012 [-0.030, 0.007] & 72 & 1.000\\
SSE 50 /\allowbreak{} Astra & 0.022 [-0.008, 0.051] & 72 & 1.000\\
CSI 300 /\allowbreak{} Astra & 0.010 [-0.005, 0.026] & 72 & 1.000\\
CSI 500 /\allowbreak{} Astra & -0.001 [-0.011, 0.008] & 72 & 1.000\\
\end{longtable}
\noindent\footnotesize Monthly partial Pearson association after controlling for the nine factors and an intercept. Pointwise 95\% six-month block intervals and a separate twelve-comparison Holm family. Undefined months remain in the calendar and are counted separately. CI = confidence interval.\par\endgroup
Table 11 and Figure 5b report the partial-correlation tests; none is significant after Holm correction. CSI 500 has constant-score responses in four Luna months and seven Terra months, leaving 68 and 65 identifiable partial-correlation months; Sol, Astra and all other pool-model combinations have 72. Table 11 averages identifiable months, while Table 10 retains the full 72-month return sample. Supplementary Material reports cost and block-length sensitivity; cross-run consistency and individual-answer rank information coefficients are retained in the replication archive.

\begin{figure}[H]\centering
\includegraphics[width=\linewidth]{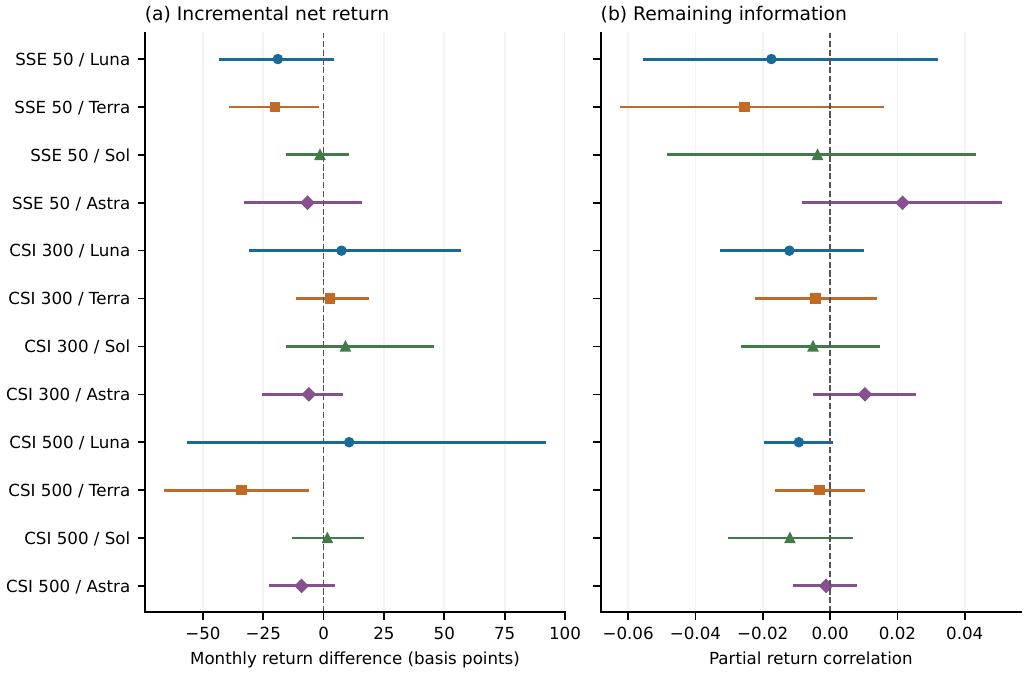}
\caption{Model-minus-rule simulated net returns and conditional return association. Panel a shows model-minus-frozen-linear monthly net returns in basis points under the stated execution engine. Panel b shows monthly partial correlations with stock forward returns after controlling for the nine original factors. Intervals are pointwise 95\% intervals; Tables 10 and 11 report the separate twelve-test Holm families}\label{fig:5}
\end{figure}
\FloatBarrier
\subsection{Input-intervention return results}
At 10 basis points per side, the primary-condition portfolios in the Luna, Terra and Sol input-intervention batch have the following mean CAGR [minimum, maximum]: Luna 7.29\% [6.31\%, 9.16\%]; Terra 9.15\% [8.64\%, 9.53\%]; Sol 8.79\% [8.28\%, 9.65\%]. Each summary uses three complete trajectories. Tables 3-11 describe the historical ranking sample.

At the primary six-month block length, all 24 paired intervention-return contrasts have bootstrap Holm-adjusted p-values above 0.05. Numerical neutralizations and semantic removal also have HAC adjusted p-values above 0.05. For Luna, reordering rows changes mean monthly net return by 0.542 percentage points. Its pointwise six-month joint-bootstrap interval is [0.154, 1.125] percentage points. The bootstrap Holm-adjusted p-value is 0.098, while the HAC Holm-adjusted p-value is 0.010 within the three row-order comparisons. Across all 24 six-month intervention contrasts as one alternative family, the smallest HAC Holm-adjusted p-value is 0.0815. The inference therefore depends on the uncertainty procedure and the comparison family; the confidence interval is pointwise.

Supplementary Material, Tables S33-S35 report all paired tests at the three block lengths, Tables S36-S39 give all four cost scenarios, and Table S40 compares repeated and altered-input rankings. These returns are estimated from 72 shared market months. Repeated responses improve measurement of model behavior within those months; the interval widths describe return uncertainty.

\FloatBarrier
\section{Discussion}
The audit treats a stock-ranking model as part of an executable economic policy. Repeated requests measure decision variation, frozen rules test reproducible numerical structure, and matched portfolios measure returns associated with preferences beyond those rules.

The frozen mapping transfers without refitting to larger opportunity sets. For both feature-group removals and matched single-feature shifts, it lowers aggregate rank-change prediction error relative to a zero-change prediction. In the matched experiment, entry-and-exit event Jaccard is 0.000-0.125, so these tests support an approximation of aggregate ranking responses more strongly than exact selection changes.

The primary 10-bp-per-side, six-month-block return tests find no Holm-adjusted model-minus-rule difference in the twelve-comparison family; the factor-controlled association tests are also nonsignificant. Three higher-cost, twelve-month-block Terra comparisons favor the rule within their twelve-test sensitivity slices, with no adjusted finding over the full 144-test grid. The separate benchmark test does not establish formal equivalence between model and equal-factor portfolios.

\FloatBarrier
\subsection{Scope and interpretation}
The evidence concerns observable policy under nine structured inputs, the declared rule families and a six-year Chinese equity sample. It does not identify an internal reasoning mechanism or cover systems that use news, retrieval or external tools. The three universes share market months. Archived altered-input conditions have one monthly trajectory, while the matched experiment contributes three responses per condition over 24 stratified months. The repeated input-intervention design includes three complete trajectories per condition for Luna, Terra and Sol, with separate within-batch controls. Service behavior can differ across collection periods, which limits direct comparisons between batches. In CSI 500, frequent selection-boundary ties make reported portfolio outcomes sensitive to the specified anonymous-identifier tie rule; Section S17 reports alternative fixed-priority results.

Because model requests and market-data collection were retrospective, the study identifies historical policy behavior rather than live forecasting ability \citep{r4,r8,r9}. Service labels do not independently verify training cutoffs or backend snapshots. The matched-response design and the final inference rule were fixed before collection.

\FloatBarrier
\section{Conclusion}
Development-fitted linear rules closely approximate the four models' aggregate rankings across stock pools. They lower aggregate rank-change prediction error relative to a zero-change prediction in twelve archived feature-group comparisons and eight matched single-feature comparisons, using the reported separate adjustment families. Prediction of individual entries and exits is less precise. At the primary 10-bp-per-side cost and six-month block length, the twelve-comparison model-minus-rule return family and factor-controlled association tests have no adjusted finding. The higher-cost Terra signals arise in twelve-month-block sensitivity slices and do not survive adjustment over the full cost-by-block-length grid. The Luna row-order return result depends on the uncertainty procedure and comparison family. Frequent score ties in CSI 500 also make some simulated returns sensitive to the tie priority. These experiments make observed rankings and simulated investment outcomes directly testable while leaving live forecasting performance for prospective evaluation.

\section*{Acknowledgments}
We thank Kevin Gifford, Daniel Acuna, Yiming Cheng, Rongbao Wu, and Xiaohua Gao for research discussions, help organizing reference materials, and comments on the presentation and manuscript.
\section*{Supplementary material}
Supplementary Material (included after the declarations): Experimental specifications, exact inputs, conventional model training, data coverage, historical comparisons, stability and risk diagnostics, precision calculations, rule transfer, matched interventions, economic sensitivity, dated replication batches and reproduction procedures.
\bibliography{references}
\section*{Declarations}
\subsection*{Funding}
This research received no external funding.
\subsection*{Competing interests}
The authors have no relevant financial or non-financial interests to disclose.
\subsection*{Author contributions}
Shuai Wu: conceptualization, methodology, software, formal analysis, investigation, visualization, writing - original draft, and project administration. Xue Li: project administration, research coordination, resources, and writing - review and editing. Zhijun Wang: methodology, formal analysis, mathematical validation, academic supervision, and writing - review and editing. Bolun Liu: quantitative finance expertise, financial data interpretation, resources, and writing - review and editing. Weilin Cai: data curation, computational analysis, validation, international research context, and writing - review and editing. All authors reviewed and approved the final manuscript and agree to be accountable for their contributions.
\subsection*{Data availability}
The accompanying research\_\allowbreak{}materials.tar.xz contains model inputs and responses, derived analysis tables, fixed specifications, analysis and authoring code, and a file-integrity manifest. The retained batches contain 10,104 unique valid model responses and 1,260 input files, with reused copies identified separately. A-share market observations come from authorized Tushare access; delisted-board quotations come from the archived Eastmoney source; corporate-event terms and dates come from exchange and issuer disclosures. Source permissions and redistribution conditions apply to each provider. Original market observations, panels and price-bearing transaction and holdings records remain in the authorized local archive. Supplementary Material describes the available derived results, source identities and materials required for portfolio reproduction. Requests concerning restricted source materials should be directed to shuai.wu@colorado.edu.
\subsection*{Code availability}
Analysis code, fixed configurations, prompts and archived model responses accompany the manuscript in research\_\allowbreak{}materials.tar.xz. Numerical analyses use stored responses. Recomputing portfolio returns from market observations additionally requires licensed source data. Supplementary Material gives the relevant files and commands, including the replication procedures.
\subsection*{Ethics approval and consent}
The study uses historical market data and model-generated responses. It does not recruit human participants or execute real securities transactions. Ethics approval and participant consent are not applicable.
\clearpage
\setcounter{table}{0}\renewcommand{\thetable}{S\arabic{table}}\renewcommand{\theHtable}{supplement.\arabic{table}}
\section*{Supplementary Material}
\section*{S1 Experimental design and inference families}
The primary comparison varies the stock-ranking rule while holding the information set and portfolio engine fixed. Table S1 summarizes the shared investment environment. The Luna, Terra and Sol historical grid contains 2,772 tasks, and Astra has 924 tasks under the same design. Source records 001 through 007 document the three-model historical configuration; records 008 and 009 document the development-planning sample and request settings. Record 010 defines the security-identity crosswalk for historical risk covariates. Table S2 summarizes these specifications and identifies their source records.

\begingroup\small\setlength{\tabcolsep}{3pt}\renewcommand{\arraystretch}{1.17}
\begin{longtable}{@{}>{\raggedright\arraybackslash}p{\dimexpr 0.285714\linewidth-1.71429pt\relax}>{\raggedright\arraybackslash}p{\dimexpr 0.714286\linewidth-4.28571pt\relax}@{}}
\caption{Shared experimental design}\label{tab:S1}\\
\toprule
\textbf{Component} & \textbf{Specification}\\\midrule\endfirsthead
\multicolumn{2}{l}{\textbf{Table S1 (continued)}}\\\toprule
\textbf{Component} & \textbf{Specification}\\\midrule\endhead
\midrule\multicolumn{2}{r}{Continued on next page}\\\endfoot
\bottomrule\endlastfoot
Universes & SSE50 (000016.SH), CSI300 (000300.SH), CSI500 (000905.SH)\\
Models & gpt-5.6-luna, gpt-5.6-terra, gpt-5.6-sol, gpt-6-astra\\
Development signals & 2014-12 to 2019-11; 60 monthly signals per universe\\
Historical test signals & 2019-12 to 2025-11; 72 monthly signals per universe\\
Main repetitions & SSE50: 5 complete trajectories per model; CSI300: 1 complete trajectory per model; CSI500: 1 complete trajectory per model; identical prompts across repetitions\\
Portfolio & Highest-scoring ceil(0.2N) eligible stocks, equally weighted; long only\\
Membership & Latest constituent snapshot at or before prior calendar month end; maximum age 99 days.\\
Signal and execution & Last SSE trading day of month after close. Second SSE trading day of next month open; exit second trading day of following month open.\\
Costs & 0, 10, 25, 50 basis points each side; primary comparison at 10\\
Comparators & Equal factor, Equal factor group, Ridge, Gradient boosting, Universe equal weight\\
Mechanisms & SSE 50 only: all 72 test months, one complete run per model in each of three factor-neutralization, hidden-semantics and reordered-row conditions; paired with primary run 1.\\
\end{longtable}
\noindent\footnotesize Repeated model trajectories (five in the SSE 50, one in each of the CSI 300 and CSI 500) measure variation in the ranking procedure without adding market months.\par\endgroup
\begingroup\small\setlength{\tabcolsep}{3pt}\renewcommand{\arraystretch}{1.17}
\begin{longtable}{@{}>{\raggedright\arraybackslash}p{\dimexpr 0.092199\linewidth-1.10638pt\relax}>{\raggedright\arraybackslash}p{\dimexpr 0.198582\linewidth-2.38298pt\relax}>{\raggedright\arraybackslash}p{\dimexpr 0.709220\linewidth-8.51064pt\relax}@{}}
\caption{Experimental specifications by source record}\label{tab:S2}\\
\toprule
\textbf{Source record} & \textbf{Design component} & \textbf{Specification}\\\midrule\endfirsthead
\multicolumn{3}{l}{\textbf{Table S2 (continued)}}\\\toprule
\textbf{Source record} & \textbf{Design component} & \textbf{Specification}\\\midrule\endhead
\midrule\multicolumn{3}{r}{Continued on next page}\\\endfoot
\bottomrule\endlastfoot
001 & Scoring & One score in [0,1] for every candidate; exact-key schema; descending scores and shared anonymous-identifier tie breaking.\\
002 & Comparison framework & Shared tie rule; development-fitted proxies; historical risk matching; full-run stability; declared mechanism contrasts and development-based sample-size rule.\\
003 & Disjoint samples & Mutually exclusive candidate subsets formed from the original universe scores; industry classifications follow taxonomy release dates.\\
004 & Industry history & Risk matching uses official dated SW membership history, taxonomy-based classifications and the recorded beta reference.\\
005 & Historical grid & Historical three-model grid of 2,772 predictions: five SSE 50 main runs, one CSI 300 and CSI 500 main run each, SSE 50 development and mechanism conditions. Model interface: https:/\allowbreak{}/\allowbreak{}api.openai.com/\allowbreak{}v1/\allowbreak{}responses.\\
006 & Request interface & Terra and Sol request configuration using https:/\allowbreak{}/\allowbreak{}api.openai.com/\allowbreak{}v1/\allowbreak{}responses.\\
007 & Request settings & All three models use the Responses interface, https:/\allowbreak{}/\allowbreak{}api.openai.com/\allowbreak{}v1/\allowbreak{}responses, with default requested tier for this batch.\\
008 & Return-precision sample & Return-precision planning uses 360 primary-condition predictions over 120 registered inputs: one run per model in CSI 300 and CSI 500 over 60 development months (2014-12 through 2019-11). The 2,772-task historical grid and the planning sample have separate inference roles. The three CSI 300 comparisons determine the sample-size calculation; CSI 500 supplies additional variance estimates.\\
009 & Planning requests & Development-only planning configuration: https:/\allowbreak{}/\allowbreak{}api.openai.com/\allowbreak{}v1/\allowbreak{}responses, default requested tier, medium reasoning, no tools, independent streamed Responses calls and all-candidate scoring.\\
010 & Security identities & Historical risk covariates link SW source code 300114 to market-panel code 302132.SZ as the continuing security, following the issuer announcement of the code change effective 2025-02-17. The three earlier SW rows keep their source codes, industry labels, effective dates and provider update timestamps. Interval ends follow the next dated record, with record 004's July 2021 taxonomy announcement as the classification boundary. Classification intervals begin at their recorded effective dates. This crosswalk governs industry attribution in the historical risk diagnostic.\\
\end{longtable}
\noindent\footnotesize Source records and their UTC timestamps are in study\_\allowbreak{}v3/\allowbreak{}historical\_\allowbreak{}protocol.json and records 001 through 010 in study\_\allowbreak{}v3/\allowbreak{}amendments/\allowbreak{}. Records 008 and 009 describe the development-planning sample and request configuration; record 010 defines the security-identity crosswalk.\par\endgroup
All historical comparisons are exploratory.

The complete four-model historical grid has 3,696 valid predictions. A separate 360-prediction development batch from the three models supplies return-precision inputs. The matched single-feature grid has 864 valid responses. Astra contributes 924 historical, 216 matched and 1,296 replication responses, totaling 2,436 unique valid responses. Four Astra hidden-semantics responses have constant scores, for which rank correlations are undefined. The 27 September replication batch contributes 3,888 responses from Luna, Terra and Sol. Total retained valid responses across the five archive categories are 10,104; the separate 360-response planning archive is used for development-based precision calculations.

\begingroup\small\setlength{\tabcolsep}{3pt}\renewcommand{\arraystretch}{1.17}
\begin{longtable}{@{}>{\raggedright\arraybackslash}p{\dimexpr 0.473684\linewidth-5.68421pt\relax}>{\raggedright\arraybackslash}p{\dimexpr 0.263158\linewidth-3.15789pt\relax}>{\raggedright\arraybackslash}p{\dimexpr 0.263158\linewidth-3.15789pt\relax}@{}}
\caption{Comparison families in the integrated evaluation}\label{tab:S3}\\
\toprule
\textbf{Family} & \textbf{Tests} & \textbf{Endpoint}\\\midrule\endfirsthead
\multicolumn{3}{l}{\textbf{Table S3 (continued)}}\\\toprule
\textbf{Family} & \textbf{Tests} & \textbf{Endpoint}\\\midrule\endhead
\midrule\multicolumn{3}{r}{Continued on next page}\\\endfoot
\bottomrule\endlastfoot
Primary return & 12 & Per-universe LLM run mean minus equal-factor return\\
Other conventional comparators & 48 & Per-universe LLM run mean minus each of the other four baselines\\
Three-model factor ablation & 9 & Ablated LLM run 1 minus primary LLM run 1\\
Three-model feature semantics & 3 & Anonymous-feature run 1 minus primary run 1\\
Three-model row order & 3 & Reordered-input run 1 minus primary run 1\\
Three-model baseline-adjusted ablation & 9 & LLM ablation change minus the corresponding equal-factor change\\
Three-model SSE 50 proxy diagnostic & 3 & Five-run SSE 50 LLM mean minus its development-selected proxy\\
Risk-matched return & 12 & Per-universe risk-matched LLM run mean minus risk-matched equal factor\\
Disjoint candidates & 12 & Per-universe LLM run mean minus equal factor within the assigned subset\\
Economic equivalence & 12 & Primary mean difference lies within +/\allowbreak{}-0.10 percentage points\\
Astra historical numerical return changes & 3 + 3 & Raw and baseline-adjusted changes; separate families\\
Astra historical semantics and row order & 1 + 1 & Separate single-contrast return tests\\
Astra three-run replication & 3 + 1 + 1 & Numerical removal family; separate semantics/\allowbreak{}order tests\\
Rule complexity & 3 + 1 & Bounded forest minus frozen-linear rank agreement\\
Archived response prediction & 9 + 3 & Zero-change MSE minus frozen predicted-change MSE\\
Matched response prediction & 6 + 2 & Matched single-feature MSE improvement\\
Final incremental investment value & 12 & Model minus frozen-linear net return across three pools\\
Final remaining information & 12 & Partial return association after nine-factor controls\\
Three-model replication & 9 + 3 + 3 + 9 & Numerical, semantics, row order and equal-factor-adjusted return families; separate from Astra\\
\end{longtable}
\noindent\footnotesize The response and complexity tests carry plus signs that separate the three-model and Astra families; each is adjusted independently. Corrections apply separately within each family and block-length specification; there is no global familywise claim. The three-model proxy diagnostic is retained for provenance. The incremental-value analysis uses its own twelve-test family and is not counted as an independent replication of the SSE 50 contrasts. Matched responses use holding-year-stratified month clusters; other primary bootstrap results use six-month blocks.\par\endgroup
\section*{S2 Exact model inputs and scoring instructions}
The following system instruction and user matrix are reproduced from study\_\allowbreak{}v3/\allowbreak{}scored/\allowbreak{}inputs/\allowbreak{}SSE50/\allowbreak{}primary/\allowbreak{}20191231.json. This archived SSE 50 input was used by all four models and all five primary runs for this signal.

You are the stock-selection decision function in a controlled financial research experiment. Use only the provided numerical feature matrix. Do not use external information or tools. The anonymous identifiers carry no meaning. Assign each candidate a desirability score from 0 to 1 for expected relative total return over the next monthly holding period. Higher scores mean stronger preferences. The equal-weight long-only portfolio will select the highest-scoring 10 candidates. These scores express relative preference, not calibrated return forecasts. Each number is a cross-sectional percentile in [0,1]. A larger number represents a stronger exposure to the named investment characteristic; a missing observation has the neutral value 0.5. Use economic judgment about factor combinations and trade-offs. Do not assume that any factor always predicts a positive return, and do not simply sort by the first column. No return labels or historical identities are provided. Return a JSON object with the key "scores", containing exactly one numeric score for every candidate identifier. Score every candidate; do not omit candidates. Use enough precision to distinguish preferences. Return no commentary or additional keys.

Feature order and definitions:

book\_\allowbreak{}to\_\allowbreak{}price: higher book equity per unit of market price.

earnings\_\allowbreak{}yield: higher trailing earnings per unit of market price.

dividend\_\allowbreak{}yield: higher trailing dividend yield.

momentum\_\allowbreak{}12\_\allowbreak{}1: higher adjusted return from 252 to 21 trading sessions ago.

momentum\_\allowbreak{}6\_\allowbreak{}1: higher adjusted return from 126 to 21 trading sessions ago.

reversal\_\allowbreak{}1: lower adjusted return over the last 21 sessions.

low\_\allowbreak{}volatility\_\allowbreak{}63: lower annualized standard deviation of daily adjusted returns over the last 63 sessions.

small\_\allowbreak{}size: lower floating market capitalization.

low\_\allowbreak{}turnover: lower turnover rate at the signal date.

Candidates in CSV format:

\begin{lstlisting}[basicstyle=\small\ttfamily,breaklines=true,columns=fullflexible,keepspaces=true]
id,book_to_price,earnings_yield,dividend_yield,momentum_12_1,momentum_6_1,reversal_1,low_volatility_63,small_size,low_turnover
\end{lstlisting}
\begin{lstlisting}[basicstyle=\small\ttfamily,breaklines=true,columns=fullflexible,keepspaces=true]
A047,0.9000,0.9000,0.8600,0.3800,0.6400,0.9300,0.9800,0.0600,0.9200
\end{lstlisting}
\begin{lstlisting}[basicstyle=\small\ttfamily,breaklines=true,columns=fullflexible,keepspaces=true]
A050,0.8200,0.5200,0.9800,0.1800,0.3600,0.7600,0.9600,0.1800,0.9000
\end{lstlisting}
\begin{lstlisting}[basicstyle=\small\ttfamily,breaklines=true,columns=fullflexible,keepspaces=true]
A042,0.2200,0.4800,0.4200,0.6200,0.6000,0.1200,0.1000,0.6600,0.4000
\end{lstlisting}
\begin{lstlisting}[basicstyle=\small\ttfamily,breaklines=true,columns=fullflexible,keepspaces=true]
A030,0.1800,0.1000,0.2600,0.6800,1.0000,0.3400,0.0200,0.8800,0.0800
\end{lstlisting}
\begin{lstlisting}[basicstyle=\small\ttfamily,breaklines=true,columns=fullflexible,keepspaces=true]
A004,0.3000,0.6400,0.7000,0.9000,0.8200,0.0600,0.4000,0.4000,0.2600
\end{lstlisting}
\begin{lstlisting}[basicstyle=\small\ttfamily,breaklines=true,columns=fullflexible,keepspaces=true]
A032,0.6600,0.7000,0.8800,0.2200,0.2200,0.4800,0.7200,0.2600,0.6600
\end{lstlisting}
\begin{lstlisting}[basicstyle=\small\ttfamily,breaklines=true,columns=fullflexible,keepspaces=true]
A029,0.4800,0.4400,0.4800,0.0600,0.1600,0.7400,0.8600,0.5000,0.7600
\end{lstlisting}
\begin{lstlisting}[basicstyle=\small\ttfamily,breaklines=true,columns=fullflexible,keepspaces=true]
A036,1.0000,0.9800,0.9400,0.2600,0.4400,0.8800,0.9200,0.3800,0.8200
\end{lstlisting}
\begin{lstlisting}[basicstyle=\small\ttfamily,breaklines=true,columns=fullflexible,keepspaces=true]
A003,0.2800,0.0800,0.5800,0.1400,0.2000,0.0400,0.1600,0.8600,0.1600
\end{lstlisting}
\begin{lstlisting}[basicstyle=\small\ttfamily,breaklines=true,columns=fullflexible,keepspaces=true]
A002,0.4400,0.5800,0.5400,0.7400,0.7200,0.7200,0.6600,0.1400,0.7200
\end{lstlisting}
\begin{lstlisting}[basicstyle=\small\ttfamily,breaklines=true,columns=fullflexible,keepspaces=true]
A034,0.8400,0.7200,0.4400,0.0800,0.1800,0.4600,0.7600,0.7000,0.8000
\end{lstlisting}
\begin{lstlisting}[basicstyle=\small\ttfamily,breaklines=true,columns=fullflexible,keepspaces=true]
A026,0.6000,0.2000,0.1600,0.2000,0.1400,0.3200,0.2800,0.9200,0.1800
\end{lstlisting}
\begin{lstlisting}[basicstyle=\small\ttfamily,breaklines=true,columns=fullflexible,keepspaces=true]
A013,0.0400,0.1800,0.3000,0.9800,0.8400,0.6600,0.5000,0.0400,0.7800
\end{lstlisting}
\begin{lstlisting}[basicstyle=\small\ttfamily,breaklines=true,columns=fullflexible,keepspaces=true]
A012,0.7200,0.7800,0.8200,0.5200,0.5800,0.9600,0.9000,0.9000,0.3400
\end{lstlisting}
\begin{lstlisting}[basicstyle=\small\ttfamily,breaklines=true,columns=fullflexible,keepspaces=true]
A014,0.5600,0.0200,0.0200,0.6400,0.4000,1.0000,0.8200,0.7600,0.6800
\end{lstlisting}
\begin{lstlisting}[basicstyle=\small\ttfamily,breaklines=true,columns=fullflexible,keepspaces=true]
A016,0.5000,0.2400,0.3400,0.3200,0.0400,0.0800,0.3000,0.5200,0.1000
\end{lstlisting}
\begin{lstlisting}[basicstyle=\small\ttfamily,breaklines=true,columns=fullflexible,keepspaces=true]
A033,0.0800,0.0600,0.0600,0.8800,0.9800,0.5800,0.0600,0.7800,0.3000
\end{lstlisting}
\begin{lstlisting}[basicstyle=\small\ttfamily,breaklines=true,columns=fullflexible,keepspaces=true]
A044,0.3800,0.5600,0.4000,0.3400,0.1000,0.2400,0.6000,0.7400,0.2800
\end{lstlisting}
\begin{lstlisting}[basicstyle=\small\ttfamily,breaklines=true,columns=fullflexible,keepspaces=true]
A020,0.6800,0.9600,0.6600,0.1200,0.2600,0.2600,0.6800,0.3400,0.5400
\end{lstlisting}
\begin{lstlisting}[basicstyle=\small\ttfamily,breaklines=true,columns=fullflexible,keepspaces=true]
A031,0.9200,0.8800,0.7800,0.5600,0.8000,0.5000,0.2600,0.4600,0.7000
\end{lstlisting}
\begin{lstlisting}[basicstyle=\small\ttfamily,breaklines=true,columns=fullflexible,keepspaces=true]
A027,0.5800,0.3200,0.3600,0.4200,0.3800,0.2800,0.5200,0.5600,0.4200
\end{lstlisting}
\begin{lstlisting}[basicstyle=\small\ttfamily,breaklines=true,columns=fullflexible,keepspaces=true]
A018,0.2600,0.6000,0.5000,0.8000,0.4600,0.8600,0.7800,0.1000,0.5600
\end{lstlisting}
\begin{lstlisting}[basicstyle=\small\ttfamily,breaklines=true,columns=fullflexible,keepspaces=true]
A023,0.8000,0.7600,0.4600,0.1000,0.5600,0.5400,0.5600,0.6800,0.4600
\end{lstlisting}
\begin{lstlisting}[basicstyle=\small\ttfamily,breaklines=true,columns=fullflexible,keepspaces=true]
A037,0.5200,0.4000,0.2400,0.4000,0.3000,0.2000,0.2200,0.8000,0.3200
\end{lstlisting}
\begin{lstlisting}[basicstyle=\small\ttfamily,breaklines=true,columns=fullflexible,keepspaces=true]
A008,0.7700,0.8600,0.7200,0.7000,0.7400,0.6000,0.6200,0.2200,0.6200
\end{lstlisting}
\begin{lstlisting}[basicstyle=\small\ttfamily,breaklines=true,columns=fullflexible,keepspaces=true]
A046,0.9600,0.9400,0.9000,0.3600,0.5000,0.9300,1.0000,0.1200,0.9800
\end{lstlisting}
\begin{lstlisting}[basicstyle=\small\ttfamily,breaklines=true,columns=fullflexible,keepspaces=true]
A038,0.8800,0.9200,0.6200,0.5800,0.6600,0.7000,0.6400,0.2400,0.8800
\end{lstlisting}
\begin{lstlisting}[basicstyle=\small\ttfamily,breaklines=true,columns=fullflexible,keepspaces=true]
A024,0.4200,0.4600,0.1200,0.4600,0.0200,0.1400,0.1200,0.9600,0.1400
\end{lstlisting}
\begin{lstlisting}[basicstyle=\small\ttfamily,breaklines=true,columns=fullflexible,keepspaces=true]
A022,0.8600,0.7400,0.5600,0.0200,0.0600,0.7800,0.3400,0.7200,0.3600
\end{lstlisting}
\begin{lstlisting}[basicstyle=\small\ttfamily,breaklines=true,columns=fullflexible,keepspaces=true]
A039,0.9800,1.0000,0.9600,0.5000,0.6800,0.9000,0.9400,0.3600,0.8400
\end{lstlisting}
\begin{lstlisting}[basicstyle=\small\ttfamily,breaklines=true,columns=fullflexible,keepspaces=true]
A021,0.2000,0.5000,0.3800,0.9400,0.9000,0.1600,0.3800,0.5400,0.2000
\end{lstlisting}
\begin{lstlisting}[basicstyle=\small\ttfamily,breaklines=true,columns=fullflexible,keepspaces=true]
A019,0.1400,0.3400,0.1800,0.8200,0.9600,0.3800,0.0800,0.9800,0.0400
\end{lstlisting}
\begin{lstlisting}[basicstyle=\small\ttfamily,breaklines=true,columns=fullflexible,keepspaces=true]
A017,0.7400,0.2600,0.6400,0.0400,0.0800,0.5600,0.8400,0.0800,1.0000
\end{lstlisting}
\begin{lstlisting}[basicstyle=\small\ttfamily,breaklines=true,columns=fullflexible,keepspaces=true]
A049,0.2400,0.3600,0.0800,0.8600,0.8800,0.4000,0.1800,0.1600,0.9400
\end{lstlisting}
\begin{lstlisting}[basicstyle=\small\ttfamily,breaklines=true,columns=fullflexible,keepspaces=true]
A006,0.5400,0.3800,0.2200,0.8400,0.5400,0.2200,0.3600,0.6200,0.1200
\end{lstlisting}
\begin{lstlisting}[basicstyle=\small\ttfamily,breaklines=true,columns=fullflexible,keepspaces=true]
A015,0.6200,0.1600,0.2000,0.4400,0.3200,0.6200,0.7000,0.6400,0.5800
\end{lstlisting}
\begin{lstlisting}[basicstyle=\small\ttfamily,breaklines=true,columns=fullflexible,keepspaces=true]
A011,0.9400,0.6800,1.0000,0.1600,0.1200,0.5200,0.5800,0.5800,0.5200
\end{lstlisting}
\begin{lstlisting}[basicstyle=\small\ttfamily,breaklines=true,columns=fullflexible,keepspaces=true]
A025,0.0600,0.1400,0.1400,0.7800,0.4200,0.4200,0.3200,0.4800,0.6000
\end{lstlisting}
\begin{lstlisting}[basicstyle=\small\ttfamily,breaklines=true,columns=fullflexible,keepspaces=true]
A045,0.1600,0.4200,0.7400,0.9200,0.8600,0.1800,0.4200,0.8400,0.2400
\end{lstlisting}
\begin{lstlisting}[basicstyle=\small\ttfamily,breaklines=true,columns=fullflexible,keepspaces=true]
A007,0.6400,0.6200,0.9200,0.2400,0.4800,0.8000,0.7400,0.2800,0.7400
\end{lstlisting}
\begin{lstlisting}[basicstyle=\small\ttfamily,breaklines=true,columns=fullflexible,keepspaces=true]
A010,0.1200,0.1200,0.1000,1.0000,0.7800,0.0200,0.0400,1.0000,0.0200
\end{lstlisting}
\begin{lstlisting}[basicstyle=\small\ttfamily,breaklines=true,columns=fullflexible,keepspaces=true]
A005,0.4600,0.6600,0.6800,0.6600,0.9200,0.3600,0.2000,0.4200,0.3800
\end{lstlisting}
\begin{lstlisting}[basicstyle=\small\ttfamily,breaklines=true,columns=fullflexible,keepspaces=true]
A028,0.7000,0.8200,0.7600,0.3000,0.7600,0.8400,0.8000,0.6000,0.8600
\end{lstlisting}
\begin{lstlisting}[basicstyle=\small\ttfamily,breaklines=true,columns=fullflexible,keepspaces=true]
A001,0.3200,0.2800,0.2800,0.2800,0.6200,0.4400,0.4400,0.9400,0.2200
\end{lstlisting}
\begin{lstlisting}[basicstyle=\small\ttfamily,breaklines=true,columns=fullflexible,keepspaces=true]
A048,0.3400,0.8400,0.8000,0.4800,0.2800,0.8200,0.5400,0.8200,0.5000
\end{lstlisting}
\begin{lstlisting}[basicstyle=\small\ttfamily,breaklines=true,columns=fullflexible,keepspaces=true]
A009,0.0200,0.0400,0.0400,0.9600,0.9400,0.6400,0.1400,0.2000,0.4400
\end{lstlisting}
\begin{lstlisting}[basicstyle=\small\ttfamily,breaklines=true,columns=fullflexible,keepspaces=true]
A041,0.3600,0.5400,0.6000,0.6000,0.5200,0.3000,0.4800,0.3200,0.6400
\end{lstlisting}
\begin{lstlisting}[basicstyle=\small\ttfamily,breaklines=true,columns=fullflexible,keepspaces=true]
A035,0.1000,0.2200,0.5200,0.7600,0.2400,0.6800,0.4600,0.4400,0.4800
\end{lstlisting}
\begin{lstlisting}[basicstyle=\small\ttfamily,breaklines=true,columns=fullflexible,keepspaces=true]
A040,0.7700,0.8000,0.8400,0.5400,0.7000,0.9800,0.8800,0.0200,0.9600
\end{lstlisting}
\begin{lstlisting}[basicstyle=\small\ttfamily,breaklines=true,columns=fullflexible,keepspaces=true]
A043,0.4000,0.3000,0.3200,0.7200,0.3400,0.1000,0.2400,0.3000,0.0600
\end{lstlisting}
The request also transmits the exact-key response schema saved in response\_\allowbreak{}schema. Its scores object requires 50 candidate keys, each with a numeric value between 0 and 1, and disallows additional keys. The server receives the system instruction, user matrix and schema. Signal dates, stock mappings and future returns are local records. Score ties use ascending anonymous identifier.

\begingroup\small\setlength{\tabcolsep}{3pt}\renewcommand{\arraystretch}{1.17}
\begin{longtable}{@{}>{\raggedright\arraybackslash}p{\dimexpr 0.375000\linewidth-6.75000pt\relax}>{\raggedright\arraybackslash}p{\dimexpr 0.208333\linewidth-3.75000pt\relax}>{\raggedright\arraybackslash}p{\dimexpr 0.208333\linewidth-3.75000pt\relax}>{\raggedright\arraybackslash}p{\dimexpr 0.208333\linewidth-3.75000pt\relax}@{}}
\caption{Exact input archive}\label{tab:S4}\\
\toprule
\textbf{Condition} & \textbf{Candidates} & \textbf{Selected} & \textbf{Example input file}\\\midrule\endfirsthead
\multicolumn{4}{l}{\textbf{Table S4 (continued)}}\\\toprule
\textbf{Condition} & \textbf{Candidates} & \textbf{Selected} & \textbf{Example input file}\\\midrule\endhead
\midrule\multicolumn{4}{r}{Continued on next page}\\\endfoot
\bottomrule\endlastfoot
Primary & 50 & 10 & SSE50/\allowbreak{}primary/\allowbreak{}20191231.json\\
Valuation removed & 50 & 10 & SSE50/\allowbreak{}no\_\allowbreak{}valuation/\allowbreak{}20191231.json\\
Momentum removed & 50 & 10 & SSE50/\allowbreak{}no\_\allowbreak{}momentum/\allowbreak{}20191231.json\\
Risk/\allowbreak{}trading removed & 50 & 10 & SSE50/\allowbreak{}no\_\allowbreak{}risk\_\allowbreak{}trading/\allowbreak{}20191231.json\\
Semantics removed & 50 & 10 & SSE50/\allowbreak{}no\_\allowbreak{}semantics/\allowbreak{}20191231.json\\
Row order changed & 50 & 10 & SSE50/\allowbreak{}row\_\allowbreak{}order/\allowbreak{}20191231.json\\
\end{longtable}
\noindent\footnotesize Paths are relative to study\_\allowbreak{}v3/\allowbreak{}scored/\allowbreak{}inputs/\allowbreak{}. Each universe and month has its own archived matrix, anonymous mapping, display order and schema. The three factor removals replace only their specified columns with 0.5000; no\_\allowbreak{}semantics changes the definitions and labels; row\_\allowbreak{}order changes display order. The primary condition also has a separate archived input for each development signal.\par\endgroup
\section*{S3 Conventional training and parameter selection}
Ridge considers alpha values [0.1, 1, 10, 100]; histogram gradient boosting considers [7, 15, 31] leaf counts and L2 values [1, 10]. The boosting settings are 200 iterations, learning rate 0.05 and minimum leaf size 40. Selection maximizes mean monthly validation rank IC over 2017, 2018, 2019, using validation labels that exit by 20191130.

\begingroup\small\setlength{\tabcolsep}{3pt}\renewcommand{\arraystretch}{1.17}
\begin{longtable}{@{}>{\raggedright\arraybackslash}p{\dimexpr 0.310345\linewidth-7.44828pt\relax}>{\raggedright\arraybackslash}p{\dimexpr 0.172414\linewidth-4.13793pt\relax}>{\raggedright\arraybackslash}p{\dimexpr 0.172414\linewidth-4.13793pt\relax}>{\raggedright\arraybackslash}p{\dimexpr 0.172414\linewidth-4.13793pt\relax}>{\raggedright\arraybackslash}p{\dimexpr 0.172414\linewidth-4.13793pt\relax}@{}}
\caption{Selected parameters in six universe-condition configurations}\label{tab:S5}\\
\toprule
\textbf{Universe} & \textbf{Condition} & \textbf{Ridge alpha} & \textbf{Boosting parameters} & \textbf{Validation IC: Ridge /\allowbreak{} boosting}\\\midrule\endfirsthead
\multicolumn{5}{l}{\textbf{Table S5 (continued)}}\\\toprule
\textbf{Universe} & \textbf{Condition} & \textbf{Ridge alpha} & \textbf{Boosting parameters} & \textbf{Validation IC: Ridge /\allowbreak{} boosting}\\\midrule\endhead
\midrule\multicolumn{5}{r}{Continued on next page}\\\endfoot
\bottomrule\endlastfoot
SSE50 & Primary & 100 & leaves=15; L2=1 & 0.0045/\allowbreak{}0.0045\\
SSE50 & Valuation removed & 100 & leaves=31; L2=10 & 0.0127/\allowbreak{}0.0060\\
SSE50 & Momentum removed & 100 & leaves=15; L2=1 & 0.0101/\allowbreak{}-0.0061\\
SSE50 & Risk/\allowbreak{}trading removed & 100 & leaves=7; L2=1 & 0.0260/\allowbreak{}0.0052\\
CSI300 & Primary & 100 & leaves=15; L2=1 & 0.0163/\allowbreak{}0.0151\\
CSI500 & Primary & 100 & leaves=31; L2=1 & 0.0316/\allowbreak{}0.0324\\
\end{longtable}
\noindent\footnotesize Every candidate, fold boundary and validation score is retained in analysis/\allowbreak{}results/\allowbreak{}<universe>/\allowbreak{}<condition>/\allowbreak{}hyperparameter\_\allowbreak{}validation.json. The reported IC fields are mean\_\allowbreak{}rank\_\allowbreak{}ic; selection never uses the historical test returns.\par\endgroup
\begingroup\small\setlength{\tabcolsep}{3pt}\renewcommand{\arraystretch}{1.17}
\begin{longtable}{@{}>{\raggedright\arraybackslash}p{\dimexpr 0.310345\linewidth-7.44828pt\relax}>{\raggedright\arraybackslash}p{\dimexpr 0.172414\linewidth-4.13793pt\relax}>{\raggedright\arraybackslash}p{\dimexpr 0.172414\linewidth-4.13793pt\relax}>{\raggedright\arraybackslash}p{\dimexpr 0.172414\linewidth-4.13793pt\relax}>{\raggedright\arraybackslash}p{\dimexpr 0.172414\linewidth-4.13793pt\relax}@{}}
\caption{Expanding training samples}\label{tab:S6}\\
\toprule
\textbf{Universe} & \textbf{Condition} & \textbf{Rows: first to last} & \textbf{Months: first to last} & \textbf{Final signal /\allowbreak{} latest label exit}\\\midrule\endfirsthead
\multicolumn{5}{l}{\textbf{Table S6 (continued)}}\\\toprule
\textbf{Universe} & \textbf{Condition} & \textbf{Rows: first to last} & \textbf{Months: first to last} & \textbf{Final signal /\allowbreak{} latest label exit}\\\midrule\endhead
\midrule\multicolumn{5}{r}{Continued on next page}\\\endfoot
\bottomrule\endlastfoot
SSE50 & Primary & 2,750 to 6,263 & 59 to 130 & 20251128 /\allowbreak{} 20251104\\
SSE50 & Valuation removed & 2,750 to 6,263 & 59 to 130 & 20251128 /\allowbreak{} 20251104\\
SSE50 & Momentum removed & 2,750 to 6,263 & 59 to 130 & 20251128 /\allowbreak{} 20251104\\
SSE50 & Risk/\allowbreak{}trading removed & 2,750 to 6,263 & 59 to 130 & 20251128 /\allowbreak{} 20251104\\
CSI300 & Primary & 15,819 to 36,966 & 59 to 130 & 20251128 /\allowbreak{} 20251104\\
CSI500 & Primary & 24,961 to 60,263 & 59 to 130 & 20251128 /\allowbreak{} 20251104\\
\end{longtable}
\noindent\footnotesize Each of the six training\_\allowbreak{}audit.csv files contains all 72 historical test signals. Training uses only labels whose exit date precedes the current signal. The corresponding baseline\_\allowbreak{}scores.parquet and ridge\_\allowbreak{}coefficients.csv preserve scores and fitted coefficients.\par\endgroup
Each validation fold trains on labels completed before its first signal. Candidate performance is the mean monthly Spearman correlation between predictions and stock returns. Gradient boosting uses no early stopping. All candidate results and selected parameters are retained separately for each universe and condition.

Share exchanges preserve the investor's economic entitlement across security codes. At the record-date close, the engine converts held adjusted units to original shares using that date's adjustment factor, applies the announced exchange ratio and records a locked claim on the successor. The registration date records completion of registration; the claim becomes tradable at 09:30 on its announced availability date. Where the successor already has an observed market price, locked claims are valued using contemporaneous successor quotations. Claims on a successor awaiting listing retain the known value of the surrendered holding until a market quote becomes available. Successor adjustment factors convert original shares into adjusted accounting units without changing the announced share entitlement.

A delisted security transferred to the two-network and delisted-company board retains its share entitlement under the successor code. The engine attempts disposal only at the 15:00 auction on a scheduled monthly rebalance day. Suspension, an absent auction, zero volume or a closing price at the lower limit leaves the position invested until a later scheduled disposal. Afternoon proceeds become available for the next monthly opening rebalance. For a previous price of n integer cents, the daily permitted movement is max(1, floor(n/\allowbreak{}20)) cents, with a one-cent minimum price; first-day limits follow the recorded listing terms. These prices and volumes come from the archived Eastmoney quotation source. The historical record contains 255 observed 400256 auction dates; an independent second quotation source is unavailable. Security 400174 remains suspended throughout its recorded period and is valued at its last actual A-share price of CNY 0.40. Shares issued to restructuring investors and creditors are assigned to those recipients rather than to the pre-existing shareholder claim. Source records distinguish the original official announcement URLs from the downloaded public issuer-disclosure mirrors for the two board-transfer notices; their byte identity with the original official files is unverified.

All stock-level forward-return labels use the same zero-cost economic-wealth calculation between the entry and exit 09:30 endpoints, including share exchanges, locked entitlements and board transfers. The initial mark uses the actual opening quotation or the last earlier observed close during suspension. These labels measure the return on a held economic claim. Strategy purchases additionally require an executable opening quotation and satisfy the engine's price-limit checks. Supervised return targets, selection hit rates and factor-controlled return associations use this common stock-wealth definition.

The archived company-event and board-price verification window ends on 06 January 2026. Later settlement windows require an explicitly dated verified event and quotation version. Each economic claim has a retained event identifier, original and successor code, entitlement quantity, valuation status and availability date. Daily snapshots at rebalances, auctions, company events and termination accompany every cost scenario.

\section*{S4 Data coverage and missing characteristics}
\begingroup\small\setlength{\tabcolsep}{3pt}\renewcommand{\arraystretch}{1.17}
\begin{longtable}{@{}>{\raggedright\arraybackslash}p{\dimexpr 0.310345\linewidth-7.44828pt\relax}>{\raggedright\arraybackslash}p{\dimexpr 0.172414\linewidth-4.13793pt\relax}>{\raggedright\arraybackslash}p{\dimexpr 0.172414\linewidth-4.13793pt\relax}>{\raggedright\arraybackslash}p{\dimexpr 0.172414\linewidth-4.13793pt\relax}>{\raggedright\arraybackslash}p{\dimexpr 0.172414\linewidth-4.13793pt\relax}@{}}
\caption{Contemporaneous panel coverage}\label{tab:S7}\\
\toprule
\textbf{Universe} & \textbf{Stock-month rows} & \textbf{Distinct stocks} & \textbf{Test stock-months} & \textbf{Test candidates per month}\\\midrule\endfirsthead
\multicolumn{5}{l}{\textbf{Table S7 (continued)}}\\\toprule
\textbf{Universe} & \textbf{Stock-month rows} & \textbf{Distinct stocks} & \textbf{Test stock-months} & \textbf{Test candidates per month}\\\midrule\endhead
\midrule\multicolumn{5}{r}{Continued on next page}\\\endfoot
\bottomrule\endlastfoot
SSE50 & 6,363 & 127 & 3563 & 47 to 50\\
CSI300 & 37,562 & 661 & 21446 & 292 to 300\\
CSI500 & 61,262 & 1311 & 35806 & 490 to 500\\
\end{longtable}
\noindent\footnotesize Each panel has 132 monthly signals: 60 development signals and 72 test signals. Monthly membership, eligibility and exclusion counts are in study\_\allowbreak{}v3/\allowbreak{}data/\allowbreak{}<universe>\_\allowbreak{}panel\_\allowbreak{}audit.json.\par\endgroup
\begingroup\small\setlength{\tabcolsep}{3pt}\renewcommand{\arraystretch}{1.17}
\begin{longtable}{@{}>{\raggedright\arraybackslash}p{\dimexpr 0.310345\linewidth-7.44828pt\relax}>{\raggedright\arraybackslash}p{\dimexpr 0.172414\linewidth-4.13793pt\relax}>{\raggedright\arraybackslash}p{\dimexpr 0.172414\linewidth-4.13793pt\relax}>{\raggedright\arraybackslash}p{\dimexpr 0.172414\linewidth-4.13793pt\relax}>{\raggedright\arraybackslash}p{\dimexpr 0.172414\linewidth-4.13793pt\relax}@{}}
\caption{Missing feature observations}\label{tab:S8}\\
\toprule
\textbf{Universe} & \textbf{Feature} & \textbf{All signals} & \textbf{Development} & \textbf{Test}\\\midrule\endfirsthead
\multicolumn{5}{l}{\textbf{Table S8 (continued)}}\\\toprule
\textbf{Universe} & \textbf{Feature} & \textbf{All signals} & \textbf{Development} & \textbf{Test}\\\midrule\endhead
\midrule\multicolumn{5}{r}{Continued on next page}\\\endfoot
\bottomrule\endlastfoot
SSE50 & book\_\allowbreak{}to\_\allowbreak{}price & 0 & 0 & 0\\
SSE50 & earnings\_\allowbreak{}yield & 121 & 40 & 81\\
SSE50 & dividend\_\allowbreak{}yield & 282 & 145 & 137\\
SSE50 & momentum\_\allowbreak{}12\_\allowbreak{}1 & 0 & 0 & 0\\
SSE50 & momentum\_\allowbreak{}6\_\allowbreak{}1 & 0 & 0 & 0\\
SSE50 & reversal\_\allowbreak{}1 & 0 & 0 & 0\\
SSE50 & low\_\allowbreak{}volatility\_\allowbreak{}63 & 0 & 0 & 0\\
SSE50 & small\_\allowbreak{}size & 0 & 0 & 0\\
SSE50 & low\_\allowbreak{}turnover & 0 & 0 & 0\\
CSI300 & book\_\allowbreak{}to\_\allowbreak{}price & 0 & 0 & 0\\
CSI300 & earnings\_\allowbreak{}yield & 1810 & 643 & 1167\\
CSI300 & dividend\_\allowbreak{}yield & 3450 & 1480 & 1970\\
CSI300 & momentum\_\allowbreak{}12\_\allowbreak{}1 & 0 & 0 & 0\\
CSI300 & momentum\_\allowbreak{}6\_\allowbreak{}1 & 0 & 0 & 0\\
CSI300 & reversal\_\allowbreak{}1 & 0 & 0 & 0\\
CSI300 & low\_\allowbreak{}volatility\_\allowbreak{}63 & 0 & 0 & 0\\
CSI300 & small\_\allowbreak{}size & 0 & 0 & 0\\
CSI300 & low\_\allowbreak{}turnover & 0 & 0 & 0\\
CSI500 & book\_\allowbreak{}to\_\allowbreak{}price & 32 & 0 & 32\\
CSI500 & earnings\_\allowbreak{}yield & 5754 & 1951 & 3803\\
CSI500 & dividend\_\allowbreak{}yield & 10500 & 4561 & 5939\\
CSI500 & momentum\_\allowbreak{}12\_\allowbreak{}1 & 0 & 0 & 0\\
CSI500 & momentum\_\allowbreak{}6\_\allowbreak{}1 & 0 & 0 & 0\\
CSI500 & reversal\_\allowbreak{}1 & 0 & 0 & 0\\
CSI500 & low\_\allowbreak{}volatility\_\allowbreak{}63 & 0 & 0 & 0\\
CSI500 & small\_\allowbreak{}size & 0 & 0 & 0\\
CSI500 & low\_\allowbreak{}turnover & 0 & 0 & 0\\
\end{longtable}
\noindent\footnotesize Counts are calculated before within-month feature ranking among eligible stocks. The declared feature treatment assigns missing ranks 0.5. These are feature availability counts, rather than missing prediction outcomes.\par\endgroup
The final quote audit covered 1,551 stock-level price/\allowbreak{}adjustment pairs and 4,105,829 price rows. Its stock-level checks are archived at study\_\allowbreak{}v3\textbackslash{}results\textbackslash{}quote\_\allowbreak{}audit\textbackslash{}20260917T133147Z\textbackslash{}stock\_\allowbreak{}checks.csv. The checks cover unique dates, the requested date range, exchange trading dates, adjustment alignment and valid price/\allowbreak{}volume observations.

The official SW history covers 1,551 inventory stocks, using 4160 dated classification records. The source was retrieved at 2026-09-14T05:36:50.5436217Z from https:/\allowbreak{}/\allowbreak{}www.swsresearch.com/\allowbreak{}swindex/\allowbreak{}pdf/\allowbreak{}SwClass2021/\allowbreak{}StockClassifyUse\_\allowbreak{}stock.xls. Identity-linked intervals and provider update timestamps are retained in study\_\allowbreak{}v3/\allowbreak{}data/\allowbreak{}official\_\allowbreak{}industry\_\allowbreak{}intervals\_\allowbreak{}linked.csv, with coverage recorded in study\_\allowbreak{}v3/\allowbreak{}data/\allowbreak{}official\_\allowbreak{}industry\_\allowbreak{}history\_\allowbreak{}linked\_\allowbreak{}audit.json; the earlier normalization remains in study\_\allowbreak{}v3/\allowbreak{}data/\allowbreak{}official\_\allowbreak{}industry\_\allowbreak{}intervals.csv. The risk analysis combines this history with the archived taxonomy checks and CSI300 beta reference; it is a retrospective risk diagnostic.

The record 010 identity crosswalk maps old\_\allowbreak{}numeric\_\allowbreak{}code 300114 to canonical\_\allowbreak{}numeric\_\allowbreak{}code 302132 on exchange SZ with code\_\allowbreak{}change\_\allowbreak{}effective\_\allowbreak{}date 20250217. The mapping is supported by the issuer announcement dated 20250214 (https:/\allowbreak{}/\allowbreak{}static.cninfo.com.cn/\allowbreak{}finalpage/\allowbreak{}2025-02-15/\allowbreak{}1222544408.PDF; archived at study\_\allowbreak{}v3/\allowbreak{}sources/\allowbreak{}sw/\allowbreak{}security\_\allowbreak{}code\_\allowbreak{}change\_\allowbreak{}300114\_\allowbreak{}302132.json). The linked intervals retain stock\_\allowbreak{}numeric\_\allowbreak{}code, industry\_\allowbreak{}code, effective\_\allowbreak{}date and provider\_\allowbreak{}updated\_\allowbreak{}at from the original SW rows alongside canonical\_\allowbreak{}numeric\_\allowbreak{}code and ts\_\allowbreak{}code.

\section*{S5 Historical return inference and intervention returns}
\begingroup\small\setlength{\tabcolsep}{3pt}\renewcommand{\arraystretch}{1.17}
\begin{longtable}{@{}>{\raggedright\arraybackslash}p{\dimexpr 0.264706\linewidth-7.94118pt\relax}>{\raggedright\arraybackslash}p{\dimexpr 0.147059\linewidth-4.41176pt\relax}>{\raggedright\arraybackslash}p{\dimexpr 0.147059\linewidth-4.41176pt\relax}>{\raggedright\arraybackslash}p{\dimexpr 0.147059\linewidth-4.41176pt\relax}>{\raggedright\arraybackslash}p{\dimexpr 0.147059\linewidth-4.41176pt\relax}>{\raggedright\arraybackslash}p{\dimexpr 0.147059\linewidth-4.41176pt\relax}@{}}
\caption{Twelve primary historical contrasts}\label{tab:S9}\\
\toprule
\textbf{Universe /\allowbreak{} model} & \textbf{Mean diff. pp} & \textbf{95\% interval pp} & \textbf{HAC Holm p} & \textbf{Bootstrap Holm p} & \textbf{Equivalence Holm p}\\\midrule\endfirsthead
\multicolumn{6}{l}{\textbf{Table S9 (continued)}}\\\toprule
\textbf{Universe /\allowbreak{} model} & \textbf{Mean diff. pp} & \textbf{95\% interval pp} & \textbf{HAC Holm p} & \textbf{Bootstrap Holm p} & \textbf{Equivalence Holm p}\\\midrule\endhead
\midrule\multicolumn{6}{r}{Continued on next page}\\\endfoot
\bottomrule\endlastfoot
SSE 50 /\allowbreak{} Luna & -0.070 & [-0.521, 0.331] & 1.000 & 1.000 & 1.000\\
SSE 50 /\allowbreak{} Terra & -0.010 & [-0.420, 0.342] & 1.000 & 1.000 & 1.000\\
SSE 50 /\allowbreak{} Sol & 0.088 & [-0.231, 0.401] & 1.000 & 1.000 & 1.000\\
SSE 50 /\allowbreak{} Astra & 0.018 & [-0.352, 0.338] & 1.000 & 1.000 & 1.000\\
CSI 300 /\allowbreak{} Luna & 0.172 & [-0.232, 0.724] & 1.000 & 1.000 & 1.000\\
CSI 300 /\allowbreak{} Terra & 0.130 & [-0.069, 0.372] & 1.000 & 1.000 & 1.000\\
CSI 300 /\allowbreak{} Sol & 0.115 & [-0.191, 0.563] & 1.000 & 1.000 & 1.000\\
CSI 300 /\allowbreak{} Astra & -0.067 & [-0.231, 0.113] & 1.000 & 1.000 & 1.000\\
CSI 500 /\allowbreak{} Luna & 0.065 & [-0.623, 0.890] & 1.000 & 1.000 & 1.000\\
CSI 500 /\allowbreak{} Terra & -0.323 & [-0.733, 0.014] & 1.000 & 1.000 & 1.000\\
CSI 500 /\allowbreak{} Sol & 0.026 & [-0.160, 0.211] & 1.000 & 1.000 & 1.000\\
CSI 500 /\allowbreak{} Astra & 0.001 & [-0.264, 0.257] & 1.000 & 1.000 & 1.000\\
\end{longtable}
\noindent\footnotesize The twelve comparisons use whole-run trajectories and six-month calendar blocks. Difference tests and equivalence tests have separate twelve-comparison Holm adjustments. Equivalence bounds are +/\allowbreak{}-0.10 monthly percentage points.\par\endgroup
\begingroup\small\setlength{\tabcolsep}{3pt}\renewcommand{\arraystretch}{1.17}
\begin{longtable}{@{}>{\raggedright\arraybackslash}p{\dimexpr 0.310345\linewidth-7.44828pt\relax}>{\raggedright\arraybackslash}p{\dimexpr 0.172414\linewidth-4.13793pt\relax}>{\raggedright\arraybackslash}p{\dimexpr 0.172414\linewidth-4.13793pt\relax}>{\raggedright\arraybackslash}p{\dimexpr 0.172414\linewidth-4.13793pt\relax}>{\raggedright\arraybackslash}p{\dimexpr 0.172414\linewidth-4.13793pt\relax}@{}}
\caption{Declared testing families}\label{tab:S10}\\
\toprule
\textbf{Archive} & \textbf{Family} & \textbf{Tests per block} & \textbf{HAC Holm p < .05} & \textbf{Bootstrap Holm p < .05}\\\midrule\endfirsthead
\multicolumn{5}{l}{\textbf{Table S10 (continued)}}\\\toprule
\textbf{Archive} & \textbf{Family} & \textbf{Tests per block} & \textbf{HAC Holm p < .05} & \textbf{Bootstrap Holm p < .05}\\\midrule\endhead
\midrule\multicolumn{5}{r}{Continued on next page}\\\endfoot
\bottomrule\endlastfoot
Four-model & primary\_\allowbreak{}12 & 12 & 0 & 0\\
Four-model & other\_\allowbreak{}baselines\_\allowbreak{}48 & 48 & 0 & 0\\
Three-model & factor\_\allowbreak{}ablation\_\allowbreak{}9 & 9 & 0 & 0\\
Three-model & baseline\_\allowbreak{}adjusted\_\allowbreak{}ablation\_\allowbreak{}9 & 9 & 0 & 0\\
Three-model & no\_\allowbreak{}semantics\_\allowbreak{}3 & 3 & 0 & 0\\
Three-model & row\_\allowbreak{}order\_\allowbreak{}3 & 3 & 0 & 0\\
Three-model & frozen\_\allowbreak{}proxy\_\allowbreak{}3 & 3 & 1 & 0\\
Four-model & risk\_\allowbreak{}matched\_\allowbreak{}12 & 12 & 0 & 0\\
Four-model & disjoint\_\allowbreak{}12 & 12 & 0 & 0\\
Astra historical & baseline\_\allowbreak{}adjusted\_\allowbreak{}ablation\_\allowbreak{}3 & 3 & 0 & 0\\
Astra historical & factor\_\allowbreak{}ablation\_\allowbreak{}3 & 3 & 0 & 0\\
Astra historical & no\_\allowbreak{}semantics & 1 & 0 & 0\\
Astra historical & row\_\allowbreak{}order & 1 & 0 & 0\\
\end{longtable}
\noindent\footnotesize The displayed counts use the six-month block specification and include differences in either direction. Each family is adjusted separately; there is no global familywise claim. Three- and twelve-month block results remain in the same complete CSV files.\par\endgroup
The complete numerical records are analysis/\allowbreak{}results/\allowbreak{}main\_\allowbreak{}paired\_\allowbreak{}inference.csv (135 rows), analysis/\allowbreak{}results/\allowbreak{}mechanism\_\allowbreak{}paired\_\allowbreak{}inference.csv (72 rows), analysis/\allowbreak{}results/\allowbreak{}auxiliary\_\allowbreak{}paired\_\allowbreak{}inference.csv (36 rows), and analysis/\allowbreak{}results/\allowbreak{}disjoint\_\allowbreak{}paired\_\allowbreak{}inference.csv (27 rows). Every record identifies its universe, model, contrast and block length, and includes the mean difference, HAC standard error, raw tests, pointwise interval and the relevant Holm-adjusted tests. The main file also includes all four additional baselines and equivalence results.

Mechanism contrasts subtract the prespecified primary run 1 from altered-input run 1. Positive values indicate higher returns for the altered condition. The baseline-adjusted ablation contrast additionally subtracts the corresponding EqualFactor change. Factor removals, semantics removal, row-order changes and baseline-adjusted removals form separate families. There is one altered-input trajectory per condition, so these comparisons do not estimate independent altered-run variance.

\begingroup\small\setlength{\tabcolsep}{3pt}\renewcommand{\arraystretch}{1.17}
\begin{longtable}{@{}>{\raggedright\arraybackslash}p{\dimexpr 0.264706\linewidth-7.94118pt\relax}>{\raggedright\arraybackslash}p{\dimexpr 0.147059\linewidth-4.41176pt\relax}>{\raggedright\arraybackslash}p{\dimexpr 0.147059\linewidth-4.41176pt\relax}>{\raggedright\arraybackslash}p{\dimexpr 0.147059\linewidth-4.41176pt\relax}>{\raggedright\arraybackslash}p{\dimexpr 0.147059\linewidth-4.41176pt\relax}>{\raggedright\arraybackslash}p{\dimexpr 0.147059\linewidth-4.41176pt\relax}@{}}
\caption{Proxy and risk-matched contrasts}\label{tab:S11}\\
\toprule
\textbf{Universe /\allowbreak{} model} & \textbf{Comparison} & \textbf{Mean diff. pp} & \textbf{95\% interval pp} & \textbf{HAC Holm p} & \textbf{Bootstrap Holm p}\\\midrule\endfirsthead
\multicolumn{6}{l}{\textbf{Table S11 (continued)}}\\\toprule
\textbf{Universe /\allowbreak{} model} & \textbf{Comparison} & \textbf{Mean diff. pp} & \textbf{95\% interval pp} & \textbf{HAC Holm p} & \textbf{Bootstrap Holm p}\\\midrule\endhead
\midrule\multicolumn{6}{r}{Continued on next page}\\\endfoot
\bottomrule\endlastfoot
SSE50 /\allowbreak{} Luna & Ridge\_\allowbreak{}development\_\allowbreak{}selected\_\allowbreak{}proxy & -0.19 & [-0.44, 0.05] & 0.197 & 0.235\\
SSE50 /\allowbreak{} Terra & Ridge\_\allowbreak{}development\_\allowbreak{}selected\_\allowbreak{}proxy & -0.20 & [-0.39, -0.02] & 0.048 & 0.104\\
SSE50 /\allowbreak{} Sol & Ridge\_\allowbreak{}development\_\allowbreak{}selected\_\allowbreak{}proxy & -0.02 & [-0.15, 0.10] & 0.821 & 0.813\\
CSI 300 /\allowbreak{} Luna & Risk-matched equal factor & -0.236 & [-0.602, 0.143] & 1.000 & 1.000\\
CSI 300 /\allowbreak{} Terra & Risk-matched equal factor & -0.242 & [-0.579, 0.112] & 1.000 & 1.000\\
CSI 300 /\allowbreak{} Sol & Risk-matched equal factor & -0.106 & [-0.443, 0.258] & 1.000 & 1.000\\
CSI 300 /\allowbreak{} Astra & Risk-matched equal factor & -0.055 & [-0.309, 0.199] & 1.000 & 1.000\\
CSI 500 /\allowbreak{} Luna & Risk-matched equal factor & -0.021 & [-0.486, 0.481] & 1.000 & 1.000\\
CSI 500 /\allowbreak{} Terra & Risk-matched equal factor & -0.250 & [-0.637, 0.183] & 1.000 & 1.000\\
CSI 500 /\allowbreak{} Sol & Risk-matched equal factor & 0.002 & [-0.329, 0.357] & 1.000 & 1.000\\
CSI 500 /\allowbreak{} Astra & Risk-matched equal factor & 0.068 & [-0.211, 0.347] & 1.000 & 1.000\\
SSE 50 /\allowbreak{} Luna & Risk-matched equal factor & -0.097 & [-0.375, 0.213] & 1.000 & 1.000\\
SSE 50 /\allowbreak{} Terra & Risk-matched equal factor & -0.096 & [-0.286, 0.109] & 1.000 & 1.000\\
SSE 50 /\allowbreak{} Sol & Risk-matched equal factor & 0.033 & [-0.166, 0.253] & 1.000 & 1.000\\
SSE 50 /\allowbreak{} Astra & Risk-matched equal factor & 0.050 & [-0.187, 0.326] & 1.000 & 1.000\\
\end{longtable}
\noindent\footnotesize The three-model SSE 50 model-minus-proxy comparisons form a three-test family. Risk-matched model-minus-equal-factor comparisons form a separate twelve-test family. All use the scheduled model trajectories, paired calendar months and 10 basis points per side. The final twelve-pool-model proxy family is reported separately in the main text and Section S13.\par\endgroup
\begingroup\small\setlength{\tabcolsep}{3pt}\renewcommand{\arraystretch}{1.17}
\begin{longtable}{@{}>{\raggedright\arraybackslash}p{\dimexpr 0.310345\linewidth-7.44828pt\relax}>{\raggedright\arraybackslash}p{\dimexpr 0.172414\linewidth-4.13793pt\relax}>{\raggedright\arraybackslash}p{\dimexpr 0.172414\linewidth-4.13793pt\relax}>{\raggedright\arraybackslash}p{\dimexpr 0.172414\linewidth-4.13793pt\relax}>{\raggedright\arraybackslash}p{\dimexpr 0.172414\linewidth-4.13793pt\relax}@{}}
\caption{Disjoint-candidate contrasts}\label{tab:S12}\\
\toprule
\textbf{Universe /\allowbreak{} model} & \textbf{Mean diff. pp} & \textbf{95\% interval pp} & \textbf{HAC Holm p} & \textbf{Bootstrap Holm p}\\\midrule\endfirsthead
\multicolumn{5}{l}{\textbf{Table S12 (continued)}}\\\toprule
\textbf{Universe /\allowbreak{} model} & \textbf{Mean diff. pp} & \textbf{95\% interval pp} & \textbf{HAC Holm p} & \textbf{Bootstrap Holm p}\\\midrule\endhead
\midrule\multicolumn{5}{r}{Continued on next page}\\\endfoot
\bottomrule\endlastfoot
CSI 300 /\allowbreak{} Luna & 0.170 & [-0.236, 0.744] & 1.000 & 1.000\\
CSI 300 /\allowbreak{} Terra & 0.039 & [-0.250, 0.363] & 1.000 & 1.000\\
CSI 300 /\allowbreak{} Sol & 0.127 & [-0.234, 0.615] & 1.000 & 1.000\\
CSI 300 /\allowbreak{} Astra & -0.047 & [-0.292, 0.193] & 1.000 & 1.000\\
CSI 500 /\allowbreak{} Luna & 0.065 & [-0.623, 0.890] & 1.000 & 1.000\\
CSI 500 /\allowbreak{} Terra & -0.323 & [-0.733, 0.014] & 1.000 & 1.000\\
CSI 500 /\allowbreak{} Sol & 0.026 & [-0.160, 0.211] & 1.000 & 1.000\\
CSI 500 /\allowbreak{} Astra & 0.001 & [-0.264, 0.257] & 1.000 & 1.000\\
SSE 50 /\allowbreak{} Luna & -0.070 & [-0.521, 0.331] & 1.000 & 1.000\\
SSE 50 /\allowbreak{} Terra & -0.010 & [-0.420, 0.342] & 1.000 & 1.000\\
SSE 50 /\allowbreak{} Sol & 0.088 & [-0.231, 0.401] & 1.000 & 1.000\\
SSE 50 /\allowbreak{} Astra & 0.018 & [-0.352, 0.338] & 1.000 & 1.000\\
\end{longtable}
\noindent\footnotesize Candidate sets are assigned in SSE50, CSI300, CSI500 order at each signal. Scores, factor transformations and training retain their parent-universe context. These are conditional portfolio comparisons using existing scores; disjoint stocks do not imply independent market returns.\par\endgroup
\begingroup\small\setlength{\tabcolsep}{3pt}\renewcommand{\arraystretch}{1.17}
\begin{longtable}{@{}>{\raggedright\arraybackslash}p{\dimexpr 0.264706\linewidth-7.94118pt\relax}>{\raggedright\arraybackslash}p{\dimexpr 0.147059\linewidth-4.41176pt\relax}>{\raggedright\arraybackslash}p{\dimexpr 0.147059\linewidth-4.41176pt\relax}>{\raggedright\arraybackslash}p{\dimexpr 0.147059\linewidth-4.41176pt\relax}>{\raggedright\arraybackslash}p{\dimexpr 0.147059\linewidth-4.41176pt\relax}>{\raggedright\arraybackslash}p{\dimexpr 0.147059\linewidth-4.41176pt\relax}@{}}
\caption{Mean monthly return changes under input interventions}\label{tab:S13}\\
\toprule
\textbf{Universe /\allowbreak{} model} & \textbf{Valuation removed} & \textbf{Momentum removed} & \textbf{Risk/\allowbreak{}trading removed} & \textbf{Semantics removed} & \textbf{Rows reordered}\\\midrule\endfirsthead
\multicolumn{6}{l}{\textbf{Table S13 (continued)}}\\\toprule
\textbf{Universe /\allowbreak{} model} & \textbf{Valuation removed} & \textbf{Momentum removed} & \textbf{Risk/\allowbreak{}trading removed} & \textbf{Semantics removed} & \textbf{Rows reordered}\\\midrule\endhead
\midrule\multicolumn{6}{r}{Continued on next page}\\\endfoot
\bottomrule\endlastfoot
SSE 50 /\allowbreak{} Luna & 0.402 & 0.354 & 0.422 & 0.153 & 0.002\\
SSE 50 /\allowbreak{} Terra & 0.211 & 0.535 & -0.004 & 0.300 & 0.239\\
SSE 50 /\allowbreak{} Sol & 0.085 & 0.105 & -0.066 & -0.062 & 0.053\\
SSE 50 /\allowbreak{} Astra & 0.023 & 0.102 & -0.121 & -0.344 & 0.052\\
\end{longtable}
\noindent\footnotesize All entries are percentage points per month, altered-input run 1 minus primary run 1. The first three columns additionally subtract the matching change in equal-factor returns. The final two columns show unadjusted LLM changes. An asterisk denotes a block-bootstrap Holm p-value below 0.05 within the declared family; no mark implies no such finding. Raw ablation changes, both testing procedures, intervals, and all block lengths are reported in the archive. Each altered condition has one complete run, so these contrasts do not separately estimate its request-to-request variance. Astra uses three-comparison numerical families and separate single semantics/\allowbreak{}order contrasts; the three-model families remain unchanged.\par\endgroup
\section*{S6 Preference fits, decision stability and risk diagnostics}
\begingroup\small\setlength{\tabcolsep}{3pt}\renewcommand{\arraystretch}{1.17}
\begin{longtable}{@{}>{\raggedright\arraybackslash}p{\dimexpr 0.310345\linewidth-7.44828pt\relax}>{\raggedright\arraybackslash}p{\dimexpr 0.172414\linewidth-4.13793pt\relax}>{\raggedright\arraybackslash}p{\dimexpr 0.172414\linewidth-4.13793pt\relax}>{\raggedright\arraybackslash}p{\dimexpr 0.172414\linewidth-4.13793pt\relax}>{\raggedright\arraybackslash}p{\dimexpr 0.172414\linewidth-4.13793pt\relax}@{}}
\caption{Frozen preference proxies}\label{tab:S14}\\
\toprule
\textbf{Universe} & \textbf{Source model} & \textbf{Selected family} & \textbf{Alpha} & \textbf{Valid development months}\\\midrule\endfirsthead
\multicolumn{5}{l}{\textbf{Table S14 (continued)}}\\\toprule
\textbf{Universe} & \textbf{Source model} & \textbf{Selected family} & \textbf{Alpha} & \textbf{Valid development months}\\\midrule\endhead
\midrule\multicolumn{5}{r}{Continued on next page}\\\endfoot
\bottomrule\endlastfoot
SSE50 & Luna & Ridge & 0.1 & 60\\
SSE50 & Terra & Ridge & 0.1 & 60\\
SSE50 & Sol & Ridge & 0.1 & 60\\
SSE50 & Astra & Ridge & 0.1 & 60\\
\end{longtable}
\noindent\footnotesize All preference fits use development months only. Archived coefficients are preserved; the Astra fit uses the same Ridge/\allowbreak{}Lasso selection procedure, with its complete fit record under astra\_\allowbreak{}20260921/\allowbreak{}rules/\allowbreak{}fit\_\allowbreak{}complete.json.\par\endgroup
\begingroup\small\setlength{\tabcolsep}{3pt}\renewcommand{\arraystretch}{1.17}
\begin{longtable}{@{}>{\raggedright\arraybackslash}p{\dimexpr 0.310345\linewidth-7.44828pt\relax}>{\raggedright\arraybackslash}p{\dimexpr 0.172414\linewidth-4.13793pt\relax}>{\raggedright\arraybackslash}p{\dimexpr 0.172414\linewidth-4.13793pt\relax}>{\raggedright\arraybackslash}p{\dimexpr 0.172414\linewidth-4.13793pt\relax}>{\raggedright\arraybackslash}p{\dimexpr 0.172414\linewidth-4.13793pt\relax}@{}}
\caption{Identical-input decision stability}\label{tab:S15}\\
\toprule
\textbf{Universe /\allowbreak{} model} & \textbf{Valid /\allowbreak{} scheduled pairs} & \textbf{Defined correlations} & \textbf{Mean rank correlation} & \textbf{Mean top-set overlap}\\\midrule\endfirsthead
\multicolumn{5}{l}{\textbf{Table S15 (continued)}}\\\toprule
\textbf{Universe /\allowbreak{} model} & \textbf{Valid /\allowbreak{} scheduled pairs} & \textbf{Defined correlations} & \textbf{Mean rank correlation} & \textbf{Mean top-set overlap}\\\midrule\endhead
\midrule\multicolumn{5}{r}{Continued on next page}\\\endfoot
\bottomrule\endlastfoot
SSE 50 /\allowbreak{} Luna & 720 /\allowbreak{} 720 & 720 & 0.892 & 83.36\%\\
SSE 50 /\allowbreak{} Terra & 720 /\allowbreak{} 720 & 720 & 0.966 & 90.50\%\\
SSE 50 /\allowbreak{} Sol & 720 /\allowbreak{} 720 & 720 & 0.971 & 91.43\%\\
SSE 50 /\allowbreak{} Astra & 720 /\allowbreak{} 720 & 720 & 0.991 & 94.63\%\\
\end{longtable}
\noindent\footnotesize Five SSE 50 primary runs yield ten pairwise comparisons per model, each over 72 signals. Constant score vectors have undefined rank correlation. The full archive also compares primary run 1 with each altered-input condition.\par\endgroup
\begingroup\small\setlength{\tabcolsep}{3pt}\renewcommand{\arraystretch}{1.17}
\begin{longtable}{@{}>{\raggedright\arraybackslash}p{\dimexpr 0.375000\linewidth-6.75000pt\relax}>{\raggedright\arraybackslash}p{\dimexpr 0.208333\linewidth-3.75000pt\relax}>{\raggedright\arraybackslash}p{\dimexpr 0.208333\linewidth-3.75000pt\relax}>{\raggedright\arraybackslash}p{\dimexpr 0.208333\linewidth-3.75000pt\relax}@{}}
\caption{Historical risk-data coverage and realized exposures}\label{tab:S16}\\
\toprule
\textbf{Universe} & \textbf{Required stock-months} & \textbf{Resolved stock-months} & \textbf{Unresolved issues}\\\midrule\endfirsthead
\multicolumn{4}{l}{\textbf{Table S16 (continued)}}\\\toprule
\textbf{Universe} & \textbf{Required stock-months} & \textbf{Resolved stock-months} & \textbf{Unresolved issues}\\\midrule\endhead
\midrule\multicolumn{4}{r}{Continued on next page}\\\endfoot
\bottomrule\endlastfoot
SSE50 & 3563 & 3563 & 0\\
CSI300 & 21446 & 21446 & 0\\
CSI500 & 35806 & 35806 & 0\\
\end{longtable}
\noindent\footnotesize Signal-time risk-data coverage is counted among current candidates. The exposure readings below use actual 09:30 holdings after monthly A-share rebalancing at 10 bp, once per strategy-month. Cash, locked successor claims and retained board shares are included in portfolio wealth. The stated industry and standardized-risk deviations cover holdings with observed candidate covariates; outside-candidate counts and weights identify the remainder. The target optimization constraints and post-trade exposures are distinct measurements.\par\endgroup
Three-model cohort: SSE50: maximum target constraint violation 9.99e-16; at 10 bp, the largest industry deviation among holdings with observed covariates was 5.319 percentage points and the largest standardized risk deviation was 0.11207. Holdings outside the candidate set occurred in 11/\allowbreak{}1,440 strategy-months; complete exposure coverage was unavailable in 11/\allowbreak{}1,440, with a maximum outside-candidate portfolio weight of 3.555\%. Astra: 360 strategy-months; maximum observed industry deviation 2.406 percentage points and standardized risk deviation 0.07780; incomplete exposure coverage in 2/\allowbreak{}360 strategy-months.

Three-model cohort: CSI300: maximum target constraint violation 9.54e-16; at 10 bp, the largest industry deviation among holdings with observed covariates was 3.999 percentage points and the largest standardized risk deviation was 0.09453. Holdings outside the candidate set occurred in 33/\allowbreak{}576 strategy-months; complete exposure coverage was unavailable in 33/\allowbreak{}576, with a maximum outside-candidate portfolio weight of 2.266\%. Astra: 72 strategy-months; maximum observed industry deviation 3.998 percentage points and standardized risk deviation 0.07075; incomplete exposure coverage in 2/\allowbreak{}72 strategy-months.

Three-model cohort: CSI500: maximum target constraint violation 7.39e-16; at 10 bp, the largest industry deviation among holdings with observed covariates was 4.142 percentage points and the largest standardized risk deviation was 0.13931. Holdings outside the candidate set occurred in 146/\allowbreak{}576 strategy-months; complete exposure coverage was unavailable in 146/\allowbreak{}576, with a maximum outside-candidate portfolio weight of 6.890\%. Astra: 72 strategy-months; maximum observed industry deviation 2.981 percentage points and standardized risk deviation 0.06898; incomplete exposure coverage in 5/\allowbreak{}72 strategy-months.

For each universe, analysis/\allowbreak{}results/\allowbreak{}<universe>/\allowbreak{}risk\_\allowbreak{}matched/\allowbreak{} contains risk\_\allowbreak{}covariates.parquet, optimizer\_\allowbreak{}checks.json, target\_\allowbreak{}weights.parquet and realized\_\allowbreak{}exposures.csv. The target checks assess the optimization constraints; post-trade deviations and incomplete coverage record the portfolios produced by execution. The primary and mechanism directories also retain score\_\allowbreak{}ties.csv, selected\_\allowbreak{}feature\_\allowbreak{}exposures.csv, execution ledgers and terminal\_\allowbreak{}positions.json. Astra records use analysis/\allowbreak{}astra\_\allowbreak{}20260921/\allowbreak{}controls/\allowbreak{}<universe>/\allowbreak{}risk\_\allowbreak{}matched/\allowbreak{}run\_\allowbreak{}XX/\allowbreak{}, including optimizer\_\allowbreak{}checks.csv and realized\_\allowbreak{}exposures.csv.

Risk matching forms a separate portfolio experiment over all eligible candidates. Every conventional method and each universe's complete LLM runs (five in the SSE 50, one in each of the CSI 300 and CSI 500) use the same equal-weight-universe risk reference. Scores are converted to within-month percentile ranks with average ties. The objective maximizes the portfolio-weighted ranking score minus 0.1 times the squared distance of its weights from the equal-weight reference. Target weights are nonnegative and sum to one; the single-stock ceiling is max(0.02, 2/\allowbreak{}N). Each target industry weight can differ from its reference by at most 0.02. Target standardized size, volatility, and beta exposures can each differ by at most 0.05.

Size is log floating market capitalization; volatility is the trailing 63-session adjusted-return sample standard deviation annualized by sqrt(252). Beta is the slope with an intercept from 252 daily adjusted stock returns against CSI 300 close-to-close returns through the signal date, calculated as close[t] /\allowbreak{} close[t-1] - 1 on consecutive exchange sessions. Risk variables are standardized within the cross-section using the population standard deviation. Constant columns impose no differential exposure constraint and are recorded. Industry membership uses dated intervals from the official Shenwan history, checked against the corresponding taxonomy. These provider records support a retrospective exposure diagnostic; original release snapshots and revision histories are not reconstructed.

Sequential least-squares quadratic programming solves the objective with an analytical gradient, at most 1,000 iterations and function tolerance 1e-10. Solver success and an independently measured maximum constraint violation of at most 1e-7 are required. Executed exposure diagnostics use the actual 09:30 holdings snapshot after each monthly A-share rebalance, including locked successor claims, retained board shares and cash. Multiple claims under a code are aggregated for covariate matching. Signal-time covariates describe holdings present in the current candidate set; the weight and count of holdings outside that set quantify incomplete portfolio coverage. Target feasibility and realized exposure are reported separately. All four cost scenarios have their own holdings and exposure records; twelve primary contrasts compare models with risk-matched equal-factor returns.

A monthly partition retains SSE 50 candidates, removes those candidates from CSI 300, and removes the union of SSE 50 and CSI 300 candidates from CSI 500. The subsets are nonempty, mutually exclusive, and preserve the union of the original eligible sets. Rankings are restricted to each assigned subset using parent-universe conventional scores and each universe's complete primary LLM runs (five in the SSE 50, one in the CSI 300 and CSI 500). Factor percentiles and fitted models retain their original universe context. Top-fraction and universe-equal-weight portfolios use the common engine and all four costs. Twelve LLM-minus-equal-factor contrasts at 10 basis points form the disjoint-candidate family. This experiment measures the portfolio effect of removing stock overlap while retaining common market-month shocks.

\section*{S7 Development-based return power and precision}
This archived planning calculation concerns only the Luna, Terra and Sol models. Development-period LLM-minus-EqualFactor returns supply the variance estimates for the sample-size calculation. Each model contributes one 60-month development trajectory per universe. The planning specification covers 360 CSI 300 and CSI 500 development predictions over December 2014 through November 2019, separate from the 2,772 historical tasks and all historical test families. CSI 500 provides additional variance estimates; the three CSI 300 comparisons determine the archived duration calculation. The calculation uses HAC lag-3 long-run variance, a monthly effect of 0.10 percentage points, and conservative alpha/\allowbreak{}3 planning for three CSI300 comparisons.

\begingroup\small\setlength{\tabcolsep}{3pt}\renewcommand{\arraystretch}{1.17}
\begin{longtable}{@{}>{\raggedright\arraybackslash}p{\dimexpr 0.310345\linewidth-7.44828pt\relax}>{\raggedright\arraybackslash}p{\dimexpr 0.172414\linewidth-4.13793pt\relax}>{\raggedright\arraybackslash}p{\dimexpr 0.172414\linewidth-4.13793pt\relax}>{\raggedright\arraybackslash}p{\dimexpr 0.172414\linewidth-4.13793pt\relax}>{\raggedright\arraybackslash}p{\dimexpr 0.172414\linewidth-4.13793pt\relax}@{}}
\caption{Development-based power}\label{tab:S17}\\
\toprule
\textbf{Universe} & \textbf{Model} & \textbf{Power at 24 months} & \textbf{Power at 36 months} & \textbf{Months for 80\%}\\\midrule\endfirsthead
\multicolumn{5}{l}{\textbf{Table S17 (continued)}}\\\toprule
\textbf{Universe} & \textbf{Model} & \textbf{Power at 24 months} & \textbf{Power at 36 months} & \textbf{Months for 80\%}\\\midrule\endhead
\midrule\multicolumn{5}{r}{Continued on next page}\\\endfoot
\bottomrule\endlastfoot
SSE50 & Luna & 1.81\% & 1.88\% & 9776\\
SSE50 & Terra & 2.63\% & 3.13\% & 1480\\
SSE50 & Sol & 2.07\% & 2.28\% & 3423\\
CSI300 & Luna & 2.71\% & 3.27\% & 1358\\
CSI300 & Terra & 2.44\% & 2.84\% & 1827\\
CSI300 & Sol & 3.26\% & 4.11\% & 913\\
CSI500 & Luna & 2.26\% & 2.57\% & 2354\\
CSI500 & Terra & 2.31\% & 2.64\% & 2195\\
CSI500 & Sol & 2.45\% & 2.86\% & 1796\\
\end{longtable}
\noindent\footnotesize All nine planning comparisons, empirical variances and precision measures are in analysis/\allowbreak{}results/\allowbreak{}power/\allowbreak{}prospective\_\allowbreak{}power.csv. One development trajectory supplies each variance estimate; the displayed power is a planning calculation.\par\endgroup
\begingroup\small\setlength{\tabcolsep}{3pt}\renewcommand{\arraystretch}{1.17}
\begin{longtable}{@{}>{\raggedright\arraybackslash}p{\dimexpr 0.310345\linewidth-7.44828pt\relax}>{\raggedright\arraybackslash}p{\dimexpr 0.172414\linewidth-4.13793pt\relax}>{\raggedright\arraybackslash}p{\dimexpr 0.172414\linewidth-4.13793pt\relax}>{\raggedright\arraybackslash}p{\dimexpr 0.172414\linewidth-4.13793pt\relax}>{\raggedright\arraybackslash}p{\dimexpr 0.172414\linewidth-4.13793pt\relax}@{}}
\caption{Planned CSI 300 precision}\label{tab:S18}\\
\toprule
\textbf{Model} & \textbf{Long-run variance} & \textbf{24-month half-width pp} & \textbf{36-month half-width pp} & \textbf{Selected-duration power}\\\midrule\endfirsthead
\multicolumn{5}{l}{\textbf{Table S18 (continued)}}\\\toprule
\textbf{Model} & \textbf{Long-run variance} & \textbf{24-month half-width pp} & \textbf{36-month half-width pp} & \textbf{Selected-duration power}\\\midrule\endhead
\midrule\multicolumn{5}{r}{Continued on next page}\\\endfoot
\bottomrule\endlastfoot
Luna & 0.000129685 & 0.56 & 0.45 & 3.27\%\\
Terra & 0.00017445 & 0.65 & 0.53 & 2.84\%\\
Sol & 8.71466e-05 & 0.46 & 0.37 & 4.11\%\\
\end{longtable}
\noindent\footnotesize Half-widths use simultaneous normal planning intervals for the three CSI300 comparisons.\par\endgroup
The fixed duration rule uses 24 months only if all three CSI 300 comparisons reach 80\% planned power; otherwise, it selects 36 months. It selects 36 months, where model-specific planning power is 2.84\%-4.11\%. Table S17 also reports the implied sample lengths needed for the 80\% target. The duration selection and the attained power are separate outputs of the planning calculation.

This financial-return planning exercise is separate from the matched-response conditional measurement-noise calculation in Section S12. Its selected duration does not change the observed 72-month historical sample or the 24-month controlled-response grid.

\section*{S8 Model configuration and completed prediction coverage}
\begingroup\small\setlength{\tabcolsep}{3pt}\renewcommand{\arraystretch}{1.17}
\begin{longtable}{@{}>{\raggedright\arraybackslash}p{\dimexpr 0.473684\linewidth-5.68421pt\relax}>{\raggedright\arraybackslash}p{\dimexpr 0.263158\linewidth-3.15789pt\relax}>{\raggedright\arraybackslash}p{\dimexpr 0.263158\linewidth-3.15789pt\relax}@{}}
\caption{Model configuration}\label{tab:S19}\\
\toprule
\textbf{Model} & \textbf{Interface} & \textbf{Requested tier}\\\midrule\endfirsthead
\multicolumn{3}{l}{\textbf{Table S19 (continued)}}\\\toprule
\textbf{Model} & \textbf{Interface} & \textbf{Requested tier}\\\midrule\endhead
\midrule\multicolumn{3}{r}{Continued on next page}\\\endfoot
\bottomrule\endlastfoot
Luna & api.openai.com & default; priority\\
Terra & api.openai.com & default; fast; priority\\
Sol & api.openai.com & default; fast; priority\\
Astra & api.openai.com & default\\
\end{longtable}
\noindent\footnotesize This table records the historical model configurations. The repeated input-intervention design uses the same official interface; its execution settings are in Section S15. All four historical model configurations use the official OpenAI Responses API with medium reasoning, no external tools and at most three concurrent calls. Requested service tiers identify the settings used for each collection batch; every response also retains its returned tier.\par\endgroup
\begingroup\small\setlength{\tabcolsep}{3pt}\renewcommand{\arraystretch}{1.17}
\begin{longtable}{@{}>{\raggedright\arraybackslash}p{\dimexpr 0.310345\linewidth-7.44828pt\relax}>{\raggedright\arraybackslash}p{\dimexpr 0.172414\linewidth-4.13793pt\relax}>{\raggedright\arraybackslash}p{\dimexpr 0.172414\linewidth-4.13793pt\relax}>{\raggedright\arraybackslash}p{\dimexpr 0.172414\linewidth-4.13793pt\relax}>{\raggedright\arraybackslash}p{\dimexpr 0.172414\linewidth-4.13793pt\relax}@{}}
\caption{Prediction-job coverage}\label{tab:S20}\\
\toprule
\textbf{Universe} & \textbf{Model} & \textbf{Stage} & \textbf{Planned} & \textbf{Valid}\\\midrule\endfirsthead
\multicolumn{5}{l}{\textbf{Table S20 (continued)}}\\\toprule
\textbf{Universe} & \textbf{Model} & \textbf{Stage} & \textbf{Planned} & \textbf{Valid}\\\midrule\endhead
\midrule\multicolumn{5}{r}{Continued on next page}\\\endfoot
\bottomrule\endlastfoot
SSE50 & Luna & Historical main & 360 & 360\\
SSE50 & Luna & Historical development: 1 run & 60 & 60\\
SSE50 & Luna & Historical mechanisms: 5 conditions & 360 & 360\\
SSE50 & Terra & Historical main & 360 & 360\\
SSE50 & Terra & Historical development: 1 run & 60 & 60\\
SSE50 & Terra & Historical mechanisms: 5 conditions & 360 & 360\\
SSE50 & Sol & Historical main & 360 & 360\\
SSE50 & Sol & Historical development: 1 run & 60 & 60\\
SSE50 & Sol & Historical mechanisms: 5 conditions & 360 & 360\\
CSI300 & Luna & Historical main & 72 & 72\\
CSI300 & Terra & Historical main & 72 & 72\\
CSI300 & Sol & Historical main & 72 & 72\\
CSI500 & Luna & Historical main & 72 & 72\\
CSI500 & Terra & Historical main & 72 & 72\\
CSI500 & Sol & Historical main & 72 & 72\\
CSI300 & Luna & Separate planning: 1 run & 60 & 60\\
CSI300 & Terra & Separate planning: 1 run & 60 & 60\\
CSI300 & Sol & Separate planning: 1 run & 60 & 60\\
CSI500 & Luna & Separate planning: 1 run & 60 & 60\\
CSI500 & Terra & Separate planning: 1 run & 60 & 60\\
CSI500 & Sol & Separate planning: 1 run & 60 & 60\\
SSE50 & Astra & Historical main & 360 & 360\\
SSE50 & Astra & Historical development & 60 & 60\\
SSE50 & Astra & Historical mechanisms: 5 conditions & 360 & 360\\
CSI300 & Astra & Historical main & 72 & 72\\
CSI500 & Astra & Historical main & 72 & 72\\
\end{longtable}
\noindent\footnotesize The historical grid contains 3,696 valid tasks, including 924 Astra tasks. The separate 360-task return-planning batch covers the Luna, Terra and Sol models. Matched interventions and the three-run replication batch are reported separately.\par\endgroup
The historical coverage table excludes the matched-response grid. That grid contributes exactly 216 valid responses for each of Luna, Terra, Sol and Astra, with 24 months, three conditions and three answers per condition. It is documented in Section S12.

\section*{S9 Frozen-rule agreement and annual transfer}
All summaries preserve scheduled months and report defined rank-correlation months. Disjoint subsets use original full-pool answers restricted to contemporaneously nonoverlapping stocks, with new ranks computed inside the retained set. The restricted ranks are computed from stored responses without additional model calls.

\begingroup\small\setlength{\tabcolsep}{3pt}\renewcommand{\arraystretch}{1.17}
\begin{longtable}{@{}>{\raggedright\arraybackslash}p{\dimexpr 0.264706\linewidth-7.94118pt\relax}>{\raggedright\arraybackslash}p{\dimexpr 0.147059\linewidth-4.41176pt\relax}>{\raggedright\arraybackslash}p{\dimexpr 0.147059\linewidth-4.41176pt\relax}>{\raggedright\arraybackslash}p{\dimexpr 0.147059\linewidth-4.41176pt\relax}>{\raggedright\arraybackslash}p{\dimexpr 0.147059\linewidth-4.41176pt\relax}>{\raggedright\arraybackslash}p{\dimexpr 0.147059\linewidth-4.41176pt\relax}@{}}
\caption{All rule and pool agreement summaries}\label{tab:S21}\\
\toprule
\textbf{Pool /\allowbreak{} model} & \textbf{Rule} & \textbf{Scope} & \textbf{Spearman} & \textbf{Overlap (\%)} & \textbf{Defined months}\\\midrule\endfirsthead
\multicolumn{6}{l}{\textbf{Table S21 (continued)}}\\\toprule
\textbf{Pool /\allowbreak{} model} & \textbf{Rule} & \textbf{Scope} & \textbf{Spearman} & \textbf{Overlap (\%)} & \textbf{Defined months}\\\midrule\endhead
\midrule\multicolumn{6}{r}{Continued on next page}\\\endfoot
\bottomrule\endlastfoot
CSI 300 /\allowbreak{} Luna & Equal factor & Disjoint subset & 0.750 & 73.1 & 72\\
CSI 300 /\allowbreak{} Luna & Equal factor & Original pool & 0.767 & 73.8 & 72\\
CSI 300 /\allowbreak{} Luna & Frozen linear & Disjoint subset & 0.861 & 81.7 & 72\\
CSI 300 /\allowbreak{} Luna & Frozen linear & Original pool & 0.870 & 82.3 & 72\\
CSI 300 /\allowbreak{} Luna & Random forest & Disjoint subset & 0.784 & 72.1 & 72\\
CSI 300 /\allowbreak{} Luna & Random forest & Original pool & 0.798 & 74.1 & 72\\
CSI 300 /\allowbreak{} Luna & Single factor & Disjoint subset & 0.645 & 65.1 & 72\\
CSI 300 /\allowbreak{} Luna & Single factor & Original pool & 0.672 & 65.8 & 72\\
CSI 300 /\allowbreak{} Sol & Equal factor & Disjoint subset & 0.853 & 78.1 & 72\\
CSI 300 /\allowbreak{} Sol & Equal factor & Original pool & 0.862 & 78.1 & 72\\
CSI 300 /\allowbreak{} Sol & Frozen linear & Disjoint subset & 0.920 & 85.2 & 72\\
CSI 300 /\allowbreak{} Sol & Frozen linear & Original pool & 0.926 & 85.3 & 72\\
CSI 300 /\allowbreak{} Sol & Random forest & Disjoint subset & 0.827 & 77.2 & 72\\
CSI 300 /\allowbreak{} Sol & Random forest & Original pool & 0.840 & 77.9 & 72\\
CSI 300 /\allowbreak{} Sol & Single factor & Disjoint subset & 0.706 & 67.1 & 72\\
CSI 300 /\allowbreak{} Sol & Single factor & Original pool & 0.732 & 68.2 & 72\\
CSI 300 /\allowbreak{} Terra & Equal factor & Disjoint subset & 0.808 & 75.5 & 72\\
CSI 300 /\allowbreak{} Terra & Equal factor & Original pool & 0.822 & 75.7 & 72\\
CSI 300 /\allowbreak{} Terra & Frozen linear & Disjoint subset & 0.910 & 84.3 & 72\\
CSI 300 /\allowbreak{} Terra & Frozen linear & Original pool & 0.917 & 84.8 & 72\\
CSI 300 /\allowbreak{} Terra & Random forest & Disjoint subset & 0.824 & 74.2 & 72\\
CSI 300 /\allowbreak{} Terra & Random forest & Original pool & 0.838 & 74.7 & 72\\
CSI 300 /\allowbreak{} Terra & Single factor & Disjoint subset & 0.706 & 68.4 & 72\\
CSI 300 /\allowbreak{} Terra & Single factor & Original pool & 0.732 & 69.3 & 72\\
CSI 500 /\allowbreak{} Luna & Equal factor & Disjoint subset & 0.672 & 62.0 & 68\\
CSI 500 /\allowbreak{} Luna & Equal factor & Original pool & 0.672 & 62.0 & 68\\
CSI 500 /\allowbreak{} Luna & Frozen linear & Disjoint subset & 0.795 & 70.2 & 68\\
CSI 500 /\allowbreak{} Luna & Frozen linear & Original pool & 0.795 & 70.2 & 68\\
CSI 500 /\allowbreak{} Luna & Random forest & Disjoint subset & 0.728 & 64.8 & 68\\
CSI 500 /\allowbreak{} Luna & Random forest & Original pool & 0.728 & 64.8 & 68\\
CSI 500 /\allowbreak{} Luna & Single factor & Disjoint subset & 0.592 & 59.0 & 68\\
CSI 500 /\allowbreak{} Luna & Single factor & Original pool & 0.592 & 59.0 & 68\\
CSI 500 /\allowbreak{} Sol & Equal factor & Disjoint subset & 0.868 & 76.2 & 72\\
CSI 500 /\allowbreak{} Sol & Equal factor & Original pool & 0.868 & 76.2 & 72\\
CSI 500 /\allowbreak{} Sol & Frozen linear & Disjoint subset & 0.933 & 83.6 & 72\\
CSI 500 /\allowbreak{} Sol & Frozen linear & Original pool & 0.933 & 83.6 & 72\\
CSI 500 /\allowbreak{} Sol & Random forest & Disjoint subset & 0.804 & 74.5 & 72\\
CSI 500 /\allowbreak{} Sol & Random forest & Original pool & 0.804 & 74.5 & 72\\
CSI 500 /\allowbreak{} Sol & Single factor & Disjoint subset & 0.682 & 66.3 & 72\\
CSI 500 /\allowbreak{} Sol & Single factor & Original pool & 0.682 & 66.3 & 72\\
CSI 500 /\allowbreak{} Terra & Equal factor & Disjoint subset & 0.803 & 67.9 & 65\\
CSI 500 /\allowbreak{} Terra & Equal factor & Original pool & 0.803 & 67.9 & 65\\
CSI 500 /\allowbreak{} Terra & Frozen linear & Disjoint subset & 0.892 & 75.2 & 65\\
CSI 500 /\allowbreak{} Terra & Frozen linear & Original pool & 0.892 & 75.2 & 65\\
CSI 500 /\allowbreak{} Terra & Random forest & Disjoint subset & 0.791 & 67.3 & 65\\
CSI 500 /\allowbreak{} Terra & Random forest & Original pool & 0.791 & 67.3 & 65\\
CSI 500 /\allowbreak{} Terra & Single factor & Disjoint subset & 0.719 & 65.2 & 65\\
CSI 500 /\allowbreak{} Terra & Single factor & Original pool & 0.719 & 65.2 & 65\\
SSE 50 /\allowbreak{} Luna & Equal factor & Disjoint subset & 0.827 & 75.4 & 72\\
SSE 50 /\allowbreak{} Luna & Equal factor & Original pool & 0.827 & 75.4 & 72\\
SSE 50 /\allowbreak{} Luna & Frozen linear & Disjoint subset & 0.923 & 85.8 & 72\\
SSE 50 /\allowbreak{} Luna & Frozen linear & Original pool & 0.923 & 85.8 & 72\\
SSE 50 /\allowbreak{} Luna & Random forest & Disjoint subset & 0.845 & 77.9 & 72\\
SSE 50 /\allowbreak{} Luna & Random forest & Original pool & 0.845 & 77.9 & 72\\
SSE 50 /\allowbreak{} Luna & Single factor & Disjoint subset & 0.774 & 68.5 & 72\\
SSE 50 /\allowbreak{} Luna & Single factor & Original pool & 0.774 & 68.5 & 72\\
SSE 50 /\allowbreak{} Sol & Equal factor & Disjoint subset & 0.947 & 84.9 & 72\\
SSE 50 /\allowbreak{} Sol & Equal factor & Original pool & 0.947 & 84.9 & 72\\
SSE 50 /\allowbreak{} Sol & Frozen linear & Disjoint subset & 0.975 & 90.3 & 72\\
SSE 50 /\allowbreak{} Sol & Frozen linear & Original pool & 0.975 & 90.3 & 72\\
SSE 50 /\allowbreak{} Sol & Random forest & Disjoint subset & 0.863 & 78.8 & 72\\
SSE 50 /\allowbreak{} Sol & Random forest & Original pool & 0.863 & 78.8 & 72\\
SSE 50 /\allowbreak{} Sol & Single factor & Disjoint subset & 0.749 & 65.2 & 72\\
SSE 50 /\allowbreak{} Sol & Single factor & Original pool & 0.749 & 65.2 & 72\\
SSE 50 /\allowbreak{} Terra & Equal factor & Disjoint subset & 0.909 & 79.7 & 72\\
SSE 50 /\allowbreak{} Terra & Equal factor & Original pool & 0.909 & 79.7 & 72\\
SSE 50 /\allowbreak{} Terra & Frozen linear & Disjoint subset & 0.962 & 87.3 & 72\\
SSE 50 /\allowbreak{} Terra & Frozen linear & Original pool & 0.962 & 87.3 & 72\\
SSE 50 /\allowbreak{} Terra & Random forest & Disjoint subset & 0.880 & 79.5 & 72\\
SSE 50 /\allowbreak{} Terra & Random forest & Original pool & 0.880 & 79.5 & 72\\
SSE 50 /\allowbreak{} Terra & Single factor & Disjoint subset & 0.809 & 68.0 & 72\\
SSE 50 /\allowbreak{} Terra & Single factor & Original pool & 0.809 & 68.0 & 72\\
CSI 300 /\allowbreak{} Astra & Equal factor & Disjoint subset & 0.886 & 77.7 & 72\\
CSI 300 /\allowbreak{} Astra & Equal factor & Original pool & 0.892 & 78.9 & 72\\
CSI 300 /\allowbreak{} Astra & Frozen linear & Disjoint subset & 0.984 & 92.2 & 72\\
CSI 300 /\allowbreak{} Astra & Frozen linear & Original pool & 0.985 & 92.6 & 72\\
CSI 300 /\allowbreak{} Astra & Random forest & Disjoint subset & 0.885 & 78.8 & 72\\
CSI 300 /\allowbreak{} Astra & Random forest & Original pool & 0.893 & 79.2 & 72\\
CSI 300 /\allowbreak{} Astra & Single factor & Disjoint subset & 0.680 & 62.3 & 72\\
CSI 300 /\allowbreak{} Astra & Single factor & Original pool & 0.703 & 63.5 & 72\\
CSI 500 /\allowbreak{} Astra & Equal factor & Disjoint subset & 0.842 & 72.4 & 72\\
CSI 500 /\allowbreak{} Astra & Equal factor & Original pool & 0.842 & 72.4 & 72\\
CSI 500 /\allowbreak{} Astra & Frozen linear & Disjoint subset & 0.978 & 90.1 & 72\\
CSI 500 /\allowbreak{} Astra & Frozen linear & Original pool & 0.978 & 90.1 & 72\\
CSI 500 /\allowbreak{} Astra & Random forest & Disjoint subset & 0.862 & 76.2 & 72\\
CSI 500 /\allowbreak{} Astra & Random forest & Original pool & 0.862 & 76.2 & 72\\
CSI 500 /\allowbreak{} Astra & Single factor & Disjoint subset & 0.683 & 63.2 & 72\\
CSI 500 /\allowbreak{} Astra & Single factor & Original pool & 0.683 & 63.2 & 72\\
SSE 50 /\allowbreak{} Astra & Equal factor & Disjoint subset & 0.894 & 80.1 & 72\\
SSE 50 /\allowbreak{} Astra & Equal factor & Original pool & 0.894 & 80.1 & 72\\
SSE 50 /\allowbreak{} Astra & Frozen linear & Disjoint subset & 0.984 & 91.3 & 72\\
SSE 50 /\allowbreak{} Astra & Frozen linear & Original pool & 0.984 & 91.3 & 72\\
SSE 50 /\allowbreak{} Astra & Random forest & Disjoint subset & 0.904 & 79.4 & 72\\
SSE 50 /\allowbreak{} Astra & Random forest & Original pool & 0.904 & 79.4 & 72\\
SSE 50 /\allowbreak{} Astra & Single factor & Disjoint subset & 0.695 & 62.6 & 72\\
SSE 50 /\allowbreak{} Astra & Single factor & Original pool & 0.695 & 62.6 & 72\\
\end{longtable}
\noindent\footnotesize Parameters are frozen after development. Reported means are calculated across months after averaging the available scheduled runs within each month.\par\endgroup
\begingroup\small\setlength{\tabcolsep}{3pt}\renewcommand{\arraystretch}{1.17}
\begin{longtable}{@{}>{\raggedright\arraybackslash}p{\dimexpr 0.310345\linewidth-7.44828pt\relax}>{\raggedright\arraybackslash}p{\dimexpr 0.172414\linewidth-4.13793pt\relax}>{\raggedright\arraybackslash}p{\dimexpr 0.172414\linewidth-4.13793pt\relax}>{\raggedright\arraybackslash}p{\dimexpr 0.172414\linewidth-4.13793pt\relax}>{\raggedright\arraybackslash}p{\dimexpr 0.172414\linewidth-4.13793pt\relax}@{}}
\caption{Annual frozen-linear agreement}\label{tab:S22}\\
\toprule
\textbf{Pool /\allowbreak{} model} & \textbf{Holding year} & \textbf{Spearman} & \textbf{Overlap (\%)} & \textbf{Defined months}\\\midrule\endfirsthead
\multicolumn{5}{l}{\textbf{Table S22 (continued)}}\\\toprule
\textbf{Pool /\allowbreak{} model} & \textbf{Holding year} & \textbf{Spearman} & \textbf{Overlap (\%)} & \textbf{Defined months}\\\midrule\endhead
\midrule\multicolumn{5}{r}{Continued on next page}\\\endfoot
\bottomrule\endlastfoot
CSI 300 /\allowbreak{} Luna & 2020 & 0.682 & 68.7 & 12\\
CSI 300 /\allowbreak{} Luna & 2021 & 0.881 & 85.0 & 12\\
CSI 300 /\allowbreak{} Luna & 2022 & 0.951 & 86.5 & 12\\
CSI 300 /\allowbreak{} Luna & 2023 & 0.948 & 87.1 & 12\\
CSI 300 /\allowbreak{} Luna & 2024 & 0.952 & 86.2 & 12\\
CSI 300 /\allowbreak{} Luna & 2025 & 0.807 & 80.1 & 12\\
CSI 300 /\allowbreak{} Sol & 2020 & 0.824 & 76.8 & 12\\
CSI 300 /\allowbreak{} Sol & 2021 & 0.939 & 85.7 & 12\\
CSI 300 /\allowbreak{} Sol & 2022 & 0.940 & 86.9 & 12\\
CSI 300 /\allowbreak{} Sol & 2023 & 0.938 & 84.4 & 12\\
CSI 300 /\allowbreak{} Sol & 2024 & 0.963 & 88.9 & 12\\
CSI 300 /\allowbreak{} Sol & 2025 & 0.953 & 88.9 & 12\\
CSI 300 /\allowbreak{} Terra & 2020 & 0.774 & 76.5 & 12\\
CSI 300 /\allowbreak{} Terra & 2021 & 0.894 & 81.9 & 12\\
CSI 300 /\allowbreak{} Terra & 2022 & 0.962 & 86.2 & 12\\
CSI 300 /\allowbreak{} Terra & 2023 & 0.955 & 86.8 & 12\\
CSI 300 /\allowbreak{} Terra & 2024 & 0.957 & 88.2 & 12\\
CSI 300 /\allowbreak{} Terra & 2025 & 0.959 & 89.0 & 12\\
CSI 500 /\allowbreak{} Luna & 2020 & 0.505 & 48.3 & 11\\
CSI 500 /\allowbreak{} Luna & 2021 & 0.792 & 74.3 & 12\\
CSI 500 /\allowbreak{} Luna & 2022 & 0.895 & 78.2 & 11\\
CSI 500 /\allowbreak{} Luna & 2023 & 0.882 & 76.3 & 12\\
CSI 500 /\allowbreak{} Luna & 2024 & 0.910 & 77.3 & 11\\
CSI 500 /\allowbreak{} Luna & 2025 & 0.780 & 67.0 & 11\\
CSI 500 /\allowbreak{} Sol & 2020 & 0.824 & 74.7 & 12\\
CSI 500 /\allowbreak{} Sol & 2021 & 0.939 & 83.6 & 12\\
CSI 500 /\allowbreak{} Sol & 2022 & 0.965 & 85.9 & 12\\
CSI 500 /\allowbreak{} Sol & 2023 & 0.960 & 84.8 & 12\\
CSI 500 /\allowbreak{} Sol & 2024 & 0.972 & 88.8 & 12\\
CSI 500 /\allowbreak{} Sol & 2025 & 0.940 & 83.7 & 12\\
CSI 500 /\allowbreak{} Terra & 2020 & 0.821 & 70.6 & 11\\
CSI 500 /\allowbreak{} Terra & 2021 & 0.939 & 74.2 & 10\\
CSI 500 /\allowbreak{} Terra & 2022 & 0.887 & 73.3 & 10\\
CSI 500 /\allowbreak{} Terra & 2023 & 0.871 & 70.0 & 10\\
CSI 500 /\allowbreak{} Terra & 2024 & 0.920 & 85.4 & 12\\
CSI 500 /\allowbreak{} Terra & 2025 & 0.911 & 77.7 & 12\\
SSE 50 /\allowbreak{} Luna & 2020 & 0.902 & 88.0 & 12\\
SSE 50 /\allowbreak{} Luna & 2021 & 0.937 & 85.5 & 12\\
SSE 50 /\allowbreak{} Luna & 2022 & 0.942 & 87.2 & 12\\
SSE 50 /\allowbreak{} Luna & 2023 & 0.948 & 86.7 & 12\\
SSE 50 /\allowbreak{} Luna & 2024 & 0.952 & 82.8 & 12\\
SSE 50 /\allowbreak{} Luna & 2025 & 0.854 & 84.8 & 12\\
SSE 50 /\allowbreak{} Sol & 2020 & 0.948 & 85.3 & 12\\
SSE 50 /\allowbreak{} Sol & 2021 & 0.971 & 88.7 & 12\\
SSE 50 /\allowbreak{} Sol & 2022 & 0.986 & 91.5 & 12\\
SSE 50 /\allowbreak{} Sol & 2023 & 0.985 & 92.0 & 12\\
SSE 50 /\allowbreak{} Sol & 2024 & 0.989 & 94.3 & 12\\
SSE 50 /\allowbreak{} Sol & 2025 & 0.972 & 89.8 & 12\\
SSE 50 /\allowbreak{} Terra & 2020 & 0.928 & 88.7 & 12\\
SSE 50 /\allowbreak{} Terra & 2021 & 0.951 & 82.8 & 12\\
SSE 50 /\allowbreak{} Terra & 2022 & 0.979 & 88.2 & 12\\
SSE 50 /\allowbreak{} Terra & 2023 & 0.978 & 87.8 & 12\\
SSE 50 /\allowbreak{} Terra & 2024 & 0.977 & 84.2 & 12\\
SSE 50 /\allowbreak{} Terra & 2025 & 0.959 & 92.0 & 12\\
CSI 300 /\allowbreak{} Astra & 2020 & 0.979 & 89.8 & 12\\
CSI 300 /\allowbreak{} Astra & 2021 & 0.985 & 92.5 & 12\\
CSI 300 /\allowbreak{} Astra & 2022 & 0.989 & 92.6 & 12\\
CSI 300 /\allowbreak{} Astra & 2023 & 0.988 & 93.5 & 12\\
CSI 300 /\allowbreak{} Astra & 2024 & 0.990 & 95.1 & 12\\
CSI 300 /\allowbreak{} Astra & 2025 & 0.979 & 92.1 & 12\\
CSI 500 /\allowbreak{} Astra & 2020 & 0.972 & 88.9 & 12\\
CSI 500 /\allowbreak{} Astra & 2021 & 0.975 & 89.8 & 12\\
CSI 500 /\allowbreak{} Astra & 2022 & 0.981 & 90.8 & 12\\
CSI 500 /\allowbreak{} Astra & 2023 & 0.980 & 90.5 & 12\\
CSI 500 /\allowbreak{} Astra & 2024 & 0.987 & 92.4 & 12\\
CSI 500 /\allowbreak{} Astra & 2025 & 0.974 & 88.2 & 12\\
SSE 50 /\allowbreak{} Astra & 2020 & 0.970 & 87.0 & 12\\
SSE 50 /\allowbreak{} Astra & 2021 & 0.985 & 92.3 & 12\\
SSE 50 /\allowbreak{} Astra & 2022 & 0.987 & 86.3 & 12\\
SSE 50 /\allowbreak{} Astra & 2023 & 0.989 & 92.3 & 12\\
SSE 50 /\allowbreak{} Astra & 2024 & 0.992 & 93.7 & 12\\
SSE 50 /\allowbreak{} Astra & 2025 & 0.983 & 95.8 & 12\\
\end{longtable}
\noindent\footnotesize Years refer to portfolio entry, not the signal date. Yearly results are descriptive.\par\endgroup
\clearpage
\section*{S10 Rule-complexity inference}
\begingroup\small\setlength{\tabcolsep}{3pt}\renewcommand{\arraystretch}{1.17}
\begin{longtable}{@{}>{\raggedright\arraybackslash}p{\dimexpr 0.375000\linewidth-6.75000pt\relax}>{\raggedright\arraybackslash}p{\dimexpr 0.208333\linewidth-3.75000pt\relax}>{\raggedright\arraybackslash}p{\dimexpr 0.208333\linewidth-3.75000pt\relax}>{\raggedright\arraybackslash}p{\dimexpr 0.208333\linewidth-3.75000pt\relax}@{}}
\caption{Forest-minus-linear rank correlation}\label{tab:S23}\\
\toprule
\textbf{Model} & \textbf{Block months} & \textbf{Difference [95\% CI]} & \textbf{Holm p}\\\midrule\endfirsthead
\multicolumn{4}{l}{\textbf{Table S23 (continued)}}\\\toprule
\textbf{Model} & \textbf{Block months} & \textbf{Difference [95\% CI]} & \textbf{Holm p}\\\midrule\endhead
\midrule\multicolumn{4}{r}{Continued on next page}\\\endfoot
\bottomrule\endlastfoot
Luna & 6 & -0.077 [-0.087, -0.067] & <0.001\\
Luna & 3 & -0.077 [-0.086, -0.068] & <0.001\\
Luna & 12 & -0.077 [-0.087, -0.066] & <0.001\\
Terra & 6 & -0.082 [-0.095, -0.067] & <0.001\\
Terra & 3 & -0.082 [-0.093, -0.071] & <0.001\\
Terra & 12 & -0.082 [-0.097, -0.065] & <0.001\\
Sol & 6 & -0.112 [-0.123, -0.100] & <0.001\\
Sol & 3 & -0.112 [-0.122, -0.102] & <0.001\\
Sol & 12 & -0.112 [-0.123, -0.099] & <0.001\\
Astra & 3 & -0.080 [-0.091, -0.069] & <0.001\\
Astra & 6 & -0.080 [-0.094, -0.067] & <0.001\\
Astra & 12 & -0.080 [-0.095, -0.067] & <0.001\\
\end{longtable}
\noindent\footnotesize The reported difference subtracts frozen-linear rank agreement from bounded-forest rank agreement. The three models form one Holm family; Astra is a separate single contrast. Intervals are pointwise block-bootstrap intervals at each declared block length.\par\endgroup
\section*{S11 Archived response prediction and directional diagnostics}
\begingroup\small\setlength{\tabcolsep}{3pt}\renewcommand{\arraystretch}{1.17}
\begin{longtable}{@{}>{\raggedright\arraybackslash}p{\dimexpr 0.375000\linewidth-6.75000pt\relax}>{\raggedright\arraybackslash}p{\dimexpr 0.208333\linewidth-3.75000pt\relax}>{\raggedright\arraybackslash}p{\dimexpr 0.208333\linewidth-3.75000pt\relax}>{\raggedright\arraybackslash}p{\dimexpr 0.208333\linewidth-3.75000pt\relax}@{}}
\caption{Archived response-prediction inference}\label{tab:S24}\\
\toprule
\textbf{Model /\allowbreak{} intervention} & \textbf{Block months} & \textbf{MSE gain x 1,000 [CI]} & \textbf{Holm p}\\\midrule\endfirsthead
\multicolumn{4}{l}{\textbf{Table S24 (continued)}}\\\toprule
\textbf{Model /\allowbreak{} intervention} & \textbf{Block months} & \textbf{MSE gain x 1,000 [CI]} & \textbf{Holm p}\\\midrule\endhead
\midrule\multicolumn{4}{r}{Continued on next page}\\\endfoot
\bottomrule\endlastfoot
Luna /\allowbreak{} Momentum & 6 & 15.156 [12.870, 17.534] & <0.001\\
Luna /\allowbreak{} Momentum & 3 & 15.156 [12.486, 18.026] & <0.001\\
Luna /\allowbreak{} Momentum & 12 & 15.156 [13.010, 17.417] & <0.001\\
Luna /\allowbreak{} Risk/\allowbreak{}trading & 6 & 5.658 [1.438, 10.692] & 0.016\\
Luna /\allowbreak{} Risk/\allowbreak{}trading & 3 & 5.658 [2.140, 9.569] & 0.003\\
Luna /\allowbreak{} Risk/\allowbreak{}trading & 12 & 5.658 [0.350, 11.864] & 0.052\\
Luna /\allowbreak{} Valuation & 6 & 52.763 [42.505, 64.459] & <0.001\\
Luna /\allowbreak{} Valuation & 3 & 52.763 [43.421, 63.290] & <0.001\\
Luna /\allowbreak{} Valuation & 12 & 52.763 [40.152, 66.551] & <0.001\\
Sol /\allowbreak{} Momentum & 6 & 24.858 [21.591, 28.285] & <0.001\\
Sol /\allowbreak{} Momentum & 3 & 24.858 [22.102, 27.744] & <0.001\\
Sol /\allowbreak{} Momentum & 12 & 24.858 [21.063, 28.837] & <0.001\\
Sol /\allowbreak{} Risk/\allowbreak{}trading & 6 & 12.879 [8.485, 18.078] & <0.001\\
Sol /\allowbreak{} Risk/\allowbreak{}trading & 3 & 12.879 [9.090, 17.322] & <0.001\\
Sol /\allowbreak{} Risk/\allowbreak{}trading & 12 & 12.879 [7.485, 18.988] & <0.001\\
Sol /\allowbreak{} Valuation & 6 & 46.366 [34.186, 61.217] & <0.001\\
Sol /\allowbreak{} Valuation & 3 & 46.366 [36.332, 58.458] & <0.001\\
Sol /\allowbreak{} Valuation & 12 & 46.366 [31.066, 64.710] & <0.001\\
Terra /\allowbreak{} Momentum & 6 & 17.090 [14.599, 19.961] & <0.001\\
Terra /\allowbreak{} Momentum & 3 & 17.090 [15.018, 19.356] & <0.001\\
Terra /\allowbreak{} Momentum & 12 & 17.090 [14.064, 20.686] & <0.001\\
Terra /\allowbreak{} Risk/\allowbreak{}trading & 6 & 3.204 [2.145, 4.483] & <0.001\\
Terra /\allowbreak{} Risk/\allowbreak{}trading & 3 & 3.204 [2.175, 4.372] & <0.001\\
Terra /\allowbreak{} Risk/\allowbreak{}trading & 12 & 3.204 [2.124, 4.361] & <0.001\\
Terra /\allowbreak{} Valuation & 6 & 55.958 [42.799, 70.796] & <0.001\\
Terra /\allowbreak{} Valuation & 3 & 55.958 [44.983, 68.867] & <0.001\\
Terra /\allowbreak{} Valuation & 12 & 55.958 [39.730, 73.739] & <0.001\\
Astra /\allowbreak{} Valuation & 3 & 29.017 [23.850, 35.162] & <0.001\\
Astra /\allowbreak{} Valuation & 6 & 29.017 [23.086, 35.709] & <0.001\\
Astra /\allowbreak{} Valuation & 12 & 29.017 [22.148, 36.510] & <0.001\\
Astra /\allowbreak{} Momentum & 3 & 41.406 [35.572, 48.047] & <0.001\\
Astra /\allowbreak{} Momentum & 6 & 41.406 [34.206, 49.770] & <0.001\\
Astra /\allowbreak{} Momentum & 12 & 41.406 [32.219, 52.005] & <0.001\\
Astra /\allowbreak{} Risk/\allowbreak{}trading & 3 & 5.275 [4.308, 6.409] & <0.001\\
Astra /\allowbreak{} Risk/\allowbreak{}trading & 6 & 5.275 [4.352, 6.277] & <0.001\\
Astra /\allowbreak{} Risk/\allowbreak{}trading & 12 & 5.275 [4.292, 6.306] & <0.001\\
\end{longtable}
\noindent\footnotesize The nine historical comparisons and three Astra comparisons are adjusted separately at each block length. Intervals are pointwise.\par\endgroup
\begingroup\small\setlength{\tabcolsep}{3pt}\renewcommand{\arraystretch}{1.17}
\begin{longtable}{@{}>{\raggedright\arraybackslash}p{\dimexpr 0.310345\linewidth-7.44828pt\relax}>{\raggedright\arraybackslash}p{\dimexpr 0.172414\linewidth-4.13793pt\relax}>{\raggedright\arraybackslash}p{\dimexpr 0.172414\linewidth-4.13793pt\relax}>{\raggedright\arraybackslash}p{\dimexpr 0.172414\linewidth-4.13793pt\relax}>{\raggedright\arraybackslash}p{\dimexpr 0.172414\linewidth-4.13793pt\relax}@{}}
\caption{Direction and discrete-selection diagnostics}\label{tab:S25}\\
\toprule
\textbf{Model /\allowbreak{} condition} & \textbf{Direction agreement} & \textbf{Coverage} & \textbf{Entry Jaccard} & \textbf{Exit Jaccard}\\\midrule\endfirsthead
\multicolumn{5}{l}{\textbf{Table S25 (continued)}}\\\toprule
\textbf{Model /\allowbreak{} condition} & \textbf{Direction agreement} & \textbf{Coverage} & \textbf{Entry Jaccard} & \textbf{Exit Jaccard}\\\midrule\endhead
\midrule\multicolumn{5}{r}{Continued on next page}\\\endfoot
\bottomrule\endlastfoot
Luna /\allowbreak{} Momentum & 0.818 & 0.889 & 0.353 & 0.462\\
Luna /\allowbreak{} Risk/\allowbreak{}trading & 0.687 & 0.799 & 0.165 & 0.170\\
Luna /\allowbreak{} Hidden semantics & Undefined & 0.000 & 0.000 & 0.000\\
Luna /\allowbreak{} Valuation & 0.840 & 0.925 & 0.384 & 0.460\\
Luna /\allowbreak{} Row order & Undefined & 0.000 & 0.000 & 0.000\\
Sol /\allowbreak{} Momentum & 0.946 & 0.900 & 0.594 & 0.628\\
Sol /\allowbreak{} Risk/\allowbreak{}trading & 0.882 & 0.821 & 0.388 & 0.366\\
Sol /\allowbreak{} Hidden semantics & Undefined & 0.000 & 0.000 & 0.000\\
Sol /\allowbreak{} Valuation & 0.921 & 0.906 & 0.592 & 0.548\\
Sol /\allowbreak{} Row order & Undefined & 0.000 & 0.000 & 0.000\\
Terra /\allowbreak{} Momentum & 0.891 & 0.888 & 0.419 & 0.466\\
Terra /\allowbreak{} Risk/\allowbreak{}trading & 0.742 & 0.707 & 0.101 & 0.140\\
Terra /\allowbreak{} Hidden semantics & Undefined & 0.000 & 0.000 & 0.000\\
Terra /\allowbreak{} Valuation & 0.880 & 0.924 & 0.464 & 0.462\\
Terra /\allowbreak{} Row order & Undefined & 0.000 & 0.000 & 0.000\\
Astra /\allowbreak{} Momentum & 0.960 & 0.925 & 0.666 & 0.693\\
Astra /\allowbreak{} Risk/\allowbreak{}trading & 0.869 & 0.786 & 0.471 & 0.335\\
Astra /\allowbreak{} Hidden semantics & Undefined & 0.000 & 0.000 & 0.000\\
Astra /\allowbreak{} Valuation & 0.937 & 0.894 & 0.623 & 0.643\\
Astra /\allowbreak{} Row order & Undefined & 0.000 & 0.000 & 0.000\\
\end{longtable}
\noindent\footnotesize Direction agreement uses only nonzero actual and predicted changes, with its coverage reported. Jaccard is intersection divided by union. Empty unions have undefined Jaccard; no value is imputed.\par\endgroup
\begingroup\small\setlength{\tabcolsep}{3pt}\renewcommand{\arraystretch}{1.17}
\begin{longtable}{@{}>{\raggedright\arraybackslash}p{\dimexpr 0.375000\linewidth-6.75000pt\relax}>{\raggedright\arraybackslash}p{\dimexpr 0.208333\linewidth-3.75000pt\relax}>{\raggedright\arraybackslash}p{\dimexpr 0.208333\linewidth-3.75000pt\relax}>{\raggedright\arraybackslash}p{\dimexpr 0.208333\linewidth-3.75000pt\relax}@{}}
\caption{Archived change magnitude and repeated-input reference}\label{tab:S26}\\
\toprule
\textbf{Model /\allowbreak{} condition} & \textbf{Actual MSE} & \textbf{Repeat MSE} & \textbf{MSE difference}\\\midrule\endfirsthead
\multicolumn{4}{l}{\textbf{Table S26 (continued)}}\\\toprule
\textbf{Model /\allowbreak{} condition} & \textbf{Actual MSE} & \textbf{Repeat MSE} & \textbf{MSE difference}\\\midrule\endhead
\midrule\multicolumn{4}{r}{Continued on next page}\\\endfoot
\bottomrule\endlastfoot
Luna /\allowbreak{} Momentum & 0.03147 & 0.01795 & 0.01352\\
Luna /\allowbreak{} Risk/\allowbreak{}trading & 0.02405 & 0.01795 & 0.00610\\
Luna /\allowbreak{} Hidden semantics & 0.02736 & 0.01795 & 0.00941\\
Luna /\allowbreak{} Valuation & 0.09011 & 0.01795 & 0.07216\\
Luna /\allowbreak{} Row order & 0.02029 & 0.01795 & 0.00234\\
Sol /\allowbreak{} Momentum & 0.03015 & 0.00482 & 0.02532\\
Sol /\allowbreak{} Risk/\allowbreak{}trading & 0.01962 & 0.00482 & 0.01480\\
Sol /\allowbreak{} Hidden semantics & 0.02062 & 0.00482 & 0.01580\\
Sol /\allowbreak{} Valuation & 0.05470 & 0.00482 & 0.04988\\
Sol /\allowbreak{} Row order & 0.00561 & 0.00482 & 0.00078\\
Terra /\allowbreak{} Momentum & 0.02455 & 0.00568 & 0.01886\\
Terra /\allowbreak{} Risk/\allowbreak{}trading & 0.01258 & 0.00568 & 0.00689\\
Terra /\allowbreak{} Hidden semantics & 0.01851 & 0.00568 & 0.01283\\
Terra /\allowbreak{} Valuation & 0.07602 & 0.00568 & 0.07033\\
Terra /\allowbreak{} Row order & 0.00929 & 0.00568 & 0.00361\\
Astra /\allowbreak{} Momentum & 0.04496 & 0.00154 & 0.04343\\
Astra /\allowbreak{} Risk/\allowbreak{}trading & 0.00904 & 0.00154 & 0.00751\\
Astra /\allowbreak{} Hidden semantics & 0.10069 & 0.00154 & 0.09916\\
Astra /\allowbreak{} Valuation & 0.03512 & 0.00154 & 0.03359\\
Astra /\allowbreak{} Row order & 0.00243 & 0.00154 & 0.00089\\
\end{longtable}
\noindent\footnotesize The repeat reference averages all ten normal-input run pairs within month. Comparisons are descriptive and do not count pairs as independent market observations.\par\endgroup
\section*{S12 Matched single-feature experiment}
The fixed grid comprises 24 months, four models, three conditions and three responses per condition: 864 distinct valid responses. The fixed grid supplies 648 responses from the three models; Astra contributes 216 on exactly the same registered inputs and months. The predeclared conditional-noise precision calculation and inference specification apply to all four models. Astra collection is specified in astra\_\allowbreak{}20260921/\allowbreak{}lock.json and schedule.json.

\begingroup\small\setlength{\tabcolsep}{3pt}\renewcommand{\arraystretch}{1.17}
\begin{longtable}{@{}>{\raggedright\arraybackslash}p{\dimexpr 0.473684\linewidth-5.68421pt\relax}>{\raggedright\arraybackslash}p{\dimexpr 0.263158\linewidth-3.15789pt\relax}>{\raggedright\arraybackslash}p{\dimexpr 0.263158\linewidth-3.15789pt\relax}@{}}
\caption{Matched response-prediction inference}\label{tab:S27}\\
\toprule
\textbf{Model /\allowbreak{} intervention} & \textbf{MSE gain x 1,000 [CI]} & \textbf{Holm p}\\\midrule\endfirsthead
\multicolumn{3}{l}{\textbf{Table S27 (continued)}}\\\toprule
\textbf{Model /\allowbreak{} intervention} & \textbf{MSE gain x 1,000 [CI]} & \textbf{Holm p}\\\midrule\endhead
\midrule\multicolumn{3}{r}{Continued on next page}\\\endfoot
\bottomrule\endlastfoot
Luna /\allowbreak{} Valuation shift & 0.438 [0.182, 0.687] & 0.001\\
Luna /\allowbreak{} Momentum shift & 0.476 [0.178, 0.842] & 0.006\\
Terra /\allowbreak{} Valuation shift & 0.358 [0.169, 0.550] & 0.002\\
Terra /\allowbreak{} Momentum shift & 0.268 [0.126, 0.416] & 0.002\\
Sol /\allowbreak{} Valuation shift & 0.248 [0.094, 0.408] & 0.003\\
Sol /\allowbreak{} Momentum shift & 0.731 [0.547, 0.915] & <0.001\\
Astra /\allowbreak{} Valuation shift & 0.174 [0.095, 0.261] & <0.001\\
Astra /\allowbreak{} Momentum shift & 1.245 [0.940, 1.559] & <0.001\\
\end{longtable}
\noindent\footnotesize Squared percentile-rank units are multiplied by 1,000. There are three responses per condition and month. Intervals use month clusters stratified by holding year; the six historical and two Astra comparisons retain separate Holm families.\par\endgroup
\begingroup\small\setlength{\tabcolsep}{3pt}\renewcommand{\arraystretch}{1.17}
\begin{longtable}{@{}>{\raggedright\arraybackslash}p{\dimexpr 0.375000\linewidth-6.75000pt\relax}>{\raggedright\arraybackslash}p{\dimexpr 0.208333\linewidth-3.75000pt\relax}>{\raggedright\arraybackslash}p{\dimexpr 0.208333\linewidth-3.75000pt\relax}>{\raggedright\arraybackslash}p{\dimexpr 0.208333\linewidth-3.75000pt\relax}@{}}
\caption{Within-condition repeat agreement}\label{tab:S28}\\
\toprule
\textbf{Model /\allowbreak{} condition} & \textbf{Spearman} & \textbf{Selection overlap} & \textbf{Rank-change MSE}\\\midrule\endfirsthead
\multicolumn{4}{l}{\textbf{Table S28 (continued)}}\\\toprule
\textbf{Model /\allowbreak{} condition} & \textbf{Spearman} & \textbf{Selection overlap} & \textbf{Rank-change MSE}\\\midrule\endhead
\midrule\multicolumn{4}{r}{Continued on next page}\\\endfoot
\bottomrule\endlastfoot
Luna /\allowbreak{} Momentum shift & 0.840 & 0.801 & 0.02658\\
Luna /\allowbreak{} Original & 0.889 & 0.832 & 0.01845\\
Luna /\allowbreak{} Valuation shift & 0.860 & 0.822 & 0.02327\\
Sol /\allowbreak{} Momentum shift & 0.952 & 0.915 & 0.00795\\
Sol /\allowbreak{} Original & 0.960 & 0.929 & 0.00673\\
Sol /\allowbreak{} Valuation shift & 0.955 & 0.918 & 0.00743\\
Terra /\allowbreak{} Momentum shift & 0.957 & 0.872 & 0.00714\\
Terra /\allowbreak{} Original & 0.964 & 0.906 & 0.00595\\
Terra /\allowbreak{} Valuation shift & 0.957 & 0.869 & 0.00725\\
Astra /\allowbreak{} Momentum shift & 0.989 & 0.943 & 0.00183\\
Astra /\allowbreak{} Original & 0.989 & 0.960 & 0.00186\\
Astra /\allowbreak{} Valuation shift & 0.989 & 0.957 & 0.00181\\
\end{longtable}
\noindent\footnotesize Three response-pair comparisons are averaged within month before summarizing across months.\par\endgroup
\begingroup\small\setlength{\tabcolsep}{3pt}\renewcommand{\arraystretch}{1.17}
\begin{longtable}{@{}>{\raggedright\arraybackslash}p{\dimexpr 0.310345\linewidth-7.44828pt\relax}>{\raggedright\arraybackslash}p{\dimexpr 0.172414\linewidth-4.13793pt\relax}>{\raggedright\arraybackslash}p{\dimexpr 0.172414\linewidth-4.13793pt\relax}>{\raggedright\arraybackslash}p{\dimexpr 0.172414\linewidth-4.13793pt\relax}>{\raggedright\arraybackslash}p{\dimexpr 0.172414\linewidth-4.13793pt\relax}@{}}
\caption{Direction and selection under matched feature shifts}\label{tab:S29}\\
\toprule
\textbf{Model /\allowbreak{} condition} & \textbf{Signed response (pp)} & \textbf{Direction agreement /\allowbreak{} coverage} & \textbf{Selection overlap} & \textbf{Event Jaccard}\\\midrule\endfirsthead
\multicolumn{5}{l}{\textbf{Table S29 (continued)}}\\\toprule
\textbf{Model /\allowbreak{} condition} & \textbf{Signed response (pp)} & \textbf{Direction agreement /\allowbreak{} coverage} & \textbf{Selection overlap} & \textbf{Event Jaccard}\\\midrule\endhead
\midrule\multicolumn{5}{r}{Continued on next page}\\\endfoot
\bottomrule\endlastfoot
Luna /\allowbreak{} Momentum shift & 6.12 & 0.642 /\allowbreak{} 0.413 & 0.888 & 0.085\\
Luna /\allowbreak{} Valuation shift & 6.16 & 0.605 /\allowbreak{} 0.425 & 0.875 & 0.021\\
Sol /\allowbreak{} Momentum shift & 6.92 & 0.703 /\allowbreak{} 0.400 & 0.921 & 0.031\\
Sol /\allowbreak{} Valuation shift & 4.84 & 0.667 /\allowbreak{} 0.364 & 0.938 & 0.008\\
Terra /\allowbreak{} Momentum shift & 5.80 & 0.662 /\allowbreak{} 0.407 & 0.896 & 0.125\\
Terra /\allowbreak{} Valuation shift & 6.35 & 0.659 /\allowbreak{} 0.443 & 0.908 & 0.027\\
Astra /\allowbreak{} Momentum shift & 8.30 & 0.808 /\allowbreak{} 0.451 & 0.954 & 0.095\\
Astra /\allowbreak{} Valuation shift & 3.81 & 0.750 /\allowbreak{} 0.263 & 0.958 & 0.000\\
\end{longtable}
\noindent\footnotesize Signed response is the perturbed-stock percentile-rank change multiplied by its intervention sign, averaged over eight stocks and then months. Direction agreement requires nonzero actual and predicted changes; coverage is the corresponding share of candidates. Selection overlap compares original and altered LLM top sets. Event Jaccard compares predicted and actual entries/\allowbreak{}exits, conditional on a nonempty event union. These are descriptive endpoints.\par\endgroup
The matched experiment selects 24 months equally across six holding years. Eight stocks per month have both valuation and momentum percentiles in [0.2,0.8]. Four receive +0.2 and four -0.2 on a single feature, using identical stocks and signs for book-to-price and 12-to-1-month momentum. Every other numerical cell, identifier, row order and prompt component is held fixed. The conditions are original input, valuation shift and momentum shift, with three fresh independent responses per model and condition, totaling 864 valid responses across four models (648 primary responses and 216 Astra replication responses). Selection and call order are seeded. These are controlled numerical input changes, not simulated changes in company fundamentals.

Existing five-run variability determines the sample size using a predeclared conditional measurement-noise precision target for signed rank response. This calculation does not estimate financial-return power or total heterogeneity of treatment responses. Ranks are averaged over three responses per condition before computing changes. Inference resamples month clusters within holding-year strata, keeping all responses, models and conditions together within each sampled month.

\clearpage
\clearpage
\section*{S13 Incremental investment value and remaining-information sensitivity}
\begingroup\small\setlength{\tabcolsep}{3pt}\renewcommand{\arraystretch}{1.17}
\begin{longtable}{@{}>{\raggedright\arraybackslash}p{\dimexpr 0.310345\linewidth-7.44828pt\relax}>{\raggedright\arraybackslash}p{\dimexpr 0.172414\linewidth-4.13793pt\relax}>{\raggedright\arraybackslash}p{\dimexpr 0.172414\linewidth-4.13793pt\relax}>{\raggedright\arraybackslash}p{\dimexpr 0.172414\linewidth-4.13793pt\relax}>{\raggedright\arraybackslash}p{\dimexpr 0.172414\linewidth-4.13793pt\relax}@{}}
\caption{All net-return comparisons}\label{tab:S30}\\
\toprule
\textbf{Pool /\allowbreak{} model} & \textbf{Cost bps/\allowbreak{}side} & \textbf{Block months} & \textbf{Difference bps [CI]} & \textbf{Holm p}\\\midrule\endfirsthead
\multicolumn{5}{l}{\textbf{Table S30 (continued)}}\\\toprule
\textbf{Pool /\allowbreak{} model} & \textbf{Cost bps/\allowbreak{}side} & \textbf{Block months} & \textbf{Difference bps [CI]} & \textbf{Holm p}\\\midrule\endhead
\midrule\multicolumn{5}{r}{Continued on next page}\\\endfoot
\bottomrule\endlastfoot
SSE 50 /\allowbreak{} Luna & 0 & 3 & -16.623 [-43.960, 9.890] & 1.000\\
SSE 50 /\allowbreak{} Luna & 10 & 3 & -18.968 [-46.774, 7.590] & 1.000\\
SSE 50 /\allowbreak{} Luna & 25 & 3 & -22.475 [-50.575, 4.249] & 1.000\\
SSE 50 /\allowbreak{} Luna & 50 & 3 & -28.295 [-56.968, -1.365] & 0.469\\
SSE 50 /\allowbreak{} Terra & 0 & 3 & -18.438 [-39.455, 1.295] & 0.896\\
SSE 50 /\allowbreak{} Terra & 10 & 3 & -20.094 [-41.096, -0.375] & 0.623\\
SSE 50 /\allowbreak{} Terra & 25 & 3 & -22.572 [-43.546, -2.881] & 0.343\\
SSE 50 /\allowbreak{} Terra & 50 & 3 & -26.684 [-47.646, -6.759] & 0.151\\
SSE 50 /\allowbreak{} Sol & 0 & 3 & -0.732 [-16.207, 15.205] & 1.000\\
SSE 50 /\allowbreak{} Sol & 10 & 3 & -1.529 [-16.815, 14.220] & 1.000\\
SSE 50 /\allowbreak{} Sol & 25 & 3 & -2.722 [-17.804, 12.784] & 1.000\\
SSE 50 /\allowbreak{} Sol & 50 & 3 & -4.705 [-19.671, 10.524] & 1.000\\
CSI 300 /\allowbreak{} Luna & 0 & 3 & 10.401 [-25.389, 56.025] & 1.000\\
CSI 300 /\allowbreak{} Luna & 10 & 3 & 7.383 [-28.137, 52.196] & 1.000\\
CSI 300 /\allowbreak{} Luna & 25 & 3 & 2.867 [-31.936, 46.582] & 1.000\\
CSI 300 /\allowbreak{} Luna & 50 & 3 & -4.627 [-38.332, 37.434] & 1.000\\
CSI 300 /\allowbreak{} Terra & 0 & 3 & 4.367 [-12.048, 23.259] & 1.000\\
CSI 300 /\allowbreak{} Terra & 10 & 3 & 2.717 [-13.525, 21.116] & 1.000\\
CSI 300 /\allowbreak{} Terra & 25 & 3 & 0.249 [-15.797, 18.057] & 1.000\\
CSI 300 /\allowbreak{} Terra & 50 & 3 & -3.847 [-19.561, 13.005] & 1.000\\
CSI 300 /\allowbreak{} Sol & 0 & 3 & 9.431 [-15.855, 47.610] & 1.000\\
CSI 300 /\allowbreak{} Sol & 10 & 3 & 9.034 [-15.975, 46.752] & 1.000\\
CSI 300 /\allowbreak{} Sol & 25 & 3 & 8.438 [-16.128, 45.264] & 1.000\\
CSI 300 /\allowbreak{} Sol & 50 & 3 & 7.446 [-16.426, 42.958] & 1.000\\
CSI 500 /\allowbreak{} Luna & 0 & 3 & 15.879 [-44.689, 84.837] & 1.000\\
CSI 500 /\allowbreak{} Luna & 10 & 3 & 10.583 [-49.737, 79.041] & 1.000\\
CSI 500 /\allowbreak{} Luna & 25 & 3 & 2.659 [-57.404, 70.225] & 1.000\\
CSI 500 /\allowbreak{} Luna & 50 & 3 & -10.500 [-70.116, 55.761] & 1.000\\
CSI 500 /\allowbreak{} Terra & 0 & 3 & -30.675 [-66.598, 1.487] & 0.896\\
CSI 500 /\allowbreak{} Terra & 10 & 3 & -34.013 [-70.312, -1.520] & 0.623\\
CSI 500 /\allowbreak{} Terra & 25 & 3 & -39.002 [-75.884, -6.399] & 0.337\\
CSI 500 /\allowbreak{} Terra & 50 & 3 & -47.286 [-85.912, -13.125] & 0.151\\
CSI 500 /\allowbreak{} Sol & 0 & 3 & 1.464 [-13.498, 17.472] & 1.000\\
CSI 500 /\allowbreak{} Sol & 10 & 3 & 1.550 [-13.029, 17.066] & 1.000\\
CSI 500 /\allowbreak{} Sol & 25 & 3 & 1.678 [-12.473, 16.410] & 1.000\\
CSI 500 /\allowbreak{} Sol & 50 & 3 & 1.890 [-11.524, 15.777] & 1.000\\
SSE 50 /\allowbreak{} Luna & 0 & 6 & -16.623 [-40.661, 6.613] & 1.000\\
SSE 50 /\allowbreak{} Luna & 10 & 6 & -18.968 [-43.180, 4.296] & 1.000\\
SSE 50 /\allowbreak{} Luna & 25 & 6 & -22.475 [-46.859, 1.049] & 0.656\\
SSE 50 /\allowbreak{} Luna & 50 & 6 & -28.295 [-53.190, -4.547] & 0.252\\
SSE 50 /\allowbreak{} Terra & 0 & 6 & -18.438 [-37.444, -0.352] & 0.562\\
SSE 50 /\allowbreak{} Terra & 10 & 6 & -20.094 [-39.035, -1.941] & 0.382\\
SSE 50 /\allowbreak{} Terra & 25 & 6 & -22.572 [-41.503, -4.292] & 0.195\\
SSE 50 /\allowbreak{} Terra & 50 & 6 & -26.684 [-45.823, -8.088] & 0.078\\
SSE 50 /\allowbreak{} Sol & 0 & 6 & -0.732 [-14.698, 11.493] & 1.000\\
SSE 50 /\allowbreak{} Sol & 10 & 6 & -1.529 [-15.449, 10.527] & 1.000\\
SSE 50 /\allowbreak{} Sol & 25 & 6 & -2.722 [-16.660, 9.114] & 1.000\\
SSE 50 /\allowbreak{} Sol & 50 & 6 & -4.705 [-18.602, 6.955] & 1.000\\
CSI 300 /\allowbreak{} Luna & 0 & 6 & 10.401 [-28.609, 60.948] & 1.000\\
CSI 300 /\allowbreak{} Luna & 10 & 6 & 7.383 [-30.781, 56.811] & 1.000\\
CSI 300 /\allowbreak{} Luna & 25 & 6 & 2.867 [-34.165, 50.631] & 1.000\\
CSI 300 /\allowbreak{} Luna & 50 & 6 & -4.627 [-39.922, 40.294] & 1.000\\
CSI 300 /\allowbreak{} Terra & 0 & 6 & 4.367 [-10.132, 20.971] & 1.000\\
CSI 300 /\allowbreak{} Terra & 10 & 6 & 2.717 [-11.641, 18.770] & 1.000\\
CSI 300 /\allowbreak{} Terra & 25 & 6 & 0.249 [-13.963, 15.739] & 1.000\\
CSI 300 /\allowbreak{} Terra & 50 & 6 & -3.847 [-18.018, 10.775] & 1.000\\
CSI 300 /\allowbreak{} Sol & 0 & 6 & 9.431 [-15.672, 46.877] & 1.000\\
CSI 300 /\allowbreak{} Sol & 10 & 6 & 9.034 [-15.677, 45.733] & 1.000\\
CSI 300 /\allowbreak{} Sol & 25 & 6 & 8.438 [-15.643, 44.048] & 1.000\\
CSI 300 /\allowbreak{} Sol & 50 & 6 & 7.446 [-15.804, 41.263] & 1.000\\
CSI 500 /\allowbreak{} Luna & 0 & 6 & 15.879 [-51.834, 98.404] & 1.000\\
CSI 500 /\allowbreak{} Luna & 10 & 6 & 10.583 [-56.438, 92.149] & 1.000\\
CSI 500 /\allowbreak{} Luna & 25 & 6 & 2.659 [-63.747, 83.309] & 1.000\\
CSI 500 /\allowbreak{} Luna & 50 & 6 & -10.500 [-75.937, 68.133] & 1.000\\
CSI 500 /\allowbreak{} Terra & 0 & 6 & -30.675 [-62.150, -2.945] & 0.522\\
CSI 500 /\allowbreak{} Terra & 10 & 6 & -34.013 [-66.166, -6.034] & 0.342\\
CSI 500 /\allowbreak{} Terra & 25 & 6 & -39.002 [-72.219, -10.478] & 0.184\\
CSI 500 /\allowbreak{} Terra & 50 & 6 & -47.286 [-82.350, -16.910] & 0.078\\
CSI 500 /\allowbreak{} Sol & 0 & 6 & 1.464 [-13.594, 17.355] & 1.000\\
CSI 500 /\allowbreak{} Sol & 10 & 6 & 1.550 [-13.075, 16.845] & 1.000\\
CSI 500 /\allowbreak{} Sol & 25 & 6 & 1.678 [-12.434, 16.351] & 1.000\\
CSI 500 /\allowbreak{} Sol & 50 & 6 & 1.890 [-11.355, 15.653] & 1.000\\
SSE 50 /\allowbreak{} Luna & 0 & 12 & -16.623 [-37.593, 3.677] & 1.000\\
SSE 50 /\allowbreak{} Luna & 10 & 12 & -18.968 [-40.005, 1.195] & 0.674\\
SSE 50 /\allowbreak{} Luna & 25 & 12 & -22.475 [-43.550, -2.292] & 0.340\\
SSE 50 /\allowbreak{} Luna & 50 & 12 & -28.295 [-49.386, -8.050] & 0.128\\
SSE 50 /\allowbreak{} Terra & 0 & 12 & -18.438 [-34.229, -3.111] & 0.219\\
SSE 50 /\allowbreak{} Terra & 10 & 12 & -20.094 [-35.747, -4.750] & 0.134\\
SSE 50 /\allowbreak{} Terra & 25 & 12 & -22.572 [-38.085, -7.320] & 0.047\\
SSE 50 /\allowbreak{} Terra & 50 & 12 & -26.684 [-42.188, -11.370] & 0.008\\
SSE 50 /\allowbreak{} Sol & 0 & 12 & -0.732 [-13.602, 10.444] & 1.000\\
SSE 50 /\allowbreak{} Sol & 10 & 12 & -1.529 [-14.412, 9.584] & 1.000\\
SSE 50 /\allowbreak{} Sol & 25 & 12 & -2.722 [-15.511, 8.322] & 1.000\\
SSE 50 /\allowbreak{} Sol & 50 & 12 & -4.705 [-17.379, 6.283] & 1.000\\
CSI 300 /\allowbreak{} Luna & 0 & 12 & 10.401 [-26.924, 64.793] & 1.000\\
CSI 300 /\allowbreak{} Luna & 10 & 12 & 7.383 [-28.977, 60.053] & 1.000\\
CSI 300 /\allowbreak{} Luna & 25 & 12 & 2.867 [-32.133, 53.305] & 1.000\\
CSI 300 /\allowbreak{} Luna & 50 & 12 & -4.627 [-37.600, 41.403] & 1.000\\
CSI 300 /\allowbreak{} Terra & 0 & 12 & 4.367 [-9.582, 16.093] & 1.000\\
CSI 300 /\allowbreak{} Terra & 10 & 12 & 2.717 [-11.116, 14.167] & 1.000\\
CSI 300 /\allowbreak{} Terra & 25 & 12 & 0.249 [-13.471, 11.278] & 1.000\\
CSI 300 /\allowbreak{} Terra & 50 & 12 & -3.847 [-17.411, 7.236] & 1.000\\
CSI 300 /\allowbreak{} Sol & 0 & 12 & 9.431 [-16.055, 43.574] & 1.000\\
CSI 300 /\allowbreak{} Sol & 10 & 12 & 9.034 [-15.948, 42.305] & 1.000\\
CSI 300 /\allowbreak{} Sol & 25 & 12 & 8.438 [-15.746, 40.438] & 1.000\\
CSI 300 /\allowbreak{} Sol & 50 & 12 & 7.446 [-15.360, 37.403] & 1.000\\
CSI 500 /\allowbreak{} Luna & 0 & 12 & 15.879 [-60.838, 110.097] & 1.000\\
CSI 500 /\allowbreak{} Luna & 10 & 12 & 10.583 [-65.318, 103.330] & 1.000\\
CSI 500 /\allowbreak{} Luna & 25 & 12 & 2.659 [-71.773, 93.484] & 1.000\\
CSI 500 /\allowbreak{} Luna & 50 & 12 & -10.500 [-83.049, 76.783] & 1.000\\
CSI 500 /\allowbreak{} Terra & 0 & 12 & -30.675 [-57.994, -8.541] & 0.200\\
CSI 500 /\allowbreak{} Terra & 10 & 12 & -34.013 [-62.107, -11.002] & 0.122\\
CSI 500 /\allowbreak{} Terra & 25 & 12 & -39.002 [-68.189, -14.448] & 0.062\\
CSI 500 /\allowbreak{} Terra & 50 & 12 & -47.286 [-78.737, -19.800] & 0.015\\
CSI 500 /\allowbreak{} Sol & 0 & 12 & 1.464 [-10.361, 14.276] & 1.000\\
CSI 500 /\allowbreak{} Sol & 10 & 12 & 1.550 [-9.931, 13.827] & 1.000\\
CSI 500 /\allowbreak{} Sol & 25 & 12 & 1.678 [-9.396, 13.223] & 1.000\\
CSI 500 /\allowbreak{} Sol & 50 & 12 & 1.890 [-8.464, 12.193] & 1.000\\
SSE 50 /\allowbreak{} Astra & 0 & 3 & -6.014 [-29.357, 14.616] & 1.000\\
SSE 50 /\allowbreak{} Astra & 0 & 6 & -6.014 [-32.641, 16.890] & 1.000\\
SSE 50 /\allowbreak{} Astra & 0 & 12 & -6.014 [-31.429, 17.611] & 1.000\\
SSE 50 /\allowbreak{} Astra & 10 & 3 & -6.697 [-29.802, 13.779] & 1.000\\
SSE 50 /\allowbreak{} Astra & 10 & 6 & -6.697 [-33.124, 15.971] & 1.000\\
SSE 50 /\allowbreak{} Astra & 10 & 12 & -6.697 [-31.704, 16.585] & 1.000\\
SSE 50 /\allowbreak{} Astra & 25 & 3 & -7.719 [-30.408, 12.410] & 1.000\\
SSE 50 /\allowbreak{} Astra & 25 & 6 & -7.719 [-33.714, 14.536] & 1.000\\
SSE 50 /\allowbreak{} Astra & 25 & 12 & -7.719 [-32.427, 15.019] & 1.000\\
SSE 50 /\allowbreak{} Astra & 50 & 3 & -9.419 [-31.429, 10.284] & 1.000\\
SSE 50 /\allowbreak{} Astra & 50 & 6 & -9.419 [-34.658, 12.288] & 1.000\\
SSE 50 /\allowbreak{} Astra & 50 & 12 & -9.419 [-33.653, 12.497] & 1.000\\
CSI 300 /\allowbreak{} Astra & 0 & 3 & -5.828 [-21.398, 6.752] & 1.000\\
CSI 300 /\allowbreak{} Astra & 0 & 6 & -5.828 [-25.045, 8.299] & 1.000\\
CSI 300 /\allowbreak{} Astra & 0 & 12 & -5.828 [-26.494, 6.909] & 1.000\\
CSI 300 /\allowbreak{} Astra & 10 & 3 & -6.109 [-21.662, 6.484] & 1.000\\
CSI 300 /\allowbreak{} Astra & 10 & 6 & -6.109 [-25.385, 8.050] & 1.000\\
CSI 300 /\allowbreak{} Astra & 10 & 12 & -6.109 [-26.788, 6.655] & 1.000\\
CSI 300 /\allowbreak{} Astra & 25 & 3 & -6.529 [-22.110, 6.087] & 1.000\\
CSI 300 /\allowbreak{} Astra & 25 & 6 & -6.529 [-25.840, 7.628] & 1.000\\
CSI 300 /\allowbreak{} Astra & 25 & 12 & -6.529 [-27.351, 6.270] & 1.000\\
CSI 300 /\allowbreak{} Astra & 50 & 3 & -7.226 [-22.873, 5.452] & 1.000\\
CSI 300 /\allowbreak{} Astra & 50 & 6 & -7.226 [-26.661, 6.961] & 1.000\\
CSI 300 /\allowbreak{} Astra & 50 & 12 & -7.226 [-28.209, 5.662] & 1.000\\
CSI 500 /\allowbreak{} Astra & 0 & 3 & -8.986 [-22.433, 4.588] & 1.000\\
CSI 500 /\allowbreak{} Astra & 0 & 6 & -8.986 [-22.519, 4.837] & 1.000\\
CSI 500 /\allowbreak{} Astra & 0 & 12 & -8.986 [-23.361, 4.714] & 1.000\\
CSI 500 /\allowbreak{} Astra & 10 & 3 & -9.161 [-22.584, 4.387] & 1.000\\
CSI 500 /\allowbreak{} Astra & 10 & 6 & -9.161 [-22.649, 4.622] & 1.000\\
CSI 500 /\allowbreak{} Astra & 10 & 12 & -9.161 [-23.490, 4.550] & 1.000\\
CSI 500 /\allowbreak{} Astra & 25 & 3 & -9.419 [-22.799, 4.061] & 1.000\\
CSI 500 /\allowbreak{} Astra & 25 & 6 & -9.419 [-22.853, 4.277] & 1.000\\
CSI 500 /\allowbreak{} Astra & 25 & 12 & -9.419 [-23.689, 4.245] & 1.000\\
CSI 500 /\allowbreak{} Astra & 50 & 3 & -9.848 [-23.201, 3.503] & 1.000\\
CSI 500 /\allowbreak{} Astra & 50 & 6 & -9.848 [-23.106, 3.697] & 1.000\\
CSI 500 /\allowbreak{} Astra & 50 & 12 & -9.848 [-23.995, 3.774] & 1.000\\
\end{longtable}
\noindent\footnotesize A separate twelve-comparison Holm adjustment applies at each cost and block length. All four models use the same execution engine. Holm adjustment across the complete 144-comparison cost-by-block-length grid gives a minimum adjusted p-value of 0.101; the table reports adjustments within each twelve-comparison slice.\par\endgroup
\begingroup\small\setlength{\tabcolsep}{3pt}\renewcommand{\arraystretch}{1.17}
\begin{longtable}{@{}>{\raggedright\arraybackslash}p{\dimexpr 0.310345\linewidth-7.44828pt\relax}>{\raggedright\arraybackslash}p{\dimexpr 0.172414\linewidth-4.13793pt\relax}>{\raggedright\arraybackslash}p{\dimexpr 0.172414\linewidth-4.13793pt\relax}>{\raggedright\arraybackslash}p{\dimexpr 0.172414\linewidth-4.13793pt\relax}>{\raggedright\arraybackslash}p{\dimexpr 0.172414\linewidth-4.13793pt\relax}@{}}
\caption{Remaining-information sensitivity}\label{tab:S31}\\
\toprule
\textbf{Pool /\allowbreak{} model} & \textbf{Block months} & \textbf{Partial r [CI]} & \textbf{Defined months} & \textbf{Holm p}\\\midrule\endfirsthead
\multicolumn{5}{l}{\textbf{Table S31 (continued)}}\\\toprule
\textbf{Pool /\allowbreak{} model} & \textbf{Block months} & \textbf{Partial r [CI]} & \textbf{Defined months} & \textbf{Holm p}\\\midrule\endhead
\midrule\multicolumn{5}{r}{Continued on next page}\\\endfoot
\bottomrule\endlastfoot
SSE 50 /\allowbreak{} Luna & 3 & -0.017 [-0.059, 0.031] & 72 & 1.000\\
SSE 50 /\allowbreak{} Terra & 3 & -0.025 [-0.066, 0.016] & 72 & 1.000\\
SSE 50 /\allowbreak{} Sol & 3 & -0.004 [-0.048, 0.045] & 72 & 1.000\\
CSI 300 /\allowbreak{} Luna & 3 & -0.012 [-0.031, 0.007] & 72 & 1.000\\
CSI 300 /\allowbreak{} Terra & 3 & -0.004 [-0.021, 0.013] & 72 & 1.000\\
CSI 300 /\allowbreak{} Sol & 3 & -0.005 [-0.023, 0.013] & 72 & 1.000\\
CSI 500 /\allowbreak{} Luna & 3 & -0.009 [-0.022, 0.004] & 68 & 1.000\\
CSI 500 /\allowbreak{} Terra & 3 & -0.003 [-0.016, 0.010] & 65 & 1.000\\
CSI 500 /\allowbreak{} Sol & 3 & -0.012 [-0.030, 0.006] & 72 & 1.000\\
SSE 50 /\allowbreak{} Luna & 6 & -0.017 [-0.056, 0.032] & 72 & 1.000\\
SSE 50 /\allowbreak{} Terra & 6 & -0.025 [-0.062, 0.016] & 72 & 1.000\\
SSE 50 /\allowbreak{} Sol & 6 & -0.004 [-0.048, 0.043] & 72 & 1.000\\
CSI 300 /\allowbreak{} Luna & 6 & -0.012 [-0.033, 0.010] & 72 & 1.000\\
CSI 300 /\allowbreak{} Terra & 6 & -0.004 [-0.022, 0.014] & 72 & 1.000\\
CSI 300 /\allowbreak{} Sol & 6 & -0.005 [-0.027, 0.015] & 72 & 1.000\\
CSI 500 /\allowbreak{} Luna & 6 & -0.009 [-0.020, 0.001] & 68 & 0.962\\
CSI 500 /\allowbreak{} Terra & 6 & -0.003 [-0.016, 0.010] & 65 & 1.000\\
CSI 500 /\allowbreak{} Sol & 6 & -0.012 [-0.030, 0.007] & 72 & 1.000\\
SSE 50 /\allowbreak{} Luna & 12 & -0.017 [-0.052, 0.030] & 72 & 1.000\\
SSE 50 /\allowbreak{} Terra & 12 & -0.025 [-0.063, 0.013] & 72 & 1.000\\
SSE 50 /\allowbreak{} Sol & 12 & -0.004 [-0.050, 0.038] & 72 & 1.000\\
CSI 300 /\allowbreak{} Luna & 12 & -0.012 [-0.031, 0.008] & 72 & 1.000\\
CSI 300 /\allowbreak{} Terra & 12 & -0.004 [-0.021, 0.014] & 72 & 1.000\\
CSI 300 /\allowbreak{} Sol & 12 & -0.005 [-0.030, 0.016] & 72 & 1.000\\
CSI 500 /\allowbreak{} Luna & 12 & -0.009 [-0.021, 0.001] & 68 & 1.000\\
CSI 500 /\allowbreak{} Terra & 12 & -0.003 [-0.016, 0.009] & 65 & 1.000\\
CSI 500 /\allowbreak{} Sol & 12 & -0.012 [-0.030, 0.007] & 72 & 1.000\\
SSE 50 /\allowbreak{} Astra & 3 & 0.022 [-0.010, 0.052] & 72 & 1.000\\
SSE 50 /\allowbreak{} Astra & 6 & 0.022 [-0.008, 0.051] & 72 & 1.000\\
SSE 50 /\allowbreak{} Astra & 12 & 0.022 [-0.003, 0.047] & 72 & 1.000\\
CSI 300 /\allowbreak{} Astra & 3 & 0.010 [-0.006, 0.027] & 72 & 1.000\\
CSI 300 /\allowbreak{} Astra & 6 & 0.010 [-0.005, 0.026] & 72 & 1.000\\
CSI 300 /\allowbreak{} Astra & 12 & 0.010 [-0.002, 0.023] & 72 & 1.000\\
CSI 500 /\allowbreak{} Astra & 3 & -0.001 [-0.011, 0.009] & 72 & 1.000\\
CSI 500 /\allowbreak{} Astra & 6 & -0.001 [-0.011, 0.008] & 72 & 1.000\\
CSI 500 /\allowbreak{} Astra & 12 & -0.001 [-0.009, 0.007] & 72 & 1.000\\
\end{longtable}
\noindent\footnotesize A separate twelve-comparison Holm adjustment applies at each block length. Defined-month counts accompany conditional return correlations.\par\endgroup
\section*{S14 Astra input-intervention results}
The Astra input-intervention design comprises 1,296 responses across 72 SSE 50 months, six conditions and three complete runs per condition. The six conditions match those for Luna, Terra and Sol in Section S15. Return contrasts pair each altered condition with its within-batch primary condition at 10 basis points per side. Numerical removals form a three-comparison Holm family; semantics and row order each have a single contrast. Astra and three-model estimates use their respective inference families.

\begingroup\small\setlength{\tabcolsep}{3pt}\renewcommand{\arraystretch}{1.17}
\begin{longtable}{@{}>{\raggedright\arraybackslash}p{\dimexpr 0.375000\linewidth-6.75000pt\relax}>{\raggedright\arraybackslash}p{\dimexpr 0.208333\linewidth-3.75000pt\relax}>{\raggedright\arraybackslash}p{\dimexpr 0.208333\linewidth-3.75000pt\relax}>{\raggedright\arraybackslash}p{\dimexpr 0.208333\linewidth-3.75000pt\relax}@{}}
\caption{Astra three-run replication return contrasts}\label{tab:S32}\\
\toprule
\textbf{Condition} & \textbf{Mean difference (pp)} & \textbf{95\% interval (pp)} & \textbf{Adjusted p}\\\midrule\endfirsthead
\multicolumn{4}{l}{\textbf{Table S32 (continued)}}\\\toprule
\textbf{Condition} & \textbf{Mean difference (pp)} & \textbf{95\% interval (pp)} & \textbf{Adjusted p}\\\midrule\endhead
\midrule\multicolumn{4}{r}{Continued on next page}\\\endfoot
\bottomrule\endlastfoot
Momentum & -0.201 & [-0.567, 0.136] & 0.808\\
Risk/\allowbreak{}trading & 0.105 & [-0.207, 0.513] & 1.000\\
Hidden semantics & -0.320 & [-1.160, 0.637] & 0.488\\
Valuation & 0.002 & [-0.657, 0.761] & 1.000\\
Row order & -0.002 & [-0.166, 0.157] & 0.981\\
\end{longtable}
\noindent\footnotesize Intervals are pointwise and use whole-run trajectories and six-month calendar blocks. Each estimate uses three within-batch trajectories. Historical contrasts use run 1.\par\endgroup
\section*{S15 Luna, Terra and Sol input-intervention results}
The Luna, Terra and Sol input-intervention design comprises 3,888 valid responses in 18 model-condition groups, each with 216 responses (72 months times three runs). The 432 input matrices implement the six conditions. All 54 trajectories use the common trading engine and 72-month calendar, with costs of 0, 10, 25 and 50 basis points per side. The collection specification was locked on 27 September 2026 at 10:53 UTC, and responses were collected between 10:55 and 14:30 UTC that day.

Requests were sent to https:/\allowbreak{}/\allowbreak{}api.openai.com/\allowbreak{}v1/\allowbreak{}responses using the three specified model labels, medium reasoning and independent tool-free contexts. Returned model labels matched the requested labels. These labels document the service response; backend model snapshots were not independently verified. Collection used concurrent workers while preserving the original input and task identity, and formal records retain the first complete valid answer for each scheduled task.

The first 1,117 valid responses requested the default service tier; the remaining 2,771 requested fast. All 3,888 reported default. Requested and reported tiers are stored separately. The collection lock and execution settings record the concurrency and service-tier configuration.

At six-month blocks, all 24 bootstrap tests have adjusted p-values above 0.05. For Luna, reordering rows changes mean monthly net return by 0.542 percentage points. Its pointwise six-month joint-bootstrap interval is [0.154, 1.125] percentage points. The bootstrap Holm-adjusted p-value is 0.098, while the HAC Holm-adjusted p-value is 0.010 within the three row-order comparisons. Across all 24 six-month intervention contrasts as one alternative family, the smallest HAC Holm-adjusted p-value is 0.0815. The inference therefore depends on the uncertainty procedure and the comparison family; the confidence interval is pointwise.

\begingroup\small\setlength{\tabcolsep}{3pt}\renewcommand{\arraystretch}{1.17}
\begin{longtable}{@{}>{\raggedright\arraybackslash}p{\dimexpr 0.264706\linewidth-7.94118pt\relax}>{\raggedright\arraybackslash}p{\dimexpr 0.147059\linewidth-4.41176pt\relax}>{\raggedright\arraybackslash}p{\dimexpr 0.147059\linewidth-4.41176pt\relax}>{\raggedright\arraybackslash}p{\dimexpr 0.147059\linewidth-4.41176pt\relax}>{\raggedright\arraybackslash}p{\dimexpr 0.147059\linewidth-4.41176pt\relax}>{\raggedright\arraybackslash}p{\dimexpr 0.147059\linewidth-4.41176pt\relax}@{}}
\caption{Paired return contrasts with 3-month blocks}\label{tab:S33}\\
\toprule
\textbf{Model /\allowbreak{} intervention} & \textbf{Contrast family /\allowbreak{} size} & \textbf{Mean pp} & \textbf{95\% joint interval pp} & \textbf{HAC p /\allowbreak{} Holm p} & \textbf{Bootstrap p /\allowbreak{} Holm p}\\\midrule\endfirsthead
\multicolumn{6}{l}{\textbf{Table S33 (continued)}}\\\toprule
\textbf{Model /\allowbreak{} intervention} & \textbf{Contrast family /\allowbreak{} size} & \textbf{Mean pp} & \textbf{95\% joint interval pp} & \textbf{HAC p /\allowbreak{} Holm p} & \textbf{Bootstrap p /\allowbreak{} Holm p}\\\midrule\endhead
\midrule\multicolumn{6}{r}{Continued on next page}\\\endfoot
\bottomrule\endlastfoot
Luna /\allowbreak{} Valuation neutralized & Factor neutralization /\allowbreak{} 9 & 0.592 & [-0.552, 1.887] & 0.314 /\allowbreak{} 1.000 & 0.344 /\allowbreak{} 1.000\\
Luna /\allowbreak{} Momentum neutralized & Factor neutralization /\allowbreak{} 9 & 0.043 & [-0.318, 0.418] & 0.791 /\allowbreak{} 1.000 & 0.816 /\allowbreak{} 1.000\\
Luna /\allowbreak{} Risk/\allowbreak{}trading neutralized & Factor neutralization /\allowbreak{} 9 & 0.161 & [-0.246, 0.631] & 0.355 /\allowbreak{} 1.000 & 0.433 /\allowbreak{} 1.000\\
Terra /\allowbreak{} Valuation neutralized & Factor neutralization /\allowbreak{} 9 & 0.096 & [-0.768, 1.017] & 0.807 /\allowbreak{} 1.000 & 0.828 /\allowbreak{} 1.000\\
Terra /\allowbreak{} Momentum neutralized & Factor neutralization /\allowbreak{} 9 & -0.055 & [-0.462, 0.333] & 0.779 /\allowbreak{} 1.000 & 0.791 /\allowbreak{} 1.000\\
Terra /\allowbreak{} Risk/\allowbreak{}trading neutralized & Factor neutralization /\allowbreak{} 9 & -0.012 & [-0.301, 0.289] & 0.919 /\allowbreak{} 1.000 & 0.929 /\allowbreak{} 1.000\\
Sol /\allowbreak{} Valuation neutralized & Factor neutralization /\allowbreak{} 9 & 0.055 & [-0.710, 0.952] & 0.886 /\allowbreak{} 1.000 & 0.896 /\allowbreak{} 1.000\\
Sol /\allowbreak{} Momentum neutralized & Factor neutralization /\allowbreak{} 9 & -0.154 & [-0.547, 0.221] & 0.384 /\allowbreak{} 1.000 & 0.434 /\allowbreak{} 1.000\\
Sol /\allowbreak{} Risk/\allowbreak{}trading neutralized & Factor neutralization /\allowbreak{} 9 & 0.118 & [-0.232, 0.557] & 0.499 /\allowbreak{} 1.000 & 0.552 /\allowbreak{} 1.000\\
Luna /\allowbreak{} Semantics removed & Semantics /\allowbreak{} 3 & 0.135 & [-0.221, 0.512] & 0.407 /\allowbreak{} 1.000 & 0.461 /\allowbreak{} 1.000\\
Terra /\allowbreak{} Semantics removed & Semantics /\allowbreak{} 3 & -0.042 & [-0.299, 0.227] & 0.732 /\allowbreak{} 1.000 & 0.750 /\allowbreak{} 1.000\\
Sol /\allowbreak{} Semantics removed & Semantics /\allowbreak{} 3 & 0.030 & [-0.310, 0.332] & 0.843 /\allowbreak{} 1.000 & 0.855 /\allowbreak{} 1.000\\
Luna /\allowbreak{} Rows reordered & Row order /\allowbreak{} 3 & 0.542 & [0.160, 1.059] & 0.003 /\allowbreak{} 0.010 & 0.022 /\allowbreak{} 0.066\\
Terra /\allowbreak{} Rows reordered & Row order /\allowbreak{} 3 & 0.051 & [-0.155, 0.277] & 0.620 /\allowbreak{} 0.620 & 0.644 /\allowbreak{} 0.644\\
Sol /\allowbreak{} Rows reordered & Row order /\allowbreak{} 3 & 0.102 & [-0.041, 0.252] & 0.071 /\allowbreak{} 0.142 & 0.156 /\allowbreak{} 0.312\\
Luna /\allowbreak{} Valuation neutralized & EqualFactor-adjusted /\allowbreak{} 9 & 0.617 & [-0.385, 1.737] & 0.217 /\allowbreak{} 1.000 & 0.257 /\allowbreak{} 1.000\\
Luna /\allowbreak{} Momentum neutralized & EqualFactor-adjusted /\allowbreak{} 9 & 0.334 & [-0.103, 0.806] & 0.118 /\allowbreak{} 1.000 & 0.142 /\allowbreak{} 1.000\\
Luna /\allowbreak{} Risk/\allowbreak{}trading neutralized & EqualFactor-adjusted /\allowbreak{} 9 & 0.084 & [-0.389, 0.595] & 0.696 /\allowbreak{} 1.000 & 0.724 /\allowbreak{} 1.000\\
Terra /\allowbreak{} Valuation neutralized & EqualFactor-adjusted /\allowbreak{} 9 & 0.121 & [-0.446, 0.694] & 0.642 /\allowbreak{} 1.000 & 0.672 /\allowbreak{} 1.000\\
Terra /\allowbreak{} Momentum neutralized & EqualFactor-adjusted /\allowbreak{} 9 & 0.236 & [-0.203, 0.665] & 0.290 /\allowbreak{} 1.000 & 0.277 /\allowbreak{} 1.000\\
Terra /\allowbreak{} Risk/\allowbreak{}trading neutralized & EqualFactor-adjusted /\allowbreak{} 9 & -0.088 & [-0.496, 0.315] & 0.649 /\allowbreak{} 1.000 & 0.668 /\allowbreak{} 1.000\\
Sol /\allowbreak{} Valuation neutralized & EqualFactor-adjusted /\allowbreak{} 9 & 0.080 & [-0.373, 0.519] & 0.713 /\allowbreak{} 1.000 & 0.724 /\allowbreak{} 1.000\\
Sol /\allowbreak{} Momentum neutralized & EqualFactor-adjusted /\allowbreak{} 9 & 0.137 & [-0.210, 0.475] & 0.415 /\allowbreak{} 1.000 & 0.431 /\allowbreak{} 1.000\\
Sol /\allowbreak{} Risk/\allowbreak{}trading neutralized & EqualFactor-adjusted /\allowbreak{} 9 & 0.042 & [-0.340, 0.434] & 0.796 /\allowbreak{} 1.000 & 0.824 /\allowbreak{} 1.000\\
\end{longtable}
\noindent\footnotesize Twenty-four contrasts at 10 bp: nine factor-neutralization, three semantic, three row-order, and nine EqualFactor-adjusted factor contrasts. Positive values mean higher altered-condition returns. Adjustment subtracts the corresponding archived EqualFactor change. Inference uses three whole paired trajectories and 72 shared months; HAC uses three lags, and intervals are pointwise. Each procedure and block length uses its own specified 9/\allowbreak{}3/\allowbreak{}3/\allowbreak{}9 Holm families. Undefined HAC tests remain undefined and retain their place in the planned family size.\par\endgroup
\begingroup\small\setlength{\tabcolsep}{3pt}\renewcommand{\arraystretch}{1.17}
\begin{longtable}{@{}>{\raggedright\arraybackslash}p{\dimexpr 0.264706\linewidth-7.94118pt\relax}>{\raggedright\arraybackslash}p{\dimexpr 0.147059\linewidth-4.41176pt\relax}>{\raggedright\arraybackslash}p{\dimexpr 0.147059\linewidth-4.41176pt\relax}>{\raggedright\arraybackslash}p{\dimexpr 0.147059\linewidth-4.41176pt\relax}>{\raggedright\arraybackslash}p{\dimexpr 0.147059\linewidth-4.41176pt\relax}>{\raggedright\arraybackslash}p{\dimexpr 0.147059\linewidth-4.41176pt\relax}@{}}
\caption{Paired return contrasts with 6-month blocks}\label{tab:S34}\\
\toprule
\textbf{Model /\allowbreak{} intervention} & \textbf{Contrast family /\allowbreak{} size} & \textbf{Mean pp} & \textbf{95\% joint interval pp} & \textbf{HAC p /\allowbreak{} Holm p} & \textbf{Bootstrap p /\allowbreak{} Holm p}\\\midrule\endfirsthead
\multicolumn{6}{l}{\textbf{Table S34 (continued)}}\\\toprule
\textbf{Model /\allowbreak{} intervention} & \textbf{Contrast family /\allowbreak{} size} & \textbf{Mean pp} & \textbf{95\% joint interval pp} & \textbf{HAC p /\allowbreak{} Holm p} & \textbf{Bootstrap p /\allowbreak{} Holm p}\\\midrule\endhead
\midrule\multicolumn{6}{r}{Continued on next page}\\\endfoot
\bottomrule\endlastfoot
Luna /\allowbreak{} Valuation neutralized & Factor neutralization /\allowbreak{} 9 & 0.592 & [-0.529, 2.006] & 0.314 /\allowbreak{} 1.000 & 0.374 /\allowbreak{} 1.000\\
Luna /\allowbreak{} Momentum neutralized & Factor neutralization /\allowbreak{} 9 & 0.043 & [-0.317, 0.438] & 0.791 /\allowbreak{} 1.000 & 0.813 /\allowbreak{} 1.000\\
Luna /\allowbreak{} Risk/\allowbreak{}trading neutralized & Factor neutralization /\allowbreak{} 9 & 0.161 & [-0.248, 0.622] & 0.355 /\allowbreak{} 1.000 & 0.450 /\allowbreak{} 1.000\\
Terra /\allowbreak{} Valuation neutralized & Factor neutralization /\allowbreak{} 9 & 0.096 & [-0.668, 0.926] & 0.807 /\allowbreak{} 1.000 & 0.818 /\allowbreak{} 1.000\\
Terra /\allowbreak{} Momentum neutralized & Factor neutralization /\allowbreak{} 9 & -0.055 & [-0.472, 0.338] & 0.779 /\allowbreak{} 1.000 & 0.795 /\allowbreak{} 1.000\\
Terra /\allowbreak{} Risk/\allowbreak{}trading neutralized & Factor neutralization /\allowbreak{} 9 & -0.012 & [-0.279, 0.300] & 0.919 /\allowbreak{} 1.000 & 0.929 /\allowbreak{} 1.000\\
Sol /\allowbreak{} Valuation neutralized & Factor neutralization /\allowbreak{} 9 & 0.055 & [-0.634, 0.871] & 0.886 /\allowbreak{} 1.000 & 0.889 /\allowbreak{} 1.000\\
Sol /\allowbreak{} Momentum neutralized & Factor neutralization /\allowbreak{} 9 & -0.154 & [-0.530, 0.198] & 0.384 /\allowbreak{} 1.000 & 0.400 /\allowbreak{} 1.000\\
Sol /\allowbreak{} Risk/\allowbreak{}trading neutralized & Factor neutralization /\allowbreak{} 9 & 0.118 & [-0.181, 0.526] & 0.499 /\allowbreak{} 1.000 & 0.512 /\allowbreak{} 1.000\\
Luna /\allowbreak{} Semantics removed & Semantics /\allowbreak{} 3 & 0.135 & [-0.215, 0.512] & 0.407 /\allowbreak{} 1.000 & 0.434 /\allowbreak{} 1.000\\
Terra /\allowbreak{} Semantics removed & Semantics /\allowbreak{} 3 & -0.042 & [-0.296, 0.212] & 0.732 /\allowbreak{} 1.000 & 0.744 /\allowbreak{} 1.000\\
Sol /\allowbreak{} Semantics removed & Semantics /\allowbreak{} 3 & 0.030 & [-0.322, 0.347] & 0.843 /\allowbreak{} 1.000 & 0.852 /\allowbreak{} 1.000\\
Luna /\allowbreak{} Rows reordered & Row order /\allowbreak{} 3 & 0.542 & [0.154, 1.125] & 0.003 /\allowbreak{} 0.010 & 0.033 /\allowbreak{} 0.098\\
Terra /\allowbreak{} Rows reordered & Row order /\allowbreak{} 3 & 0.051 & [-0.158, 0.289] & 0.620 /\allowbreak{} 0.620 & 0.656 /\allowbreak{} 0.656\\
Sol /\allowbreak{} Rows reordered & Row order /\allowbreak{} 3 & 0.102 & [-0.036, 0.246] & 0.071 /\allowbreak{} 0.142 & 0.139 /\allowbreak{} 0.278\\
Luna /\allowbreak{} Valuation neutralized & EqualFactor-adjusted /\allowbreak{} 9 & 0.617 & [-0.329, 1.728] & 0.217 /\allowbreak{} 1.000 & 0.239 /\allowbreak{} 1.000\\
Luna /\allowbreak{} Momentum neutralized & EqualFactor-adjusted /\allowbreak{} 9 & 0.334 & [-0.102, 0.831] & 0.118 /\allowbreak{} 1.000 & 0.159 /\allowbreak{} 1.000\\
Luna /\allowbreak{} Risk/\allowbreak{}trading neutralized & EqualFactor-adjusted /\allowbreak{} 9 & 0.084 & [-0.365, 0.601] & 0.696 /\allowbreak{} 1.000 & 0.722 /\allowbreak{} 1.000\\
Terra /\allowbreak{} Valuation neutralized & EqualFactor-adjusted /\allowbreak{} 9 & 0.121 & [-0.443, 0.659] & 0.642 /\allowbreak{} 1.000 & 0.667 /\allowbreak{} 1.000\\
Terra /\allowbreak{} Momentum neutralized & EqualFactor-adjusted /\allowbreak{} 9 & 0.236 & [-0.200, 0.674] & 0.290 /\allowbreak{} 1.000 & 0.295 /\allowbreak{} 1.000\\
Terra /\allowbreak{} Risk/\allowbreak{}trading neutralized & EqualFactor-adjusted /\allowbreak{} 9 & -0.088 & [-0.514, 0.350] & 0.649 /\allowbreak{} 1.000 & 0.682 /\allowbreak{} 1.000\\
Sol /\allowbreak{} Valuation neutralized & EqualFactor-adjusted /\allowbreak{} 9 & 0.080 & [-0.375, 0.473] & 0.713 /\allowbreak{} 1.000 & 0.706 /\allowbreak{} 1.000\\
Sol /\allowbreak{} Momentum neutralized & EqualFactor-adjusted /\allowbreak{} 9 & 0.137 & [-0.204, 0.484] & 0.415 /\allowbreak{} 1.000 & 0.433 /\allowbreak{} 1.000\\
Sol /\allowbreak{} Risk/\allowbreak{}trading neutralized & EqualFactor-adjusted /\allowbreak{} 9 & 0.042 & [-0.318, 0.381] & 0.796 /\allowbreak{} 1.000 & 0.808 /\allowbreak{} 1.000\\
\end{longtable}
\noindent\footnotesize Twenty-four contrasts at 10 bp: nine factor-neutralization, three semantic, three row-order, and nine EqualFactor-adjusted factor contrasts. Positive values mean higher altered-condition returns. Adjustment subtracts the corresponding archived EqualFactor change. Inference uses three whole paired trajectories and 72 shared months; HAC uses three lags, and intervals are pointwise. Each procedure and block length uses its own specified 9/\allowbreak{}3/\allowbreak{}3/\allowbreak{}9 Holm families. Undefined HAC tests remain undefined and retain their place in the planned family size.\par\endgroup
\begingroup\small\setlength{\tabcolsep}{3pt}\renewcommand{\arraystretch}{1.17}
\begin{longtable}{@{}>{\raggedright\arraybackslash}p{\dimexpr 0.264706\linewidth-7.94118pt\relax}>{\raggedright\arraybackslash}p{\dimexpr 0.147059\linewidth-4.41176pt\relax}>{\raggedright\arraybackslash}p{\dimexpr 0.147059\linewidth-4.41176pt\relax}>{\raggedright\arraybackslash}p{\dimexpr 0.147059\linewidth-4.41176pt\relax}>{\raggedright\arraybackslash}p{\dimexpr 0.147059\linewidth-4.41176pt\relax}>{\raggedright\arraybackslash}p{\dimexpr 0.147059\linewidth-4.41176pt\relax}@{}}
\caption{Paired return contrasts with 12-month blocks}\label{tab:S35}\\
\toprule
\textbf{Model /\allowbreak{} intervention} & \textbf{Contrast family /\allowbreak{} size} & \textbf{Mean pp} & \textbf{95\% joint interval pp} & \textbf{HAC p /\allowbreak{} Holm p} & \textbf{Bootstrap p /\allowbreak{} Holm p}\\\midrule\endfirsthead
\multicolumn{6}{l}{\textbf{Table S35 (continued)}}\\\toprule
\textbf{Model /\allowbreak{} intervention} & \textbf{Contrast family /\allowbreak{} size} & \textbf{Mean pp} & \textbf{95\% joint interval pp} & \textbf{HAC p /\allowbreak{} Holm p} & \textbf{Bootstrap p /\allowbreak{} Holm p}\\\midrule\endhead
\midrule\multicolumn{6}{r}{Continued on next page}\\\endfoot
\bottomrule\endlastfoot
Luna /\allowbreak{} Valuation neutralized & Factor neutralization /\allowbreak{} 9 & 0.592 & [-0.588, 2.219] & 0.314 /\allowbreak{} 1.000 & 0.429 /\allowbreak{} 1.000\\
Luna /\allowbreak{} Momentum neutralized & Factor neutralization /\allowbreak{} 9 & 0.043 & [-0.318, 0.439] & 0.791 /\allowbreak{} 1.000 & 0.816 /\allowbreak{} 1.000\\
Luna /\allowbreak{} Risk/\allowbreak{}trading neutralized & Factor neutralization /\allowbreak{} 9 & 0.161 & [-0.247, 0.683] & 0.355 /\allowbreak{} 1.000 & 0.482 /\allowbreak{} 1.000\\
Terra /\allowbreak{} Valuation neutralized & Factor neutralization /\allowbreak{} 9 & 0.096 & [-0.721, 1.057] & 0.807 /\allowbreak{} 1.000 & 0.842 /\allowbreak{} 1.000\\
Terra /\allowbreak{} Momentum neutralized & Factor neutralization /\allowbreak{} 9 & -0.055 & [-0.409, 0.278] & 0.779 /\allowbreak{} 1.000 & 0.749 /\allowbreak{} 1.000\\
Terra /\allowbreak{} Risk/\allowbreak{}trading neutralized & Factor neutralization /\allowbreak{} 9 & -0.012 & [-0.216, 0.263] & 0.919 /\allowbreak{} 1.000 & 0.925 /\allowbreak{} 1.000\\
Sol /\allowbreak{} Valuation neutralized & Factor neutralization /\allowbreak{} 9 & 0.055 & [-0.661, 0.983] & 0.886 /\allowbreak{} 1.000 & 0.901 /\allowbreak{} 1.000\\
Sol /\allowbreak{} Momentum neutralized & Factor neutralization /\allowbreak{} 9 & -0.154 & [-0.485, 0.187] & 0.384 /\allowbreak{} 1.000 & 0.373 /\allowbreak{} 1.000\\
Sol /\allowbreak{} Risk/\allowbreak{}trading neutralized & Factor neutralization /\allowbreak{} 9 & 0.118 & [-0.132, 0.487] & 0.499 /\allowbreak{} 1.000 & 0.454 /\allowbreak{} 1.000\\
Luna /\allowbreak{} Semantics removed & Semantics /\allowbreak{} 3 & 0.135 & [-0.188, 0.473] & 0.407 /\allowbreak{} 1.000 & 0.388 /\allowbreak{} 1.000\\
Terra /\allowbreak{} Semantics removed & Semantics /\allowbreak{} 3 & -0.042 & [-0.225, 0.156] & 0.732 /\allowbreak{} 1.000 & 0.643 /\allowbreak{} 1.000\\
Sol /\allowbreak{} Semantics removed & Semantics /\allowbreak{} 3 & 0.030 & [-0.242, 0.290] & 0.843 /\allowbreak{} 1.000 & 0.813 /\allowbreak{} 1.000\\
Luna /\allowbreak{} Rows reordered & Row order /\allowbreak{} 3 & 0.542 & [0.142, 1.169] & 0.003 /\allowbreak{} 0.010 & 0.040 /\allowbreak{} 0.121\\
Terra /\allowbreak{} Rows reordered & Row order /\allowbreak{} 3 & 0.051 & [-0.150, 0.271] & 0.620 /\allowbreak{} 0.620 & 0.611 /\allowbreak{} 0.611\\
Sol /\allowbreak{} Rows reordered & Row order /\allowbreak{} 3 & 0.102 & [-0.025, 0.243] & 0.071 /\allowbreak{} 0.142 & 0.132 /\allowbreak{} 0.264\\
Luna /\allowbreak{} Valuation neutralized & EqualFactor-adjusted /\allowbreak{} 9 & 0.617 & [-0.188, 1.572] & 0.217 /\allowbreak{} 1.000 & 0.174 /\allowbreak{} 1.000\\
Luna /\allowbreak{} Momentum neutralized & EqualFactor-adjusted /\allowbreak{} 9 & 0.334 & [-0.070, 0.762] & 0.118 /\allowbreak{} 1.000 & 0.118 /\allowbreak{} 1.000\\
Luna /\allowbreak{} Risk/\allowbreak{}trading neutralized & EqualFactor-adjusted /\allowbreak{} 9 & 0.084 & [-0.341, 0.569] & 0.696 /\allowbreak{} 1.000 & 0.708 /\allowbreak{} 1.000\\
Terra /\allowbreak{} Valuation neutralized & EqualFactor-adjusted /\allowbreak{} 9 & 0.121 & [-0.430, 0.634] & 0.642 /\allowbreak{} 1.000 & 0.668 /\allowbreak{} 1.000\\
Terra /\allowbreak{} Momentum neutralized & EqualFactor-adjusted /\allowbreak{} 9 & 0.236 & [-0.116, 0.636] & 0.290 /\allowbreak{} 1.000 & 0.226 /\allowbreak{} 1.000\\
Terra /\allowbreak{} Risk/\allowbreak{}trading neutralized & EqualFactor-adjusted /\allowbreak{} 9 & -0.088 & [-0.468, 0.321] & 0.649 /\allowbreak{} 1.000 & 0.665 /\allowbreak{} 1.000\\
Sol /\allowbreak{} Valuation neutralized & EqualFactor-adjusted /\allowbreak{} 9 & 0.080 & [-0.306, 0.408] & 0.713 /\allowbreak{} 1.000 & 0.652 /\allowbreak{} 1.000\\
Sol /\allowbreak{} Momentum neutralized & EqualFactor-adjusted /\allowbreak{} 9 & 0.137 & [-0.154, 0.470] & 0.415 /\allowbreak{} 1.000 & 0.389 /\allowbreak{} 1.000\\
Sol /\allowbreak{} Risk/\allowbreak{}trading neutralized & EqualFactor-adjusted /\allowbreak{} 9 & 0.042 & [-0.292, 0.354] & 0.796 /\allowbreak{} 1.000 & 0.802 /\allowbreak{} 1.000\\
\end{longtable}
\noindent\footnotesize Twenty-four contrasts at 10 bp: nine factor-neutralization, three semantic, three row-order, and nine EqualFactor-adjusted factor contrasts. Positive values mean higher altered-condition returns. Adjustment subtracts the corresponding archived EqualFactor change. Inference uses three whole paired trajectories and 72 shared months; HAC uses three lags, and intervals are pointwise. Each procedure and block length uses its own specified 9/\allowbreak{}3/\allowbreak{}3/\allowbreak{}9 Holm families. Undefined HAC tests remain undefined and retain their place in the planned family size.\par\endgroup
\begingroup\small\setlength{\tabcolsep}{3pt}\renewcommand{\arraystretch}{1.17}
\begin{longtable}{@{}>{\raggedright\arraybackslash}p{\dimexpr 0.230769\linewidth-8.30769pt\relax}>{\raggedright\arraybackslash}p{\dimexpr 0.128205\linewidth-4.61538pt\relax}>{\raggedright\arraybackslash}p{\dimexpr 0.128205\linewidth-4.61538pt\relax}>{\raggedright\arraybackslash}p{\dimexpr 0.128205\linewidth-4.61538pt\relax}>{\raggedright\arraybackslash}p{\dimexpr 0.128205\linewidth-4.61538pt\relax}>{\raggedright\arraybackslash}p{\dimexpr 0.128205\linewidth-4.61538pt\relax}>{\raggedright\arraybackslash}p{\dimexpr 0.128205\linewidth-4.61538pt\relax}@{}}
\caption{Three-run performance distributions at 0 bp}\label{tab:S36}\\
\toprule
\textbf{Model /\allowbreak{} condition} & \textbf{CAGR} & \textbf{Annual volatility} & \textbf{Sharpe} & \textbf{Max drawdown} & \textbf{Monthly return} & \textbf{Total return}\\\midrule\endfirsthead
\multicolumn{7}{l}{\textbf{Table S36 (continued)}}\\\toprule
\textbf{Model /\allowbreak{} condition} & \textbf{CAGR} & \textbf{Annual volatility} & \textbf{Sharpe} & \textbf{Max drawdown} & \textbf{Monthly return} & \textbf{Total return}\\\midrule\endhead
\midrule\multicolumn{7}{r}{Continued on next page}\\\endfoot
\bottomrule\endlastfoot
Luna /\allowbreak{} Primary & 7.89\% [6.92\%, 9.75\%] & 15.61\% [15.41\%, 15.79\%] & 0.563 [0.506, 0.668] & -19.11\% [-23.13\%, -16.95\%] & 0.73\% [0.66\%, 0.88\%] & 58.04\% [49.43\%, 74.74\%]\\
Luna /\allowbreak{} Valuation neutralized & 15.20\% [13.03\%, 19.35\%] & 22.07\% [21.94\%, 22.23\%] & 0.748 [0.665, 0.913] & -31.76\% [-36.18\%, -29.42\%] & 1.38\% [1.22\%, 1.68\%] & 136.06\% [108.51\%, 188.99\%]\\
Luna /\allowbreak{} Momentum neutralized & 8.49\% [7.53\%, 9.25\%] & 15.55\% [15.49\%, 15.60\%] & 0.602 [0.544, 0.649] & -19.23\% [-21.74\%, -16.79\%] & 0.78\% [0.71\%, 0.84\%] & 63.15\% [54.60\%, 70.01\%]\\
Luna /\allowbreak{} Risk/\allowbreak{}trading neutralized & 9.71\% [6.48\%, 12.04\%] & 17.53\% [16.27\%, 19.45\%] & 0.610 [0.467, 0.683] & -19.83\% [-20.61\%, -18.72\%] & 0.90\% [0.63\%, 1.10\%] & 75.58\% [45.73\%, 97.75\%]\\
Luna /\allowbreak{} Semantics removed & 9.72\% [9.30\%, 10.10\%] & 15.80\% [15.35\%, 16.08\%] & 0.667 [0.637, 0.684] & -18.16\% [-19.22\%, -17.31\%] & 0.88\% [0.85\%, 0.91\%] & 74.46\% [70.51\%, 78.11\%]\\
Luna /\allowbreak{} Rows reordered & 15.14\% [13.00\%, 17.51\%] & 16.01\% [15.22\%, 17.13\%] & 0.961 [0.860, 1.030] & -15.81\% [-17.55\%, -14.77\%] & 1.28\% [1.12\%, 1.47\%] & 133.88\% [108.22\%, 163.31\%]\\
Terra /\allowbreak{} Primary & 9.73\% [9.24\%, 10.08\%] & 16.05\% [15.60\%, 16.32\%] & 0.660 [0.626, 0.695] & -16.51\% [-17.13\%, -15.65\%] & 0.88\% [0.85\%, 0.90\%] & 74.61\% [69.96\%, 77.97\%]\\
Terra /\allowbreak{} Valuation neutralized & 11.57\% [10.37\%, 12.85\%] & 16.51\% [16.16\%, 17.03\%] & 0.747 [0.687, 0.831] & -26.01\% [-28.98\%, -20.76\%] & 1.03\% [0.94\%, 1.12\%] & 93.07\% [80.81\%, 106.56\%]\\
Terra /\allowbreak{} Momentum neutralized & 9.15\% [8.36\%, 10.21\%] & 15.24\% [15.01\%, 15.47\%] & 0.651 [0.603, 0.706] & -20.43\% [-22.33\%, -19.30\%] & 0.83\% [0.77\%, 0.91\%] & 69.27\% [61.90\%, 79.19\%]\\
Terra /\allowbreak{} Risk/\allowbreak{}trading neutralized & 9.54\% [8.39\%, 11.13\%] & 16.51\% [15.90\%, 16.86\%] & 0.635 [0.563, 0.714] & -21.11\% [-23.53\%, -19.18\%] & 0.87\% [0.79\%, 1.00\%] & 73.09\% [62.19\%, 88.36\%]\\
Terra /\allowbreak{} Semantics removed & 9.52\% [8.62\%, 10.65\%] & 15.64\% [15.36\%, 16.01\%] & 0.659 [0.609, 0.713] & -16.72\% [-17.95\%, -15.09\%] & 0.86\% [0.79\%, 0.95\%] & 72.76\% [64.25\%, 83.54\%]\\
Terra /\allowbreak{} Rows reordered & 10.35\% [9.86\%, 11.29\%] & 16.29\% [16.14\%, 16.52\%] & 0.686 [0.663, 0.731] & -17.97\% [-20.88\%, -16.23\%] & 0.93\% [0.89\%, 1.01\%] & 80.62\% [75.85\%, 89.95\%]\\
Sol /\allowbreak{} Primary & 9.43\% [8.93\%, 10.30\%] & 15.57\% [15.30\%, 15.92\%] & 0.657 [0.630, 0.696] & -17.01\% [-18.52\%, -14.56\%] & 0.85\% [0.81\%, 0.92\%] & 71.82\% [67.02\%, 80.05\%]\\
Sol /\allowbreak{} Valuation neutralized & 10.57\% [10.03\%, 11.28\%] & 16.89\% [16.36\%, 17.18\%] & 0.680 [0.662, 0.711] & -28.16\% [-29.89\%, -25.41\%] & 0.96\% [0.91\%, 1.02\%] & 82.84\% [77.47\%, 89.94\%]\\
Sol /\allowbreak{} Momentum neutralized & 7.40\% [7.28\%, 7.61\%] & 15.08\% [15.06\%, 15.10\%] & 0.549 [0.542, 0.562] & -19.37\% [-19.72\%, -19.13\%] & 0.69\% [0.68\%, 0.71\%] & 53.50\% [52.42\%, 55.25\%]\\
Sol /\allowbreak{} Risk/\allowbreak{}trading neutralized & 10.90\% [9.19\%, 12.79\%] & 17.00\% [16.83\%, 17.11\%] & 0.695 [0.600, 0.793] & -20.62\% [-21.16\%, -19.88\%] & 0.98\% [0.86\%, 1.13\%] & 86.53\% [69.50\%, 105.93\%]\\
Sol /\allowbreak{} Semantics removed & 10.15\% [9.51\%, 10.96\%] & 15.17\% [14.83\%, 15.36\%] & 0.713 [0.688, 0.754] & -16.63\% [-17.82\%, -16.02\%] & 0.90\% [0.85\%, 0.97\%] & 78.66\% [72.47\%, 86.60\%]\\
Sol /\allowbreak{} Rows reordered & 10.76\% [10.29\%, 11.14\%] & 15.67\% [15.41\%, 16.00\%] & 0.732 [0.707, 0.764] & -15.70\% [-16.19\%, -15.03\%] & 0.96\% [0.92\%, 0.98\%] & 84.68\% [79.94\%, 88.45\%]\\
\end{longtable}
\noindent\footnotesize Entries show the arithmetic mean [minimum, maximum] of the three separately computed full-run metrics. CAGR and total return are compounded within each run before summarization; monthly return is the arithmetic mean of its 72 holding-period returns. Volatility is annualized, Sharpe uses a zero reference rate, and drawdown uses daily NAV. Undefined ratios remain explicit, with the number of defined runs shown where needed. Medians, sample standard deviations, and all 216 individual cost/\allowbreak{}run rows remain available in the saved performance tables.\par\endgroup
\begingroup\small\setlength{\tabcolsep}{3pt}\renewcommand{\arraystretch}{1.17}
\begin{longtable}{@{}>{\raggedright\arraybackslash}p{\dimexpr 0.230769\linewidth-8.30769pt\relax}>{\raggedright\arraybackslash}p{\dimexpr 0.128205\linewidth-4.61538pt\relax}>{\raggedright\arraybackslash}p{\dimexpr 0.128205\linewidth-4.61538pt\relax}>{\raggedright\arraybackslash}p{\dimexpr 0.128205\linewidth-4.61538pt\relax}>{\raggedright\arraybackslash}p{\dimexpr 0.128205\linewidth-4.61538pt\relax}>{\raggedright\arraybackslash}p{\dimexpr 0.128205\linewidth-4.61538pt\relax}>{\raggedright\arraybackslash}p{\dimexpr 0.128205\linewidth-4.61538pt\relax}@{}}
\caption{Three-run performance distributions at 10 bp}\label{tab:S37}\\
\toprule
\textbf{Model /\allowbreak{} condition} & \textbf{CAGR} & \textbf{Annual volatility} & \textbf{Sharpe} & \textbf{Max drawdown} & \textbf{Monthly return} & \textbf{Total return}\\\midrule\endfirsthead
\multicolumn{7}{l}{\textbf{Table S37 (continued)}}\\\toprule
\textbf{Model /\allowbreak{} condition} & \textbf{CAGR} & \textbf{Annual volatility} & \textbf{Sharpe} & \textbf{Max drawdown} & \textbf{Monthly return} & \textbf{Total return}\\\midrule\endhead
\midrule\multicolumn{7}{r}{Continued on next page}\\\endfoot
\bottomrule\endlastfoot
Luna /\allowbreak{} Primary & 7.29\% [6.31\%, 9.16\%] & 15.62\% [15.41\%, 15.80\%] & 0.528 [0.469, 0.634] & -19.58\% [-24.31\%, -17.07\%] & 0.69\% [0.61\%, 0.83\%] & 52.86\% [44.36\%, 69.15\%]\\
Luna /\allowbreak{} Valuation neutralized & 13.89\% [11.79\%, 18.01\%] & 22.04\% [21.90\%, 22.22\%] & 0.697 [0.614, 0.862] & -32.80\% [-37.27\%, -30.11\%] & 1.28\% [1.12\%, 1.58\%] & 120.45\% [95.15\%, 170.06\%]\\
Luna /\allowbreak{} Momentum neutralized & 7.86\% [6.98\%, 8.59\%] & 15.55\% [15.49\%, 15.59\%] & 0.564 [0.510, 0.609] & -19.91\% [-22.20\%, -17.98\%] & 0.73\% [0.66\%, 0.79\%] & 57.51\% [49.88\%, 63.94\%]\\
Luna /\allowbreak{} Risk/\allowbreak{}trading neutralized & 9.05\% [5.85\%, 11.38\%] & 17.53\% [16.27\%, 19.44\%] & 0.575 [0.430, 0.651] & -20.34\% [-20.89\%, -19.36\%] & 0.85\% [0.58\%, 1.05\%] & 69.30\% [40.63\%, 90.96\%]\\
Luna /\allowbreak{} Semantics removed & 9.00\% [8.56\%, 9.39\%] & 15.79\% [15.35\%, 16.07\%] & 0.625 [0.594, 0.642] & -19.12\% [-20.72\%, -18.13\%] & 0.82\% [0.79\%, 0.85\%] & 67.73\% [63.73\%, 71.31\%]\\
Luna /\allowbreak{} Rows reordered & 14.39\% [12.33\%, 16.66\%] & 15.99\% [15.20\%, 17.10\%] & 0.921 [0.823, 0.988] & -15.99\% [-17.67\%, -14.91\%] & 1.23\% [1.07\%, 1.41\%] & 124.80\% [100.94\%, 152.05\%]\\
Terra /\allowbreak{} Primary & 9.15\% [8.64\%, 9.53\%] & 16.05\% [15.60\%, 16.31\%] & 0.627 [0.593, 0.662] & -17.01\% [-17.66\%, -15.94\%] & 0.84\% [0.80\%, 0.86\%] & 69.14\% [64.44\%, 72.64\%]\\
Terra /\allowbreak{} Valuation neutralized & 10.34\% [9.20\%, 11.60\%] & 16.48\% [16.13\%, 17.01\%] & 0.680 [0.622, 0.762] & -26.66\% [-30.40\%, -20.93\%] & 0.93\% [0.85\%, 1.02\%] & 80.64\% [69.53\%, 93.15\%]\\
Terra /\allowbreak{} Momentum neutralized & 8.58\% [7.78\%, 9.64\%] & 15.24\% [15.02\%, 15.47\%] & 0.616 [0.568, 0.673] & -20.78\% [-22.69\%, -19.68\%] & 0.78\% [0.72\%, 0.87\%] & 63.98\% [56.75\%, 73.72\%]\\
Terra /\allowbreak{} Risk/\allowbreak{}trading neutralized & 8.92\% [7.77\%, 10.53\%] & 16.50\% [15.89\%, 16.86\%] & 0.601 [0.528, 0.682] & -21.71\% [-24.58\%, -19.76\%] & 0.83\% [0.74\%, 0.95\%] & 67.30\% [56.65\%, 82.31\%]\\
Terra /\allowbreak{} Semantics removed & 8.69\% [7.77\%, 9.81\%] & 15.62\% [15.34\%, 15.99\%] & 0.610 [0.559, 0.665] & -17.53\% [-18.69\%, -15.23\%] & 0.80\% [0.72\%, 0.89\%] & 65.00\% [56.67\%, 75.28\%]\\
Terra /\allowbreak{} Rows reordered & 9.78\% [9.31\%, 10.71\%] & 16.28\% [16.14\%, 16.52\%] & 0.655 [0.632, 0.699] & -18.36\% [-21.37\%, -16.77\%] & 0.89\% [0.85\%, 0.96\%] & 75.16\% [70.57\%, 84.15\%]\\
Sol /\allowbreak{} Primary & 8.79\% [8.28\%, 9.65\%] & 15.57\% [15.29\%, 15.92\%] & 0.619 [0.592, 0.659] & -17.45\% [-18.92\%, -14.75\%] & 0.80\% [0.76\%, 0.87\%] & 65.84\% [61.21\%, 73.82\%]\\
Sol /\allowbreak{} Valuation neutralized & 9.28\% [8.77\%, 9.94\%] & 16.85\% [16.32\%, 17.16\%] & 0.611 [0.596, 0.641] & -29.18\% [-31.26\%, -26.67\%] & 0.86\% [0.81\%, 0.91\%] & 70.39\% [65.56\%, 76.55\%]\\
Sol /\allowbreak{} Momentum neutralized & 6.89\% [6.77\%, 7.09\%] & 15.09\% [15.07\%, 15.11\%] & 0.517 [0.510, 0.529] & -19.67\% [-20.01\%, -19.47\%] & 0.65\% [0.64\%, 0.67\%] & 49.14\% [48.15\%, 50.80\%]\\
Sol /\allowbreak{} Risk/\allowbreak{}trading neutralized & 10.08\% [8.41\%, 11.91\%] & 16.97\% [16.80\%, 17.08\%] & 0.652 [0.559, 0.748] & -20.80\% [-21.32\%, -20.05\%] & 0.92\% [0.80\%, 1.06\%] & 78.40\% [62.34\%, 96.41\%]\\
Sol /\allowbreak{} Semantics removed & 9.26\% [8.57\%, 10.03\%] & 15.15\% [14.81\%, 15.33\%] & 0.660 [0.630, 0.700] & -17.46\% [-18.85\%, -16.74\%] & 0.83\% [0.78\%, 0.89\%] & 70.19\% [63.81\%, 77.47\%]\\
Sol /\allowbreak{} Rows reordered & 10.10\% [9.63\%, 10.48\%] & 15.66\% [15.40\%, 15.99\%] & 0.694 [0.668, 0.726] & -15.87\% [-16.35\%, -15.23\%] & 0.91\% [0.87\%, 0.93\%] & 78.19\% [73.57\%, 81.88\%]\\
\end{longtable}
\noindent\footnotesize Entries show the arithmetic mean [minimum, maximum] of the three separately computed full-run metrics. CAGR and total return are compounded within each run before summarization; monthly return is the arithmetic mean of its 72 holding-period returns. Volatility is annualized, Sharpe uses a zero reference rate, and drawdown uses daily NAV. Undefined ratios remain explicit, with the number of defined runs shown where needed. Medians, sample standard deviations, and all 216 individual cost/\allowbreak{}run rows remain available in the saved performance tables.\par\endgroup
\begingroup\small\setlength{\tabcolsep}{3pt}\renewcommand{\arraystretch}{1.17}
\begin{longtable}{@{}>{\raggedright\arraybackslash}p{\dimexpr 0.230769\linewidth-8.30769pt\relax}>{\raggedright\arraybackslash}p{\dimexpr 0.128205\linewidth-4.61538pt\relax}>{\raggedright\arraybackslash}p{\dimexpr 0.128205\linewidth-4.61538pt\relax}>{\raggedright\arraybackslash}p{\dimexpr 0.128205\linewidth-4.61538pt\relax}>{\raggedright\arraybackslash}p{\dimexpr 0.128205\linewidth-4.61538pt\relax}>{\raggedright\arraybackslash}p{\dimexpr 0.128205\linewidth-4.61538pt\relax}>{\raggedright\arraybackslash}p{\dimexpr 0.128205\linewidth-4.61538pt\relax}@{}}
\caption{Three-run performance distributions at 25 bp}\label{tab:S38}\\
\toprule
\textbf{Model /\allowbreak{} condition} & \textbf{CAGR} & \textbf{Annual volatility} & \textbf{Sharpe} & \textbf{Max drawdown} & \textbf{Monthly return} & \textbf{Total return}\\\midrule\endfirsthead
\multicolumn{7}{l}{\textbf{Table S38 (continued)}}\\\toprule
\textbf{Model /\allowbreak{} condition} & \textbf{CAGR} & \textbf{Annual volatility} & \textbf{Sharpe} & \textbf{Max drawdown} & \textbf{Monthly return} & \textbf{Total return}\\\midrule\endhead
\midrule\multicolumn{7}{r}{Continued on next page}\\\endfoot
\bottomrule\endlastfoot
Luna /\allowbreak{} Primary & 6.40\% [5.40\%, 8.27\%] & 15.62\% [15.41\%, 15.81\%] & 0.474 [0.414, 0.582] & -20.43\% [-26.05\%, -17.25\%] & 0.62\% [0.54\%, 0.77\%] & 45.44\% [37.08\%, 61.11\%]\\
Luna /\allowbreak{} Valuation neutralized & 11.96\% [9.90\%, 16.03\%] & 22.00\% [21.86\%, 22.21\%] & 0.619 [0.533, 0.786] & -34.32\% [-38.86\%, -31.12\%] & 1.14\% [0.98\%, 1.44\%] & 98.98\% [76.24\%, 144.00\%]\\
Luna /\allowbreak{} Momentum neutralized & 6.92\% [6.15\%, 7.61\%] & 15.55\% [15.51\%, 15.58\%] & 0.507 [0.460, 0.550] & -20.90\% [-22.89\%, -19.72\%] & 0.66\% [0.60\%, 0.71\%] & 49.44\% [43.09\%, 55.25\%]\\
Luna /\allowbreak{} Risk/\allowbreak{}trading neutralized & 8.06\% [4.91\%, 10.42\%] & 17.53\% [16.28\%, 19.42\%] & 0.523 [0.375, 0.606] & -21.11\% [-22.00\%, -20.32\%] & 0.77\% [0.51\%, 0.98\%] & 60.32\% [33.32\%, 81.22\%]\\
Luna /\allowbreak{} Semantics removed & 7.94\% [7.47\%, 8.33\%] & 15.79\% [15.35\%, 16.05\%] & 0.563 [0.531, 0.579] & -20.86\% [-22.90\%, -19.32\%] & 0.74\% [0.71\%, 0.77\%] & 58.15\% [54.09\%, 61.62\%]\\
Luna /\allowbreak{} Rows reordered & 13.27\% [11.34\%, 15.39\%] & 15.96\% [15.18\%, 17.05\%] & 0.860 [0.767, 0.926] & -16.26\% [-17.84\%, -15.13\%] & 1.15\% [1.00\%, 1.32\%] & 111.87\% [90.52\%, 136.08\%]\\
Terra /\allowbreak{} Primary & 8.29\% [7.75\%, 8.70\%] & 16.04\% [15.60\%, 16.30\%] & 0.577 [0.542, 0.613] & -17.93\% [-18.45\%, -16.92\%] & 0.77\% [0.73\%, 0.80\%] & 61.28\% [56.52\%, 64.96\%]\\
Terra /\allowbreak{} Valuation neutralized & 8.52\% [7.45\%, 9.74\%] & 16.44\% [16.09\%, 16.98\%] & 0.580 [0.524, 0.659] & -27.75\% [-32.48\%, -21.63\%] & 0.79\% [0.71\%, 0.88\%] & 63.51\% [53.94\%, 74.67\%]\\
Terra /\allowbreak{} Momentum neutralized & 7.72\% [6.91\%, 8.80\%] & 15.25\% [15.03\%, 15.46\%] & 0.563 [0.514, 0.623] & -21.31\% [-23.25\%, -20.25\%] & 0.72\% [0.65\%, 0.80\%] & 56.38\% [49.35\%, 65.83\%]\\
Terra /\allowbreak{} Risk/\allowbreak{}trading neutralized & 8.00\% [6.84\%, 9.63\%] & 16.48\% [15.87\%, 16.86\%] & 0.549 [0.477, 0.634] & -22.60\% [-26.13\%, -20.62\%] & 0.75\% [0.67\%, 0.88\%] & 58.98\% [48.72\%, 73.61\%]\\
Terra /\allowbreak{} Semantics removed & 7.45\% [6.51\%, 8.55\%] & 15.59\% [15.31\%, 15.97\%] & 0.537 [0.483, 0.593] & -18.85\% [-21.17\%, -15.45\%] & 0.70\% [0.62\%, 0.79\%] & 54.02\% [45.97\%, 63.61\%]\\
Terra /\allowbreak{} Rows reordered & 8.94\% [8.48\%, 9.86\%] & 16.28\% [16.14\%, 16.51\%] & 0.607 [0.585, 0.652] & -18.96\% [-22.09\%, -17.20\%] & 0.82\% [0.79\%, 0.90\%] & 67.29\% [62.95\%, 75.79\%]\\
Sol /\allowbreak{} Primary & 7.83\% [7.33\%, 8.69\%] & 15.57\% [15.27\%, 15.91\%] & 0.562 [0.534, 0.604] & -18.48\% [-20.35\%, -15.66\%] & 0.73\% [0.69\%, 0.80\%] & 57.28\% [52.87\%, 64.89\%]\\
Sol /\allowbreak{} Valuation neutralized & 7.38\% [6.89\%, 7.95\%] & 16.81\% [16.27\%, 17.12\%] & 0.507 [0.491, 0.535] & -31.19\% [-33.27\%, -28.82\%] & 0.71\% [0.67\%, 0.76\%] & 53.30\% [49.19\%, 58.24\%]\\
Sol /\allowbreak{} Momentum neutralized & 6.12\% [6.02\%, 6.31\%] & 15.10\% [15.08\%, 15.13\%] & 0.469 [0.462, 0.481] & -20.39\% [-20.79\%, -19.92\%] & 0.59\% [0.58\%, 0.60\%] & 42.86\% [41.97\%, 44.38\%]\\
Sol /\allowbreak{} Risk/\allowbreak{}trading neutralized & 8.87\% [7.25\%, 10.59\%] & 16.93\% [16.75\%, 17.04\%] & 0.587 [0.496, 0.679] & -21.07\% [-21.56\%, -20.30\%] & 0.83\% [0.70\%, 0.96\%] & 66.88\% [52.17\%, 82.96\%]\\
Sol /\allowbreak{} Semantics removed & 7.94\% [7.18\%, 8.66\%] & 15.11\% [14.77\%, 15.29\%] & 0.581 [0.544, 0.619] & -19.09\% [-20.60\%, -18.04\%] & 0.73\% [0.67\%, 0.79\%] & 58.25\% [51.64\%, 64.63\%]\\
Sol /\allowbreak{} Rows reordered & 9.12\% [8.64\%, 9.51\%] & 15.65\% [15.39\%, 15.98\%] & 0.637 [0.610, 0.669] & -16.12\% [-16.58\%, -15.53\%] & 0.83\% [0.79\%, 0.86\%] & 68.88\% [64.45\%, 72.48\%]\\
\end{longtable}
\noindent\footnotesize Entries show the arithmetic mean [minimum, maximum] of the three separately computed full-run metrics. CAGR and total return are compounded within each run before summarization; monthly return is the arithmetic mean of its 72 holding-period returns. Volatility is annualized, Sharpe uses a zero reference rate, and drawdown uses daily NAV. Undefined ratios remain explicit, with the number of defined runs shown where needed. Medians, sample standard deviations, and all 216 individual cost/\allowbreak{}run rows remain available in the saved performance tables.\par\endgroup
\begingroup\small\setlength{\tabcolsep}{3pt}\renewcommand{\arraystretch}{1.17}
\begin{longtable}{@{}>{\raggedright\arraybackslash}p{\dimexpr 0.230769\linewidth-8.30769pt\relax}>{\raggedright\arraybackslash}p{\dimexpr 0.128205\linewidth-4.61538pt\relax}>{\raggedright\arraybackslash}p{\dimexpr 0.128205\linewidth-4.61538pt\relax}>{\raggedright\arraybackslash}p{\dimexpr 0.128205\linewidth-4.61538pt\relax}>{\raggedright\arraybackslash}p{\dimexpr 0.128205\linewidth-4.61538pt\relax}>{\raggedright\arraybackslash}p{\dimexpr 0.128205\linewidth-4.61538pt\relax}>{\raggedright\arraybackslash}p{\dimexpr 0.128205\linewidth-4.61538pt\relax}@{}}
\caption{Three-run performance distributions at 50 bp}\label{tab:S39}\\
\toprule
\textbf{Model /\allowbreak{} condition} & \textbf{CAGR} & \textbf{Annual volatility} & \textbf{Sharpe} & \textbf{Max drawdown} & \textbf{Monthly return} & \textbf{Total return}\\\midrule\endfirsthead
\multicolumn{7}{l}{\textbf{Table S39 (continued)}}\\\toprule
\textbf{Model /\allowbreak{} condition} & \textbf{CAGR} & \textbf{Annual volatility} & \textbf{Sharpe} & \textbf{Max drawdown} & \textbf{Monthly return} & \textbf{Total return}\\\midrule\endhead
\midrule\multicolumn{7}{r}{Continued on next page}\\\endfoot
\bottomrule\endlastfoot
Luna /\allowbreak{} Primary & 4.94\% [3.90\%, 6.82\%] & 15.64\% [15.42\%, 15.83\%] & 0.385 [0.322, 0.496] & -22.52\% [-28.85\%, -17.54\%] & 0.50\% [0.42\%, 0.65\%] & 33.89\% [25.80\%, 48.60\%]\\
Luna /\allowbreak{} Valuation neutralized & 8.82\% [6.69\%, 12.81\%] & 21.94\% [21.78\%, 22.20\%] & 0.490 [0.399, 0.658] & -36.76\% [-41.41\%, -32.77\%] & 0.90\% [0.74\%, 1.20\%] & 67.80\% [47.49\%, 106.08\%]\\
Luna /\allowbreak{} Momentum neutralized & 5.37\% [4.80\%, 6.00\%] & 15.56\% [15.53\%, 15.58\%] & 0.413 [0.377, 0.452] & -22.53\% [-24.01\%, -21.03\%] & 0.54\% [0.49\%, 0.58\%] & 36.94\% [32.47\%, 41.84\%]\\
Luna /\allowbreak{} Risk/\allowbreak{}trading neutralized & 6.44\% [3.37\%, 8.83\%] & 17.53\% [16.31\%, 19.40\%] & 0.436 [0.284, 0.531] & -22.40\% [-23.79\%, -21.40\%] & 0.65\% [0.39\%, 0.86\%] & 46.43\% [22.01\%, 66.11\%]\\
Luna /\allowbreak{} Semantics removed & 6.19\% [5.68\%, 6.60\%] & 15.78\% [15.36\%, 16.03\%] & 0.459 [0.425, 0.478] & -24.25\% [-26.41\%, -22.43\%] & 0.60\% [0.57\%, 0.64\%] & 43.43\% [39.31\%, 46.71\%]\\
Luna /\allowbreak{} Rows reordered & 11.44\% [9.71\%, 13.32\%] & 15.91\% [15.16\%, 16.98\%] & 0.759 [0.674, 0.821] & -16.71\% [-18.13\%, -15.49\%] & 1.01\% [0.88\%, 1.16\%] & 92.02\% [74.39\%, 111.75\%]\\
Terra /\allowbreak{} Primary & 6.87\% [6.29\%, 7.34\%] & 16.02\% [15.61\%, 16.27\%] & 0.495 [0.458, 0.532] & -19.48\% [-20.06\%, -18.65\%] & 0.66\% [0.62\%, 0.69\%] & 49.01\% [44.20\%, 52.95\%]\\
Terra /\allowbreak{} Valuation neutralized & 5.56\% [4.62\%, 6.72\%] & 16.38\% [16.03\%, 16.93\%] & 0.412 [0.360, 0.486] & -30.11\% [-35.80\%, -22.77\%] & 0.56\% [0.49\%, 0.65\%] & 38.52\% [31.11\%, 47.76\%]\\
Terra /\allowbreak{} Momentum neutralized & 6.31\% [5.49\%, 7.40\%] & 15.25\% [15.05\%, 15.46\%] & 0.477 [0.426, 0.539] & -23.08\% [-25.38\%, -21.17\%] & 0.61\% [0.54\%, 0.69\%] & 44.51\% [37.80\%, 53.51\%]\\
Terra /\allowbreak{} Risk/\allowbreak{}trading neutralized & 6.48\% [5.31\%, 8.16\%] & 16.47\% [15.85\%, 16.86\%] & 0.463 [0.391, 0.553] & -24.24\% [-28.72\%, -21.45\%] & 0.64\% [0.55\%, 0.77\%] & 46.07\% [36.42\%, 60.07\%]\\
Terra /\allowbreak{} Semantics removed & 5.42\% [4.44\%, 6.50\%] & 15.55\% [15.27\%, 15.93\%] & 0.415 [0.357, 0.474] & -21.57\% [-25.13\%, -16.26\%] & 0.54\% [0.46\%, 0.63\%] & 37.37\% [29.77\%, 45.91\%]\\
Terra /\allowbreak{} Rows reordered & 7.57\% [7.11\%, 8.46\%] & 16.27\% [16.14\%, 16.51\%] & 0.529 [0.506, 0.574] & -19.93\% [-23.28\%, -17.60\%] & 0.72\% [0.68\%, 0.79\%] & 54.98\% [51.04\%, 62.74\%]\\
Sol /\allowbreak{} Primary & 6.26\% [5.76\%, 7.12\%] & 15.56\% [15.25\%, 15.90\%] & 0.467 [0.439, 0.512] & -20.66\% [-22.68\%, -17.35\%] & 0.61\% [0.57\%, 0.68\%] & 44.01\% [39.97\%, 51.06\%]\\
Sol /\allowbreak{} Valuation neutralized & 4.28\% [3.85\%, 4.72\%] & 16.73\% [16.20\%, 17.06\%] & 0.333 [0.314, 0.357] & -34.51\% [-36.48\%, -32.26\%] & 0.46\% [0.42\%, 0.50\%] & 28.58\% [25.45\%, 31.89\%]\\
Sol /\allowbreak{} Momentum neutralized & 4.87\% [4.77\%, 5.04\%] & 15.13\% [15.11\%, 15.16\%] & 0.389 [0.383, 0.400] & -22.05\% [-22.95\%, -21.18\%] & 0.49\% [0.48\%, 0.50\%] & 32.99\% [32.28\%, 34.31\%]\\
Sol /\allowbreak{} Risk/\allowbreak{}trading neutralized & 6.88\% [5.34\%, 8.44\%] & 16.86\% [16.68\%, 16.97\%] & 0.479 [0.391, 0.565] & -21.90\% [-23.10\%, -20.72\%] & 0.67\% [0.55\%, 0.80\%] & 49.37\% [36.66\%, 62.62\%]\\
Sol /\allowbreak{} Semantics removed & 5.79\% [4.92\%, 6.42\%] & 15.07\% [14.72\%, 15.25\%] & 0.448 [0.399, 0.484] & -22.48\% [-23.42\%, -21.11\%] & 0.56\% [0.49\%, 0.61\%] & 40.24\% [33.37\%, 45.29\%]\\
Sol /\allowbreak{} Rows reordered & 7.52\% [7.03\%, 7.91\%] & 15.63\% [15.37\%, 15.96\%] & 0.542 [0.514, 0.573] & -17.15\% [-17.95\%, -16.53\%] & 0.71\% [0.67\%, 0.73\%] & 54.49\% [50.34\%, 57.91\%]\\
\end{longtable}
\noindent\footnotesize Entries show the arithmetic mean [minimum, maximum] of the three separately computed full-run metrics. CAGR and total return are compounded within each run before summarization; monthly return is the arithmetic mean of its 72 holding-period returns. Volatility is annualized, Sharpe uses a zero reference rate, and drawdown uses daily NAV. Undefined ratios remain explicit, with the number of defined runs shown where needed. Medians, sample standard deviations, and all 216 individual cost/\allowbreak{}run rows remain available in the saved performance tables.\par\endgroup
\begingroup\small\setlength{\tabcolsep}{3pt}\renewcommand{\arraystretch}{1.17}
\begin{longtable}{@{}>{\raggedright\arraybackslash}p{\dimexpr 0.375000\linewidth-6.75000pt\relax}>{\raggedright\arraybackslash}p{\dimexpr 0.208333\linewidth-3.75000pt\relax}>{\raggedright\arraybackslash}p{\dimexpr 0.208333\linewidth-3.75000pt\relax}>{\raggedright\arraybackslash}p{\dimexpr 0.208333\linewidth-3.75000pt\relax}@{}}
\caption{Score-rank agreement across repeats and interventions}\label{tab:S40}\\
\toprule
\textbf{Model /\allowbreak{} condition} & \textbf{Primary repeats} & \textbf{Condition repeats} & \textbf{Same-run cross-condition}\\\midrule\endfirsthead
\multicolumn{4}{l}{\textbf{Table S40 (continued)}}\\\toprule
\textbf{Model /\allowbreak{} condition} & \textbf{Primary repeats} & \textbf{Condition repeats} & \textbf{Same-run cross-condition}\\\midrule\endhead
\midrule\multicolumn{4}{r}{Continued on next page}\\\endfoot
\bottomrule\endlastfoot
Luna /\allowbreak{} Primary & 0.911 (72/\allowbreak{}72) & 0.911 (72/\allowbreak{}72) & not applicable\\
Luna /\allowbreak{} Valuation neutralized & 0.911 (72/\allowbreak{}72) & 0.821 (72/\allowbreak{}72) & 0.505 (72/\allowbreak{}72)\\
Luna /\allowbreak{} Momentum neutralized & 0.911 (72/\allowbreak{}72) & 0.941 (72/\allowbreak{}72) & 0.801 (72/\allowbreak{}72)\\
Luna /\allowbreak{} Risk/\allowbreak{}trading neutralized & 0.911 (72/\allowbreak{}72) & 0.895 (72/\allowbreak{}72) & 0.855 (72/\allowbreak{}72)\\
Luna /\allowbreak{} Semantics removed & 0.911 (72/\allowbreak{}72) & 0.952 (72/\allowbreak{}72) & 0.841 (72/\allowbreak{}72)\\
Luna /\allowbreak{} Rows reordered & 0.911 (72/\allowbreak{}72) & 0.882 (72/\allowbreak{}72) & 0.864 (72/\allowbreak{}72)\\
Terra /\allowbreak{} Primary & 0.971 (72/\allowbreak{}72) & 0.971 (72/\allowbreak{}72) & not applicable\\
Terra /\allowbreak{} Valuation neutralized & 0.971 (72/\allowbreak{}72) & 0.916 (72/\allowbreak{}72) & 0.538 (72/\allowbreak{}72)\\
Terra /\allowbreak{} Momentum neutralized & 0.971 (72/\allowbreak{}72) & 0.987 (72/\allowbreak{}72) & 0.864 (72/\allowbreak{}72)\\
Terra /\allowbreak{} Risk/\allowbreak{}trading neutralized & 0.971 (72/\allowbreak{}72) & 0.968 (72/\allowbreak{}72) & 0.923 (72/\allowbreak{}72)\\
Terra /\allowbreak{} Semantics removed & 0.971 (72/\allowbreak{}72) & 0.961 (72/\allowbreak{}72) & 0.905 (72/\allowbreak{}72)\\
Terra /\allowbreak{} Rows reordered & 0.971 (72/\allowbreak{}72) & 0.972 (72/\allowbreak{}72) & 0.951 (72/\allowbreak{}72)\\
Sol /\allowbreak{} Primary & 0.960 (72/\allowbreak{}72) & 0.960 (72/\allowbreak{}72) & not applicable\\
Sol /\allowbreak{} Valuation neutralized & 0.960 (72/\allowbreak{}72) & 0.975 (72/\allowbreak{}72) & 0.661 (72/\allowbreak{}72)\\
Sol /\allowbreak{} Momentum neutralized & 0.960 (72/\allowbreak{}72) & 0.989 (72/\allowbreak{}72) & 0.819 (72/\allowbreak{}72)\\
Sol /\allowbreak{} Risk/\allowbreak{}trading neutralized & 0.960 (72/\allowbreak{}72) & 0.978 (72/\allowbreak{}72) & 0.880 (72/\allowbreak{}72)\\
Sol /\allowbreak{} Semantics removed & 0.960 (72/\allowbreak{}72) & 0.939 (72/\allowbreak{}72) & 0.882 (72/\allowbreak{}72)\\
Sol /\allowbreak{} Rows reordered & 0.960 (72/\allowbreak{}72) & 0.963 (72/\allowbreak{}72) & 0.958 (72/\allowbreak{}72)\\
\end{longtable}
\noindent\footnotesize Spearman correlations first average defined pairs within month, then defined monthly means. Parentheses report defined monthly means out of 72. Constant-score correlations are undefined; the primary condition has no intervention comparison. These are descriptive comparisons. The monthly archive retains pair-level values and the defined denominators.\par\endgroup
\begingroup\small\setlength{\tabcolsep}{3pt}\renewcommand{\arraystretch}{1.17}
\begin{longtable}{@{}>{\raggedright\arraybackslash}p{\dimexpr 0.285714\linewidth-1.71429pt\relax}>{\raggedright\arraybackslash}p{\dimexpr 0.714286\linewidth-4.28571pt\relax}@{}}
\caption{Input-intervention data and reproduction paths}\label{tab:S41}\\
\toprule
\textbf{Artifact} & \textbf{Location}\\\midrule\endfirsthead
\multicolumn{2}{l}{\textbf{Table S41 (continued)}}\\\toprule
\textbf{Artifact} & \textbf{Location}\\\midrule\endhead
\midrule\multicolumn{2}{r}{Continued on next page}\\\endfoot
\bottomrule\endlastfoot
Registered design and task order & study\_\allowbreak{}v3/\allowbreak{}replication\_\allowbreak{}20260927/\allowbreak{}lock.json; study\_\allowbreak{}v3/\allowbreak{}replication\_\allowbreak{}20260927/\allowbreak{}schedule.json\\
Anonymous inputs and formal predictions & study\_\allowbreak{}v3/\allowbreak{}replication\_\allowbreak{}20260927/\allowbreak{}inputs/\allowbreak{}<condition>/\allowbreak{}<date>.json; study\_\allowbreak{}v3/\allowbreak{}replication\_\allowbreak{}20260927/\allowbreak{}predictions/\allowbreak{}<condition>/\allowbreak{}<model>/\allowbreak{}run\_\allowbreak{}<id>/\allowbreak{}<date>.json\\
Common market inputs and engine & Authorized local SSE50 panel and prices, company-event records and board prices; data/\allowbreak{}trading\_\allowbreak{}days.json; data/\allowbreak{}raw/\allowbreak{}stk\_\allowbreak{}limit/\allowbreak{}; scripts/\allowbreak{}v3\_\allowbreak{}backtest.py\\
EqualFactor reference portfolios & analysis/\allowbreak{}results/\allowbreak{}SSE50/\allowbreak{}<condition>/\allowbreak{}monthly\_\allowbreak{}returns.parquet; analysis/\allowbreak{}replication\_\allowbreak{}20260927/\allowbreak{}results/\allowbreak{}equalfactor\_\allowbreak{}reference\_\allowbreak{}returns.parquet\\
Paired inference and run differences & analysis/\allowbreak{}replication\_\allowbreak{}20260927/\allowbreak{}results/\allowbreak{}mechanism\_\allowbreak{}paired\_\allowbreak{}inference.csv; analysis/\allowbreak{}replication\_\allowbreak{}20260927/\allowbreak{}results/\allowbreak{}paired\_\allowbreak{}return\_\allowbreak{}differences.parquet\\
Decision agreement and parallel summaries & analysis/\allowbreak{}replication\_\allowbreak{}20260927/\allowbreak{}results/\allowbreak{}monthly\_\allowbreak{}agreement.parquet; analysis/\allowbreak{}replication\_\allowbreak{}20260927/\allowbreak{}results/\allowbreak{}monthly\_\allowbreak{}agreement\_\allowbreak{}comparison.parquet; analysis/\allowbreak{}replication\_\allowbreak{}20260927/\allowbreak{}results/\allowbreak{}agreement\_\allowbreak{}comparison\_\allowbreak{}summary.csv\\
All-run performance and distributions & analysis/\allowbreak{}replication\_\allowbreak{}20260927/\allowbreak{}results/\allowbreak{}performance\_\allowbreak{}metrics.csv; analysis/\allowbreak{}replication\_\allowbreak{}20260927/\allowbreak{}results/\allowbreak{}run\_\allowbreak{}performance\_\allowbreak{}distribution.csv\\
Monthly returns, NAV and aggregate executions; stock-level trades retained locally & analysis/\allowbreak{}replication\_\allowbreak{}20260927/\allowbreak{}results/\allowbreak{}<condition>/\allowbreak{}<model>/\allowbreak{}run\_\allowbreak{}<id>/\allowbreak{}\\
Completion and source manifests & analysis/\allowbreak{}replication\_\allowbreak{}20260927/\allowbreak{}results/\allowbreak{}analysis\_\allowbreak{}complete.json; analysis/\allowbreak{}replication\_\allowbreak{}20260927/\allowbreak{}results/\allowbreak{}source\_\allowbreak{}manifest.json; analysis/\allowbreak{}replication\_\allowbreak{}20260927/\allowbreak{}results/\allowbreak{}source\_\allowbreak{}path\_\allowbreak{}index.json\\
\end{longtable}
\noindent\footnotesize Project-relative paths identify the input-intervention batch and common market sources. Model-condition contrasts use trajectories from this batch; the EqualFactor adjustment uses the deterministic reference returns for the corresponding altered matrices. Licensed market data are subject to their redistribution permissions. The public release includes monthly returns, NAV and aggregate execution records. Stock-level trades and terminal positions are in the local study archive. The listed market inputs require licensed local access.\par\endgroup
Response-level reproduction uses the stored inputs and predictions. The publication-context command below validates packaged content and source identities and renders the journal and arXiv documents. Portfolio reproduction with v3\_\allowbreak{}replication\_\allowbreak{}analysis.py also uses the authorized SSE 50 market panel, daily observations, company-event terms and board quotations. The batch lock, schedule, inputs and predictions retain their archive locations. A local analysis manifest names an empty result directory. The program validates the complete response grid and writes results there; its default endpoint matches the collection lock.

\begin{lstlisting}[basicstyle=\small\ttfamily,breaklines=true,columns=fullflexible,keepspaces=true]
python -B scripts/v3_replication_analysis.py --release-manifest <local-analysis-manifest>
\end{lstlisting}
\begin{lstlisting}[basicstyle=\small\ttfamily,breaklines=true,columns=fullflexible,keepspaces=true]
python scripts/render_current_manuscript.py --publication-context analysis/publication_context.json --compiler <Tectonic-executable> --step all
\end{lstlisting}
\section*{S16 Reproduction and source locations}
\begingroup\small\setlength{\tabcolsep}{3pt}\renewcommand{\arraystretch}{1.17}
\begin{longtable}{@{}>{\raggedright\arraybackslash}p{\dimexpr 0.285714\linewidth-1.71429pt\relax}>{\raggedright\arraybackslash}p{\dimexpr 0.714286\linewidth-4.28571pt\relax}@{}}
\caption{Reproduction sources}\label{tab:S42}\\
\toprule
\textbf{Artifact group} & \textbf{Location relative to the project root}\\\midrule\endfirsthead
\multicolumn{2}{l}{\textbf{Table S42 (continued)}}\\\toprule
\textbf{Artifact group} & \textbf{Location relative to the project root}\\\midrule\endhead
\midrule\multicolumn{2}{r}{Continued on next page}\\\endfoot
\bottomrule\endlastfoot
Specification & study\_\allowbreak{}v3/\allowbreak{}historical\_\allowbreak{}protocol.json; study\_\allowbreak{}v3/\allowbreak{}amendments/\allowbreak{}\\
Additional development-planning specification & study\_\allowbreak{}v3/\allowbreak{}amendments/\allowbreak{}008\_\allowbreak{}power\_\allowbreak{}development\_\allowbreak{}scoring.json\\
Development-planning request configuration & study\_\allowbreak{}v3/\allowbreak{}amendments/\allowbreak{} (record 009)\\
Risk security-identity crosswalk & study\_\allowbreak{}v3/\allowbreak{}amendments/\allowbreak{}010\_\allowbreak{}security\_\allowbreak{}code\_\allowbreak{}identity.json\\
Issuer code-change announcement & study\_\allowbreak{}v3/\allowbreak{}sources/\allowbreak{}sw/\allowbreak{}security\_\allowbreak{}code\_\allowbreak{}change\_\allowbreak{}300114\_\allowbreak{}302132.json\\
SW industry source and original normalization & study\_\allowbreak{}v3/\allowbreak{}sources/\allowbreak{}sw/\allowbreak{}sw2021\_\allowbreak{}history.xls; study\_\allowbreak{}v3/\allowbreak{}data/\allowbreak{}official\_\allowbreak{}industry\_\allowbreak{}intervals.csv\\
Identity-linked industry intervals and audit & study\_\allowbreak{}v3/\allowbreak{}data/\allowbreak{}official\_\allowbreak{}industry\_\allowbreak{}intervals\_\allowbreak{}linked.csv; study\_\allowbreak{}v3/\allowbreak{}data/\allowbreak{}official\_\allowbreak{}industry\_\allowbreak{}history\_\allowbreak{}linked\_\allowbreak{}audit.json\\
Market panels & Authorized local market panels and prices, corporate\_\allowbreak{}actions.json and otc\_\allowbreak{}prices.parquet; exact versions identified by the local analysis manifest\\
Inputs and predictions & study\_\allowbreak{}v3/\allowbreak{}scored/\allowbreak{}inputs/\allowbreak{} and study\_\allowbreak{}v3/\allowbreak{}scored/\allowbreak{}predictions/\allowbreak{}\\
Conventional fits and portfolios & analysis/\allowbreak{}results/\allowbreak{}<universe>/\allowbreak{}<condition>/\allowbreak{}\\
Mechanism diagnostics & analysis/\allowbreak{}results/\allowbreak{}SSE50/\allowbreak{}stability/\allowbreak{} and surrogates/\allowbreak{}; analysis/\allowbreak{}results/\allowbreak{}<universe>/\allowbreak{}risk\_\allowbreak{}matched/\allowbreak{} and disjoint/\allowbreak{}\\
Final verification & analysis/\allowbreak{}publication\_\allowbreak{}context.json and its named file-integrity manifest\\
Development-based sample-size calculation & analysis/\allowbreak{}results/\allowbreak{}power/\allowbreak{}; study\_\allowbreak{}v3/\allowbreak{}prospective/\allowbreak{}sample\_\allowbreak{}size\_\allowbreak{}decision.json\\
\end{longtable}
\noindent\footnotesize The v3 panels reuse the original trading calendar, price-limit and daily-indicator snapshots under data/\allowbreak{}. Raw versioned price/\allowbreak{}adjustment partitions and query manifests preserve source provenance. Licensed vendor data require appropriate redistribution permission. Locations describe the complete local study directory. The issuer announcement, original SW workbook and three derived industry-history files are retained locally subject to their source permissions; they are not included in the public research archive. The archive manifest lists the included research files.\par\endgroup
The accompanying research\_\allowbreak{}materials.tar.xz preserves project-relative paths for analysis code, experimental specifications, 1,260 input files, 10,104 unique valid responses, fitted rules, development-selection records, rankings and response changes, derived portfolio accounting and paired monthly statistics. Its manifest records paths, byte sizes and SHA-256 hashes; source\_\allowbreak{}path\_\allowbreak{}index.json maps the historical absolute source-manifest labels to relative paths. The local study archive also retains licensed stock panels, constituent snapshots, original observations and price-bearing trade records. Recomputing market-data-dependent quantities requires those separately authorized materials.

Rule-study results are in proxies/\allowbreak{}, legacy\_\allowbreak{}interventions/\allowbreak{}, controlled/\allowbreak{}results/\allowbreak{} and economics/\allowbreak{}final\_\allowbreak{}20260918/\allowbreak{} within analysis/\allowbreak{}decision\_\allowbreak{}rules\_\allowbreak{}20260918/\allowbreak{}. Inputs, responses, coefficient locks and research specifications are in their study\_\allowbreak{}v3/\allowbreak{} archive locations. Response-level reproduction uses these stored records. Portfolio reproduction also uses the authorized market sources.

Reproduction uses the dependency versions in requirements.txt. Response-level summaries use the stored inputs and predictions. Portfolio reproduction uses a separate research directory containing the accompanying materials, licensed source archives and empty derived-output locations. Completed study records remain read-only; the scripts enforce this constraint for immutable JSON results. The following sequence prepares conventional results and portfolio analyses for each of SSE50, CSI300 and CSI500.

\begin{lstlisting}[basicstyle=\small\ttfamily,breaklines=true,columns=fullflexible,keepspaces=true]
python scripts/v3_panel.py --universe <universe> --release-manifest <local-analysis-manifest>
\end{lstlisting}
\begin{lstlisting}[basicstyle=\small\ttfamily,breaklines=true,columns=fullflexible,keepspaces=true]
python scripts/v3_baselines.py --universe <universe> --release-manifest <local-analysis-manifest>
\end{lstlisting}
\begin{lstlisting}[basicstyle=\small\ttfamily,breaklines=true,columns=fullflexible,keepspaces=true]
python scripts/v3_analyze.py --universe <universe> --stage main --release-manifest <local-analysis-manifest>
\end{lstlisting}
\begin{lstlisting}[basicstyle=\small\ttfamily,breaklines=true,columns=fullflexible,keepspaces=true]
python scripts/v3_analyze.py --universe SSE50 --stage mechanisms --release-manifest <local-analysis-manifest>
\end{lstlisting}
\begin{lstlisting}[basicstyle=\small\ttfamily,breaklines=true,columns=fullflexible,keepspaces=true]
python scripts/v3_surrogates.py --universe SSE50 --release-manifest <local-analysis-manifest>
\end{lstlisting}
\begin{lstlisting}[basicstyle=\small\ttfamily,breaklines=true,columns=fullflexible,keepspaces=true]
python scripts/v3_mechanisms.py --universe SSE50 --release-manifest <local-analysis-manifest>
\end{lstlisting}
\begin{lstlisting}[basicstyle=\small\ttfamily,breaklines=true,columns=fullflexible,keepspaces=true]
python scripts/v3_risk_analyze.py --universe <universe> --release-manifest <local-analysis-manifest>
\end{lstlisting}
Joint historical comparisons, the planning calculation and final validation use the three universe-specific analyses. The following commands run these stages in the reproduction directory.

\begin{lstlisting}[basicstyle=\small\ttfamily,breaklines=true,columns=fullflexible,keepspaces=true]
python scripts/v3_stats.py --release-manifest <local-analysis-manifest>
\end{lstlisting}
\begin{lstlisting}[basicstyle=\small\ttfamily,breaklines=true,columns=fullflexible,keepspaces=true]
python scripts/v3_mechanisms.py --statistics --release-manifest <local-analysis-manifest>
\end{lstlisting}
\begin{lstlisting}[basicstyle=\small\ttfamily,breaklines=true,columns=fullflexible,keepspaces=true]
python scripts/v3_disjoint.py --release-manifest <local-analysis-manifest>
\end{lstlisting}
\begin{lstlisting}[basicstyle=\small\ttfamily,breaklines=true,columns=fullflexible,keepspaces=true]
python scripts/v3_power.py --release-manifest <local-analysis-manifest>
\end{lstlisting}
\begin{lstlisting}[basicstyle=\small\ttfamily,breaklines=true,columns=fullflexible,keepspaces=true]
python scripts/v3_auxiliary_stats.py --release-manifest <local-analysis-manifest>
\end{lstlisting}
\begin{lstlisting}[basicstyle=\small\ttfamily,breaklines=true,columns=fullflexible,keepspaces=true]
python scripts/v3_validate_engine.py --release-manifest <local-analysis-manifest>
\end{lstlisting}
\begin{lstlisting}[basicstyle=\small\ttfamily,breaklines=true,columns=fullflexible,keepspaces=true]
python scripts/v3_release_validate.py finalize --release-manifest <local-analysis-manifest>
\end{lstlisting}
\begin{lstlisting}[basicstyle=\small\ttfamily,breaklines=true,columns=fullflexible,keepspaces=true]
python -m unittest discover -s scripts -p test_v3_*.py
\end{lstlisting}
The public archive includes 34 test\_\allowbreak{}v3\_\allowbreak{}*.py files. Synthetic-data tests for scoring, statistics, industry history, controlled-rule analysis and replication analysis run from the public files; the complete test discovery command above includes checks that read licensed source panels and therefore requires the local authorized data copy.

Response-based reproduction uses the stored model answers, development-fitted coefficients and controlled inputs. The development fit uses the recorded candidate grid, temporal folds, month weights and source targets. Portfolio analyses use an authorized research copy, a local analysis manifest and empty result locations. Each numerical stage accepts --release-manifest; v3\_\allowbreak{}release\_\allowbreak{}run.py records its inputs, code identities and output hashes. All these analyses operate on the stored answers.

\begin{lstlisting}[basicstyle=\small\ttfamily,breaklines=true,columns=fullflexible,keepspaces=true]
python scripts/render_current_manuscript.py --publication-context analysis/publication_context.json --compiler <Tectonic-executable> --step all
\end{lstlisting}
\begin{lstlisting}[basicstyle=\small\ttfamily,breaklines=true,columns=fullflexible,keepspaces=true]
python scripts/render_current_manuscript.py --release-manifest <local-analysis-manifest> --compiler <Tectonic-executable> --step all
\end{lstlisting}
\begin{lstlisting}[basicstyle=\small\ttfamily,breaklines=true,columns=fullflexible,keepspaces=true]
python scripts/v3_rule_proxies.py evaluate --release-manifest <local-analysis-manifest>
\end{lstlisting}
\begin{lstlisting}[basicstyle=\small\ttfamily,breaklines=true,columns=fullflexible,keepspaces=true]
python scripts/v3_rule_legacy_interventions.py --release-manifest <local-analysis-manifest>
\end{lstlisting}
\begin{lstlisting}[basicstyle=\small\ttfamily,breaklines=true,columns=fullflexible,keepspaces=true]
python scripts/v3_rule_economics.py --release-manifest <local-analysis-manifest>
\end{lstlisting}
\begin{lstlisting}[basicstyle=\small\ttfamily,breaklines=true,columns=fullflexible,keepspaces=true]
python scripts/v3_rule_controlled_analysis.py --release-manifest <local-analysis-manifest>
\end{lstlisting}
scripts/\allowbreak{}render\_\allowbreak{}current\_\allowbreak{}manuscript.py builds the journal and arXiv editions from one content model and the numerical tables identified by the analysis manifest. The public publication-context mode checks the packaged content and source identities and rebuilds figures and document files. The research materials retain the authoring helper, figures, source tables and a file-integrity manifest. Rendering uses the local Springer template and a separately supplied Tectonic executable. Portfolio recomputation additionally requires the authorized source observations and the dated economic-event and board-price records.

\begin{lstlisting}[basicstyle=\small\ttfamily,breaklines=true,columns=fullflexible,keepspaces=true]
python scripts/render_current_manuscript.py --publication-context analysis/publication_context.json --compiler <Tectonic-executable> --step all
\end{lstlisting}
The Astra input-intervention archive is study\_\allowbreak{}v3/\allowbreak{}astra\_\allowbreak{}20260921/\allowbreak{}. Collection records, schedule and audit accompany the stored answers. scripts/\allowbreak{}v3\_\allowbreak{}astra\_\allowbreak{}analysis.py, v3\_\allowbreak{}astra\_\allowbreak{}rules.py, v3\_\allowbreak{}astra\_\allowbreak{}economics.py and v3\_\allowbreak{}astra\_\allowbreak{}controls.py reproduce the numerical analyses. scripts/\allowbreak{}astra\_\allowbreak{}paper\_\allowbreak{}data.py formats the four-model tables, and scripts/\allowbreak{}v3\_\allowbreak{}astra\_\allowbreak{}figures.py generates all five figures.

\begin{lstlisting}[basicstyle=\small\ttfamily,breaklines=true,columns=fullflexible,keepspaces=true]
python scripts/v3_astra_analysis.py --release-manifest <local-analysis-manifest>
python scripts/v3_astra_rules.py --release-manifest <local-analysis-manifest>
python scripts/v3_astra_economics.py --release-manifest <local-analysis-manifest>
python scripts/v3_astra_controls.py --release-manifest <local-analysis-manifest>
python scripts/render_current_manuscript.py --release-manifest <local-analysis-manifest> --compiler <Tectonic-executable> --step all
\end{lstlisting}
The historical input audit covers 636 matrices and 2,772 predictions. The Astra audit covers 2,436 unique valid responses, and the three-model replication covers 3,888 unique valid responses in 18 model-condition groups, 432 inputs and 54 trajectories. Across the retained batches, the research materials contain 10,104 unique valid responses and 1,260 input files, including reused copies across batches. Versioned input, code and output hashes identify the numerical analyses. Portfolio records reconcile trading costs, economic entitlements, daily values and terminal wealth.

\section*{S17 Selection-boundary tie sensitivity}
Selection-boundary sensitivity varies priority within equal-score groups while holding each monthly score vector and candidate set fixed. Fifty fixed integer seeds define deterministic SHA-256 priorities from the seed, stock code, universe and signal date. Each stock-month priority is shared across models and costs. Ascending anonymous identifiers define the reference priority. Every priority uses the same self-financing Market.rebalance engine at 0, 10, 25 and 50 basis points per side. Reference-priority monthly returns agree with the stored returns to within 1e-12. The resulting variation measures portfolio sensitivity to tie priority at fixed model scores.

Table S43 summarizes the primary 10-basis-point cost. Boundary ties are common in CSI 300 and CSI 500, especially for Luna and Terra in CSI 500. Whole-cross-section constant scores occur in four CSI 500 Luna months and seven Terra months. The complete four-cost performance results, monthly net returns, selection overlaps, tie counts and source hashes are in analysis/\allowbreak{}tie\_\allowbreak{}robustness\_\allowbreak{}20260929/\allowbreak{}; the executable analysis scripts are in scripts/\allowbreak{} in the research archive.

\begingroup\small\setlength{\tabcolsep}{3pt}\renewcommand{\arraystretch}{1.17}
\begin{longtable}{@{}>{\raggedright\arraybackslash}p{\dimexpr 0.310345\linewidth-7.44828pt\relax}>{\raggedright\arraybackslash}p{\dimexpr 0.172414\linewidth-4.13793pt\relax}>{\raggedright\arraybackslash}p{\dimexpr 0.172414\linewidth-4.13793pt\relax}>{\raggedright\arraybackslash}p{\dimexpr 0.172414\linewidth-4.13793pt\relax}>{\raggedright\arraybackslash}p{\dimexpr 0.172414\linewidth-4.13793pt\relax}@{}}
\caption{Selection-boundary ties and 10-bp return sensitivity}\label{tab:S43}\\
\toprule
\textbf{Universe} & \textbf{Model} & \textbf{Boundary ties /\allowbreak{} run-months} & \textbf{Archived CAGR (\%)} & \textbf{Seeded CAGR min /\allowbreak{} median /\allowbreak{} max (\%)}\\\midrule\endfirsthead
\multicolumn{5}{l}{\textbf{Table S43 (continued)}}\\\toprule
\textbf{Universe} & \textbf{Model} & \textbf{Boundary ties /\allowbreak{} run-months} & \textbf{Archived CAGR (\%)} & \textbf{Seeded CAGR min /\allowbreak{} median /\allowbreak{} max (\%)}\\\midrule\endhead
\midrule\multicolumn{5}{r}{Continued on next page}\\\endfoot
\bottomrule\endlastfoot
SSE50 & Luna & 6/\allowbreak{}360 & 7.74 & 7.70 /\allowbreak{} 7.75 /\allowbreak{} 7.80\\
SSE50 & Terra & 1/\allowbreak{}360 & 8.51 & 8.51 /\allowbreak{} 8.51 /\allowbreak{} 8.54\\
SSE50 & Sol & 0/\allowbreak{}360 & 9.83 & 9.83 /\allowbreak{} 9.83 /\allowbreak{} 9.83\\
SSE50 & Astra & 0/\allowbreak{}360 & 8.80 & 8.80 /\allowbreak{} 8.80 /\allowbreak{} 8.80\\
CSI300 & Luna & 36/\allowbreak{}72 & 11.17 & 10.25 /\allowbreak{} 11.04 /\allowbreak{} 11.36\\
CSI300 & Terra & 35/\allowbreak{}72 & 10.66 & 10.14 /\allowbreak{} 10.53 /\allowbreak{} 10.78\\
CSI300 & Sol & 27/\allowbreak{}72 & 10.39 & 10.29 /\allowbreak{} 10.61 /\allowbreak{} 10.88\\
CSI300 & Astra & 15/\allowbreak{}72 & 8.17 & 8.15 /\allowbreak{} 8.21 /\allowbreak{} 8.32\\
CSI500 & Luna & 62/\allowbreak{}72 & 11.13 & 10.09 /\allowbreak{} 11.48 /\allowbreak{} 12.34\\
CSI500 & Terra & 60/\allowbreak{}72 & 6.26 & 5.34 /\allowbreak{} 6.14 /\allowbreak{} 7.50\\
CSI500 & Sol & 18/\allowbreak{}72 & 10.72 & 10.12 /\allowbreak{} 10.55 /\allowbreak{} 10.83\\
CSI500 & Astra & 31/\allowbreak{}72 & 10.51 & 10.39 /\allowbreak{} 10.51 /\allowbreak{} 10.57\\
\end{longtable}
\noindent\footnotesize The 50 seeded priorities change only exact score ties. SSE 50 CAGR is the mean across five stored runs for each priority; its tie count covers 360 run-months per model. CSI 300 and CSI 500 each have 72 run-months per model. Ranges across fixed priorities summarize exploratory sensitivity; sampling uncertainty is reported with block-bootstrap intervals. All portfolio paths use the same execution engine and 72-month calendar.\par\endgroup
At the primary 10-basis-point cost and six-month block length, Table S44 compares the model-minus-frozen-linear return differences under the archived tie priority with the range across the 50 alternative priorities. Each priority has its own twelve-comparison Holm correction. No alternative priority yields an adjusted p-value below 0.05 in this primary comparison family. Across all four costs at six-month blocks, the only alternative priority findings occur for CSI 500 Terra at 50 basis points: six of 50 seeded priorities have adjusted p-values of 0.036-0.047, versus 0.077 under the archived priority. These values describe sensitivity to the fixed-score tie choice; they do not add independent market observations.

\begingroup\small\setlength{\tabcolsep}{3pt}\renewcommand{\arraystretch}{1.17}
\begin{longtable}{@{}>{\raggedright\arraybackslash}p{\dimexpr 0.264706\linewidth-7.94118pt\relax}>{\raggedright\arraybackslash}p{\dimexpr 0.147059\linewidth-4.41176pt\relax}>{\raggedright\arraybackslash}p{\dimexpr 0.147059\linewidth-4.41176pt\relax}>{\raggedright\arraybackslash}p{\dimexpr 0.147059\linewidth-4.41176pt\relax}>{\raggedright\arraybackslash}p{\dimexpr 0.147059\linewidth-4.41176pt\relax}>{\raggedright\arraybackslash}p{\dimexpr 0.147059\linewidth-4.41176pt\relax}@{}}
\caption{Primary model-minus-rule inference under alternative tie priorities}\label{tab:S44}\\
\toprule
\textbf{Universe} & \textbf{Model} & \textbf{Archived difference (bp/\allowbreak{}month)} & \textbf{Seeded difference range (bp/\allowbreak{}month)} & \textbf{Archived Holm p} & \textbf{Seeded Holm p range}\\\midrule\endfirsthead
\multicolumn{6}{l}{\textbf{Table S44 (continued)}}\\\toprule
\textbf{Universe} & \textbf{Model} & \textbf{Archived difference (bp/\allowbreak{}month)} & \textbf{Seeded difference range (bp/\allowbreak{}month)} & \textbf{Archived Holm p} & \textbf{Seeded Holm p range}\\\midrule\endhead
\midrule\multicolumn{6}{r}{Continued on next page}\\\endfoot
\bottomrule\endlastfoot
SSE50 & Luna & -19.0 & -19.2 to -18.5 & 1.000 & 1.000 to 1.000\\
SSE50 & Terra & -20.1 & -20.1 to -19.9 & 0.382 & 0.382 to 0.440\\
SSE50 & Sol & -1.5 & -1.5 to -1.5 & 1.000 & 1.000 to 1.000\\
SSE50 & Astra & -6.7 & -6.7 to -6.7 & 1.000 & 1.000 to 1.000\\
CSI300 & Luna & 7.4 & -0.3 to 8.9 & 1.000 & 1.000 to 1.000\\
CSI300 & Terra & 2.7 & -1.3 to 3.6 & 1.000 & 1.000 to 1.000\\
CSI300 & Sol & 9.0 & 8.2 to 12.7 & 1.000 & 1.000 to 1.000\\
CSI300 & Astra & -6.1 & -6.3 to -4.9 & 1.000 & 1.000 to 1.000\\
CSI500 & Luna & 10.6 & 2.4 to 20.6 & 1.000 & 1.000 to 1.000\\
CSI500 & Terra & -34.0 & -41.5 to -24.2 & 0.342 & 0.164 to 1.000\\
CSI500 & Sol & 1.5 & -2.9 to 2.5 & 1.000 & 1.000 to 1.000\\
CSI500 & Astra & -9.2 & -10.1 to -8.7 & 1.000 & 1.000 to 1.000\\
\end{longtable}
\noindent\footnotesize Mean difference is model minus frozen-linear net return at 10 basis points per side. All tests use six-month circular blocks and the same twelve model-universe comparisons for Holm adjustment at each fixed priority. The ranges across 50 priorities summarize exploratory implementation sensitivity; the displayed tests use the stated block-bootstrap inference.\par\endgroup
Table S45 reports the higher-cost, twelve-month-block comparisons. The reference priority has three within-slice Terra findings. Across 50 alternative priorities, these findings occur in 26, 50 and 40 cases, respectively. CSI 500 Terra at 25 basis points crosses the same threshold under 17 priorities. All other model-universe comparisons at these two costs have zero seeded findings. Each priority implements the same score vectors and market months, so the counts describe sensitivity to the tie rule.

\begingroup\small\setlength{\tabcolsep}{3pt}\renewcommand{\arraystretch}{1.17}
\begin{longtable}{@{}>{\raggedright\arraybackslash}p{\dimexpr 0.264706\linewidth-7.94118pt\relax}>{\raggedright\arraybackslash}p{\dimexpr 0.147059\linewidth-4.41176pt\relax}>{\raggedright\arraybackslash}p{\dimexpr 0.147059\linewidth-4.41176pt\relax}>{\raggedright\arraybackslash}p{\dimexpr 0.147059\linewidth-4.41176pt\relax}>{\raggedright\arraybackslash}p{\dimexpr 0.147059\linewidth-4.41176pt\relax}>{\raggedright\arraybackslash}p{\dimexpr 0.147059\linewidth-4.41176pt\relax}@{}}
\caption{Terra high-cost inference under alternative tie priorities}\label{tab:S45}\\
\toprule
\textbf{Universe} & \textbf{Cost (bp/\allowbreak{}side)} & \textbf{Archived difference (bp/\allowbreak{}month)} & \textbf{Archived Holm p} & \textbf{Seeded Holm p range} & \textbf{Seeded findings /\allowbreak{} 50}\\\midrule\endfirsthead
\multicolumn{6}{l}{\textbf{Table S45 (continued)}}\\\toprule
\textbf{Universe} & \textbf{Cost (bp/\allowbreak{}side)} & \textbf{Archived difference (bp/\allowbreak{}month)} & \textbf{Archived Holm p} & \textbf{Seeded Holm p range} & \textbf{Seeded findings /\allowbreak{} 50}\\\midrule\endhead
\midrule\multicolumn{6}{r}{Continued on next page}\\\endfoot
\bottomrule\endlastfoot
SSE50 & 25 & -22.6 & 0.0468 & 0.0429 to 0.0552 & 26/\allowbreak{}50\\
SSE50 & 50 & -26.7 & 0.0084 & 0.0077 to 0.0084 & 50/\allowbreak{}50\\
CSI300 & 25 & 0.2 & 1.0000 & 1.0000 to 1.0000 & 0/\allowbreak{}50\\
CSI300 & 50 & -3.8 & 1.0000 & 1.0000 to 1.0000 & 0/\allowbreak{}50\\
CSI500 & 25 & -39.0 & 0.0616 & 0.0060 to 0.5329 & 17/\allowbreak{}50\\
CSI500 & 50 & -47.3 & 0.0154 & 0.0012 to 0.1970 & 40/\allowbreak{}50\\
\end{longtable}
\noindent\footnotesize Model minus frozen-linear monthly net return with twelve-month circular blocks. Holm adjustment uses the twelve model-universe tests within each cost and priority. The table displays all three Terra universes at both higher costs; no other model has a seeded finding. These sensitivity slices are exploratory; the original full four-cost-by-three-block-length grid has no adjusted finding across all 144 comparisons.\par\endgroup
The following commands reproduce the tie analysis in an authorized research copy using the licensed market panels, daily trading data and other sources recorded by source\_\allowbreak{}hashes.json. The first command creates the output directory, which must be absent at that point.

\begin{lstlisting}[basicstyle=\small\ttfamily,breaklines=true,columns=fullflexible,keepspaces=true]
python -B scripts/v3_tie_robustness.py --universes CSI500 CSI300 SSE50 --release-manifest <local-analysis-manifest>
python -B scripts/v3_tie_robustness_inference.py --release-manifest <local-analysis-manifest>
python -B scripts/v3_tie_robustness_high_cost_inference.py --release-manifest <local-analysis-manifest>
\end{lstlisting}

\end{document}